\documentclass[final,3p,times]{elsarticle}

\usepackage{epsfig}
\usepackage{amssymb}
\usepackage{amsmath}
\usepackage{subfigure}
\usepackage{dcolumn}
\usepackage{array} 
\usepackage{longtable} 
\usepackage[colorlinks]{hyperref}
\usepackage[usenames,dvipsnames]{color}
\hypersetup{
     breaklinks=true,
    pdfstartview={FitH},    
    colorlinks=true,       
    linkcolor=blue,          
    citecolor=red,        
    filecolor=magenta,      
    urlcolor=blue,           
    anchorcolor=green,      
    linktocpage=true
}
\usepackage{orcidlink}

\newcommand{\Mpl}{M_{\textrm{Pl}}}
\renewcommand{\(}{\left(}
\renewcommand{\)}{\right)}

\newcommand{\nn}{\nonumber}
\def\al{\alpha}

\def\gam{\gamma}
\def\om{\omega}
\def\Om{\Omega}
\def\sig{\sigma}
\def\Lam{\Lambda}
\def\lam{\lambda}

\def\S{\mathcal{S}}
\def\P{\mathcal{P}}

\def\B{\mathcal{B}}

\def\D{\mathcal{D}}

\def\K{\mathcal{K}}
\def\Q{\mathcal{Q}}

\def\del{\delta}

\def\doi{http://doi.org}

 \def\t{\tilde}
 \def\e{\mathrm{e}}
\def\r{\mathrm{r}}
\def\g{\mathrm{g}}

\def\m{\mathrm{m}}

\def\s{\mathrm{s}}
\def\d{\mathrm{d}}

\def\vck{\vec k}

\allowdisplaybreaks

\journal{Physics of the Dark Universe}

\begin{document}

\begin{frontmatter}



\title{Quintessential early dark energy} 


\author[1]{Sk. Sohail} 
\ead{sksohail0402@gmail.com}
\author[1]{Sonej Alam}
\ead{sonejalam36@gmail.com}
\author[1,2]{Shiriny Akthar}
\ead{shirinyakthar@gmail.com}
\author[1]{Md. Wali Hossain}
\ead{mhossain@jmi.ac.in}

\affiliation[1]{organization={Department of Physics, Jamia Millia Islamia},
            city={New Delhi},
            postcode={110025}, 
            country={India}}

\affiliation[2]{organization={Department of Astronomy, Astrophysics \& Space Engineering, Indian Institute of Technology},
            city={Indore},
            postcode={453552}, 
            country={India}}

\begin{abstract}
We introduce a unified model of early and late dark energy. We call it {\it quintessential early dark energy} model where early and late dark energy are explained by a single scalar field {\it i.e.}, two different energy scales are related by a single scalar field potential. To achieve this we introduce the modified steep exponential potential, which is chosen phenomenologically. This potential has a hilltop nature during the early time which consists of a flat region followed by a steep region. This nature of the potential plays a crucial role in achieving early dark energy solution. During recent time, the potential can almost mimic the cosmological constant which can result into late time acceleration. But, at the perturbation level the potential shows significant difference with the $\Lambda$CDM model. We also constrain and compare the models for steep exponential, modified steep exponential, axionlike and power law potentials by using the available background cosmological data from CMB, BAO (including DESI DR1 2024), supernovae (Pantheon$+$, DESY5 and Union3) and Hubble parameter measurements. Even after the presence of required EDE solution in all four potentials we don't get any significant improvement in the value of $H_0$. The maximum improvement we get in the present value of Hubble parameter compared to the standard $\Lambda$CDM model is for the axionlike potential. For other potentials the constraints are similar to the $\Lambda$CDM model. We also see that the data prefers $\Lambda$CDM model over the considered scalar field models at least for the data combinations with Pantheon$+$ and Union3.
\end{abstract}




\end{frontmatter}



\section{Introduction}
\label{sec:intro}
The simplest explanation of late time acceleration \cite{SupernovaSearchTeam:1998fmf,SupernovaCosmologyProject:1998vns,Planck:2018vyg,Brout:2022vxf,Scolnic:2021amr} is a cosmological constant (CC), $\Lambda$ which has a constant equation of state (EoS). $\Lambda$ is not only the simplest but also most favoured candidate especially by the observations on cosmic microwave background (CMB) \cite{Planck:2013pxb,Planck:2018vyg}. However, theoretically, the $\Lambda$ is plagued with two measure issues, the fine-tuning
problem \cite{Martin:2012bt} and the cosmic coincidence problem \cite{Zlatev:1998tr, Steinhardt:1999nw}. Apart form these theoretical problems, interestingly, the $\Lambda$ is also in trouble due to some observational results on the local measurement of the present value of the Hubble parameter $H_0$ \cite{Riess:2021jrx}. Recently observed $\sim 5\sig$ tension between the local measurement of $H_0$ by the SH0ES team ($H_0=73.04\pm1.04~{\rm km~sec^{-1}~Mpc^{-1}}$) \cite{Riess:2021jrx} and its constraint coming from CMB observations assuming the standard $\Lambda$CDM model ($H_0=67.4\pm 0.5~{\rm km~sec^{-1}~Mpc^{-1}}$) \cite{Planck:2018vyg} puts the standard $\Lambda$CDM model in challenge. This discrepancy in the values of $H_0$ coming from different observations is known as the Hubble tension \cite{Kamionkowski:2022pkx}. The discrepancy between the two measurements is either due to some systematic errors or some new physics is there \cite{Kamionkowski:2022pkx,Freedman:2023jcz,Bernal:2016gxb,Knox:2019rjx}. If there is new physics behind this tension then it is very clear that we have to go beyond $\Lambda$CDM. One can do that by either making the dark energy dynamical \cite{Copeland:2006wr,Sahni:1999gb} or modify the gravity sector \cite{Clifton:2011jh}. 

The tension with the $\Lambda$CDM model also arises after the DESI DR1 2024 \cite{DESI:2024mwx,DESI:2024uvr,DESI:2024lzq} data release of baryon acoustic oscillations (BAO) observations as this observation points towards a dynamical dark energy solution \cite{DESI:2024mwx,DESI:2024aqx,Giare:2024smz,Berghaus:2024kra,Qu:2024lpx,Wang:2024dka,Giare:2024gpk,Shlivko:2024llw,Ye:2024ywg,Ramadan:2024kmn,Jiang:2024xnu}. When DESI DR1 BAO data are combined with the measurements of the CMB anisotropy \cite{Planck:2018vyg} and type Ia Supernovae (SNIa) the discrepancy with the standard $\Lambda$CDM model becomes $2.5\sig$, $3.5\sig$ and $3.9\sig$ for Pantheon$+$\cite{Brout:2022vxf,Scolnic:2021amr}, Union3 \cite{Rubin:2023ovl} and DESY5 \cite{DES:2024tys} data sets respectively. This motivates us to look for dark energy models which are not only beyond $\Lambda$CDM but dynamical in nature.  To make dark energy dynamical one of the simplest way is to introduce a minimally coupled canonical scalar field $\phi$ with a potential \cite{Copeland:2006wr}. In a cosmological background, the scalar field EoS $w_\phi$ can vary from $-1$ to 1. $w_\phi=-1$ corresponds to the potential energy domination while $w_\phi=1$ represents the kinetic energy domination. The energy density of the scalar field $\rho_\phi\sim \e^{-3\int (1+w_\phi)\d \ln a}$. So, when $w_\phi=-1$ the energy density $\rho_\phi$ is constant and it varies as $a^{-6}$ for $w_\phi=1$. So, we can observe wide variation in the scalar field dynamics in a cosmological background. In fact, we can classify the dynamics in three classes, scaling-freezing \cite{Copeland:1997et,Barreiro:1999zs,Sahni:1999qe}, tracking \cite{Steinhardt:1999nw,Zlatev:1998tr} and thawing \cite{Scherrer:2007pu}. In the scaling-freezing and tracker dynamics the scalar field is frozen in the past due to large Hubble damping coming from the background energy density. During the frozen period the scalar field energy density increases and when it becomes comparable to the background the scalar field starts rolling down the potential. Now, if the slope of the potential is sufficiently large then depending upon the nature of the potential the scalar field energy density can either scale the background energy density (for scaling-freezing) or decay a little slower than the background energy density (for tracker). For tracker, since the decay of scalar field energy density is slower than the background the scalar field eventually takes over matter during the recent past which may give rise to late time acceleration \cite{Hossain:2023lxs}. For scaling-freezing the scalar field scales the background after the frozen period and then takes over matter in the recent past \cite{Barreiro:1999zs}. For both the dynamics we can get rid of initial condition dependence as the late time dynamics is an attractor solution. On the other hand the thawing dynamics is very sensitive to the initial conditions like CC. In this dynamics the scalar field is frozen from the past behaving like CC and start to thaw from the recent past giving rise to deviation from CC \cite{Scherrer:2007pu}.

When the scalar field is frozen at a higher redshift (post-inflationary) and its density parameter ($\Omega_\phi=\rho_\phi/3H^2\Mpl^2$) eventually becomes small but finite ($\sim 0.1$) the scalar field will behave like a CC with finite contribution in the total energy density at a higher redshift. This is known as the early dark energy (EDE) \cite{Kamionkowski:2022pkx, Poulin:2023lkg}. The EDE solution during the decoupling era can increase the present value of the Hubble parameter $H_0$ which can alleviate the Hubble tension \cite{Riess:2021jrx,Kamionkowski:2022pkx,Perivolaropoulos:2021jda} to some extent \cite{Poulin:2018cxd,Poulin:2018dzj,Poulin:2023lkg,Smith:2019ihp,Agrawal:2019lmo,Kamionkowski:2022pkx,Kodama:2023fjq,Lin:2019qug,Niedermann:2019olb,Niedermann:2020dwg,Berghaus:2019cls,Berghaus:2022cwf,Karwal:2021vpk}. EDE scenario can also posses some issues \cite{Vagnozzi:2023nrq,Vagnozzi:2021gjh,Vagnozzi:2019ezj}. EDE, around the decoupling era, increases the sound horizon at the decoupling. As the ratio of sound horizon and the comoving angular diameter distance is a well measured quantity from the CMB observations the ratio has to be kept constant. This can be done only if we increase the value of $H_0$ as the comoving angular diameter distance at the decoupling remains unchanged as the late universe is not affected by EDE as the EDE should decay very fast just after reaching its maximum value so that it does not affect the late time history of the universe. So, in terms of the scalar field dynamics, to achieve an EDE solution, the scalar field has to be frozen at a higher redshift until its energy density becomes of the same order of the background energy density and once it reaches this value the scalar field should roll down the potential very fast so that its kinetic energy dominates over its potential energy. This nature can be obtained in a steep potential or a hilltop potential in which the hilltop region is being followed by a steep region. Oscillatory potentials {\it e.g.}, axionlike potential $\sim (1-\cos \phi)^n$ \cite{Marsh:2015xka,Poulin:2018cxd,Poulin:2018dzj,Poulin:2023lkg,Smith:2019ihp,Hlozek:2014lca,Efstathiou:2023fbn} and power law potential $\sim \phi^{2n}$ \cite{Ratra:1987rm,Agrawal:2019lmo}, where $n$ is a constant, can successfully produce EDE which can also increase the value of $H_0$. In this paper, we study the EDE scenario in non oscillatory steep potential. Particularly we consider the steep exponential (SE) potential of the form $V(\phi)\sim\e^{-\mu \phi^n}$, where $\mu$ and $n$ are constants, introduced in Ref.~\cite{Geng:2015fla}. This potential produces viable inflation and interesting postinflationary dynamics \cite{Geng:2015fla}. Due to the steep nature, especially for $n>1$, during the postinflationary dynamics the scalar field can be frozen multiple times and the frozen period is followed by kinetic energy dominated regime. This nature of the dynamics is particularly important for EDE. We study the EDE solutions and its consequences on the value of $H_0$ for the SE potential. 

As we have discussed about the importance of the steep nature of the potential for the EDE solution it also makes the scalar field energy density decay very fast or scale the background \cite{Hossain:2023lxs,Geng:2015fla}. Both the cases do not give rise to late time acceleration. Generally, if we have EDE solution then the late time dark energy (LDE) should be achieved either by adding a CC or an another potential. So, it would be interesting if we can unify the EDE and LDE with a single scalar field potential. We explore this possibility in this paper and we modify the SE potential such a way so that it gives EDE at higher redshift and LDE during the recent time. The modified steep exponential (MSE) potential, chosen phenomenologically, behaves like the SE potential during early times and from recent past it behaves almost like a CC and thereby results LDE. While producing LDE the scalar field rolls slowly. This slowly rolling scalar field is known as quintessence \cite{Ratra:1987rm,Wetterich:1987fk,Wetterich:1987fm}. Following the concept of quintessential inflation \cite{Peebles:1998qn,WaliHossain:2014usl,deHaro:2021swo,Bettoni:2021qfs} in which inflation and LDE are described through a single scalar field we call the scenario under consideration where a single scalar field results both EDE and LDE as {\it Quintessential EDE}. It is important to note here that in the scenarios like quintessential inflation or quintessential EDE there should be one energy scale associated with the scalar field potential which can be related to another energy scale and thereby resulting the unification of the two energy scales. If any potential has two energy scales \cite{Ramadan:2023ivw,Chowdhury:2023opo,MohseniSadjadi:2024ejb} then it is trivial to represent the cosmic history of two different scales with that potential but it is not unification of two energy scales. In this regard, it is also important to mention that the potentials which can give rise to scaling solutions may not be able to give sufficient amount of EDE followed by a small amount of scalar field energy density during matter dominated era and therefor may not alleviate the $H_0$ tension for a canonical scalar field \cite{Ramadan:2023ivw,Copeland:2023zqz}.

We introduce the background cosmological equations in Sec.~\ref{sec:background}. In Sec.~\ref{sec:SE} we study the cosmological dynamics in SE potential. The MSE potential has been introduced in Sec.~\ref{sec:MSE}. The discussion on EDE and how it can improve the value of $H_0$ has been done in Sec.~\ref{sec:EDE}. The growth of matter fluctuations has been discussed in Sec.~\ref{sec:PT}. The study of observational constraints has been done in Sec.~\ref{sec:obs}. We summarise and conclude the paper in Sec.~\ref{sec:conclusion}.

\section{Background equations}
\label{sec:background}
We consider a minimally coupled canonical scalar field with the following action
\begin{align}
\S=\int \d^4x\sqrt{-\g}\Bigl [\frac{\Mpl^2}{2} R-\frac{1}{2}\partial_\mu\phi\partial^\mu\phi - V(\phi) \Bigr]+ \S_\m+\S_\r \, ,
\label{eq:action}
\end{align}
where $\Mpl=1/\sqrt{8\pi G}$ is the reduced Planck mass and $V(\phi)$ is the potential of the field. $\S_\r$ and $\S_\m$ are the actions for radiation and matter respectively.

Varying the action (\ref{eq:action}) with respect to (w.r.t.) the metric $g_{\mu\nu}$ gives the Einstein's field equation
\begin{align}
\Mpl^2 G_{\mu\nu}= T_{(\m)\mu\nu}+T_{(\r)\mu\nu}+T_{(\phi)\mu\nu} \, ,
\label{eq:ee}
\end{align}
where
\begin{align}
T_{(\phi)\mu\nu}=& \phi_{;\mu}\phi_{;\nu}-\frac{1}{2}\g_{\mu\nu}(\nabla\phi)^2 -\g_{\mu\nu}V(\phi) \, .
\label{eq:emt_phi}
\end{align}
The equation of motion of the scalar field can be calculated by varying the action (\ref{eq:action}) w.r.t. the scalar field $\phi$ an it is given by  
\begin{align}
& \Box \phi-V_\phi(\phi)=0 \, ,
\label{eq:eom_phi}
\end{align}
where subscript $\phi$ denotes the derivative wrt $\phi$.

In flat  Friedmann--Lema\^itre--Robertson--Walker (FLRW) metric, given by
\begin{align}
\label{metric0}
 ds^2 = -dt^2 +a(t)^2\delta_{ij} dx^idx^j ~,
\end{align}
where $a(t)$ is the scale factor, the Friedmann equations are given by
\begin{eqnarray}
 3H^2\Mpl^2 &=& \rho_\m+\rho_\r+\rho_\phi \,  
 \label{eq:Friedmann1}\\
 \(2\dot H+3H^2\)\Mpl^2&=&-\frac{1}{3}\rho_\r-p_\phi \, ,
 \label{eq:Friedmann2}
\end{eqnarray}
where the scalar field energy density $\rho_\phi$ and pressure $p_\phi$ are given by
\begin{eqnarray}
    \rho_\phi &=& \frac{1}{2}\dot\phi^2 + V(\phi) \\
    p_\phi &=& \frac{1}{2}\dot\phi^2 - V(\phi) \, .
\end{eqnarray}
The equation of motion of the scalar field is given by
\begin{equation}
 \label{scalareom1}
 \ddot\phi+3H\dot\phi+\frac{\d V}{\d\phi}=0 \, .
 \end{equation}

Effective equation of state (EoS) and the EoS of the scalar field are given by
\begin{eqnarray}
w_{\text{eff}}&=& -\(1+\frac{2}{3}\frac{\dot H}{H^2}\)  \; , 
\label{eq:weff}\\
w_{\phi} &=& \frac{\frac{1}{2}\dot\phi^2-V(\phi)}{\frac{1}{2}\dot\phi^2+V(\phi)}  \; .
\label{eq:wphi}
\end{eqnarray}

We define the following two functions 
\begin{eqnarray}
    \lambda &=& -\Mpl\frac{V'(\phi)}{V} \; , 
    \label{eq:lam}\\
    \Gamma &=& \frac{V''(\phi)V(\phi)}{V'(\phi)^2} \; .
    \label{eq:gam}
\end{eqnarray}
While the function $\lam$ signifies the slope of the potential the function $\Gamma$ represents the nature of the potential, {\it e.g.}, $\Gamma=1$ for an exponential potential of constant slope $\lam$. The nature of these functions determines the scalar field dynamics \cite{Bahamonde:2017ize,Hossain:2023lxs}.

\section{Steep Exponential Potential}
\label{sec:SE}

We consider the following steep exponential potentials \cite{Geng:2015fla}
\begin{eqnarray}
    V(\phi)&=&V_0\e^{-\mu(\phi/\Mpl)^n} \, ,
    \label{eq:pot}
\end{eqnarray}
where $V_0$, $\lam$ and $n$ are constants. Potential~\eqref{eq:pot} is a steep exponential (SE) potential. Introduced in \cite{Geng:2015fla}, SE is well studied \cite{Geng:2017mic,Ahmad:2017itq,Agarwal:2017wxo,Das:2019ixt,Lima:2019yyv,Rezazadeh:2015dia,Das:2020xmh,Das:2023rat,Das:2023nmm} especially in the field of quintessential inflation models \cite{WaliHossain:2014usl} where inflation and late time acceleration are achieved by a single scalar field. Since the potential~\eqref{eq:pot} is very steep we can't get late time acceleration only with the potential which eventually leads to a scaling solution \cite{Geng:2015fla}. To achieve late time acceleration we need to consider an interaction of the scalar field with the massive neutrino which plays a crucial roll to achieve late time acceleration \cite{Geng:2015fla}.

\begin{figure}[t]
\centering
\includegraphics[scale=.98]{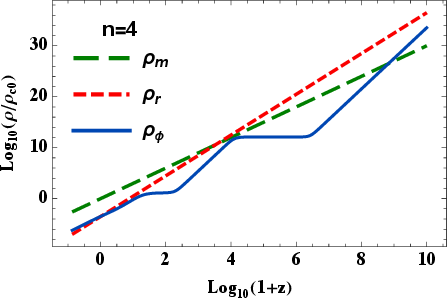}
\includegraphics[scale=.98]{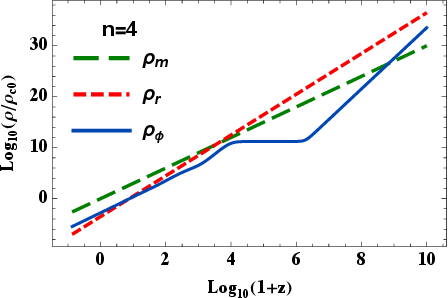}
\caption{In both the figures the solid blue, dashed green and dotted red lines represent energy densities, normalised by present critical density, of scalar field, matter and radiation respectively. Left: Initial conditions are $\phi_i=1.4$, $\phi_i'=0.15$ with parameters $n=4$, $\mu=0.606$ and $V_0=5.58\times 10^{12}$. Right: Initial conditions are $\phi_i=6$, $\phi_i'=0.15$ with parameters $n=4$, $\mu=0.012$ and $V_0=1.8\times 10^{17}$.}
\label{fig:SE_scaling}
\end{figure}

For the SE potential~\eqref{eq:pot} the functions $\lam$ and $\Gamma$ are given by \cite{Geng:2015fla}
\begin{eqnarray}
    \lam &=& \mu n \(\frac{\phi}{\Mpl}\)^{n-1} \, ,
    \label{eq:lam_SE} \\
    \Gamma &=& 1-\frac{n-1}{\mu n}\frac{\Mpl^n}{\phi^n} \, .
    \label{eq:gam_SE}
\end{eqnarray}
For $n=1$ the potential~\eqref{eq:pot} reduces to the exponential potential with $\lam=$constant and $\Gamma=1$. For $n>1$ and $\phi>0$ we see, form the Eqs.~\eqref{eq:lam_SE} and \eqref{eq:gam_SE}, that $\lam$ is ever increasing but $\Gamma$ approaches $1$ as $\phi$ increases which makes the potential effectively an exponential potential with very large slope. These natures of $\lam$ and $\Gamma$ for the SE potential gives an interesting dynamics of the scalar field. For $n>1$ and around $\phi=0$ the potential is very flat and we can have slow roll in that region. This flat region is followed by an steep region for $\phi>0$ where the scalar field will initially roll fast giving rise to $w_\phi=1$ and $\rho_\phi\sim a^{-6}$. Because of this sharp fall in the energy density the scalar field will experience a large Hubble damping due to the background and the scalar field will frozen giving rise to cosmological constant like behaviour in the scalar field dynamics. During this frozen period $\rho_\phi$ will increase and when it becomes comparable to the background energy density the scalar field again starts rolling. Now, after exiting from the frozen period the dynamics of the scalar field will be determined by the value of the field $\phi$. If $\phi$ is sufficiently large then $\Gamma$ will be close to $1$ which will lead to a scaling behaviour and if $\phi$ is not large enough then due to the steep nature of the potential the scalar field will repeat the previous dynamics \cite{Geng:2015fla} which has been depicted in the Fig.~\ref{fig:SE_scaling}. The dynamics shown in the left figure of Fig.~\ref{fig:SE_scaling} can be useful to achieve early dark energy. For early dark energy the scalar field should be frozen before the matter radiation equality and start evolving from around matter radiation equality. This behaviour will inject a small fraction of dark energy during the matter radiation equality which can alleviate the Hubble tension to some extent \cite{Kamionkowski:2022pkx,Poulin:2023lkg}. We shall discuss this in Sec.~\ref{sec:EDE}.

$\Gamma=1$ and $\lam>\sqrt{3}$ gives rise to scaling behaviour \cite{Copeland:1997et} in the scalar field dynamics where the scalar field energy density follows the background energy density. Scaling behaviour can not give rise to late time acceleration. So if a potential has large slope $\lam$ and $\Gamma=1$ then it will give rise to scaling behaviour but late time acceleration can not be achieved. Now, to get the both scaling behaviour and the late time acceleration we have to exit from the scaling behaviour in the recent past {\it i.e.}, $\Gamma$ should go away from 1 and $\lam$ should approach zero. One way to achieve this is to add a cosmological constant in the SE potential~\eqref{eq:pot}. But here we do not want to add a cosmological constant explicitly and rather we modify the SE potential such a way so that it gives rise to a cosmological constant like behaviour from the recent past. We have discussed this in the next section.

\section{Modified Steep Exponential Potential}
\label{sec:MSE}

\begin{figure}[ht]
\centering
\includegraphics[scale=1.1]{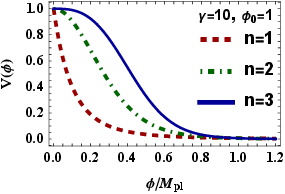} ~~~~~
\includegraphics[scale=1.1]{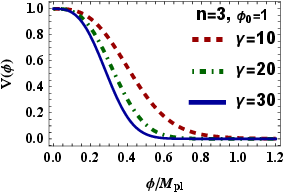}
\caption{Nature of the MSE potential~\eqref{eq:potMSE} for different values of $n$ and $\gam$.}
\label{fig:pot}
\end{figure}

As we have discussed in the previous section that the SE potential~\eqref{eq:pot} can give rise to scaling behaviour but not late time acceleration. The steep nature of the SE potential can be useful to achieve EDE solutions. Now,to get late time acceleration from a steep exponential potential while retaining the properties of scaling-like behaviour and possible EDE solutions we consider the following modified steep exponential potential (MSE)
\begin{eqnarray}
    V(\phi)&=& V_0\e^{-\gamma\frac{(\phi/\Mpl)^n}{\phi_0+(\phi/Mpl)^n}} \, ,
    \label{eq:potMSE}
\end{eqnarray}
where $V_0$, $\phi_0$, $\gamma$ and $n$ are constants. For both the potentials~\eqref{eq:pot} and \eqref{eq:potMSE}, $ V_0$ fixes the energy scale of the potential. Increasing the values of $n$ increases the flatness of the potentials around $\phi=0$. It also increases the steepness of the potential. $\gam$ is related to the slope of the potential and changes the steepness. We should note that for $\phi/\Mpl\ll\phi_0$ the MSE potential~\eqref{eq:potMSE} reduces to the potential~\eqref{eq:pot} with $\mu=\gam/\phi_0$. But in the potential~\eqref{eq:potMSE} we do not need to add any CC or an interaction as for $\phi/\Mpl\gg\phi_0$ the potential will be very flat and can give late time acceleration. Dependence of the potential~\eqref{eq:potMSE} on the $n$ and $\gam$ has been shown in Fig.~\ref{fig:pot}.

\subsection{Relating different energy scales by MSE: Unifying EDE and LDE}
The MSE potential~\eqref{eq:potMSE} can be useful to relate two different energy scales with a single canonical scalar field without considering any interaction as we have mentioned above. For small positive field values {\it i.e.}, $\phi/\Mpl\ll\phi_0$ the potential~\eqref{eq:potMSE} becomes
\begin{eqnarray}
    V(\phi)&=&V_0\e^{-\mu (\phi/\Mpl)^n} \, ,
    \label{eq:pot1}\\
    \mu&=&\gam/\phi_0 \, ,
\end{eqnarray}
which has the same form as the SE potential~\eqref{eq:pot}. For large field values let's write the MSE potential~\eqref{eq:potMSE} in the following form 
\begin{eqnarray}
    V(\phi)&=& V_0\e^{-\gam(1+\phi_0/(\phi/\Mpl)^n)^{-1}} \, .
    \label{eq:potlargephi}
\end{eqnarray}
When $\phi/\Mpl$ is sufficiently larger than $\phi_0^{1/n}$, {\it i.e.}, $\phi/\Mpl\gg\phi_0^{1/n}$, we have
\begin{eqnarray}
    V_0\e^{-\gam}=V_1 \, ,
    \label{eq:DE_Scale}
\end{eqnarray}
{\it i.e}, the MSE potential~\eqref{eq:potMSE} becomes a constant potential and one can calculate $V_1$ once we choose $V_0$ and $\gam$ and thereby we relate two energy scales $V_0$ and $V_1$ by a single scalar field. So, the potential~\eqref{eq:potMSE} can give late time acceleration without considering an interaction between scalar field and massive neutrinos \cite{Geng:2015fla} {\it i.e.}, we can achieve the unified scenarios only from a canonical scalar field with the potential~\eqref{eq:potMSE}. The last equation gives us
\begin{eqnarray}
    \gam=-\ln\frac{V_1}{V_0}\, .
    \label{eq:lam_de}
\end{eqnarray}

If $\phi/\Mpl$ is not sufficiently larger than $\phi_0^{1/n}$ at the energy scale $V_1$ then, in general, the field value $\phi_1/\Mpl$ at the energy scale $V_1$ is
\begin{eqnarray}
    \phi_1/\Mpl&=&\phi_0^{1/n}\(\frac{\gam}{\ln(V_0/V_1)}-1\)^{-1/n} \, .
    \label{eq:lam_bound}
\end{eqnarray}
While the Eq.\eqref{eq:lam_de} gives an estimate about the value of $\gam$ needed to relate $V_1$ with $V_0$ the Eq.~\eqref{eq:lam_bound} gives the bound on the value of $\gam$ which is given by
\begin{eqnarray}
    \gam>\ln(V_0/V_1)\, .
\end{eqnarray}

Since for large field values the potential becomes very flat it can behave like a CC and by choosing $V_1=\rho_{\rm DE 0}$, where $\rho_{\rm DE 0}$ is the dark energy density at present, we can relate the dark energy scale with some higher energy scale. {\it E.g.}, if we are interested to relate dark energy scale with that of the inflation scale then $V_0^{1/4}\sim 10^{16}$GeV which gives us $\gam\approx 255$. Similarly we can choose $V_0^{1/4}\sim10^{-10}$GeV which is around matter-radiation equality we get $\gam\approx 18$ which basically unifies the early and late dark energy. Important thing to notice is that the values of the parameter $\gam$ are natural and we do  not have to deal with the unnatural fine tuning of the dark energy scale. 

\subsection{Background dynamics}

\begin{figure}[t]
\centering
\includegraphics[scale=.95]{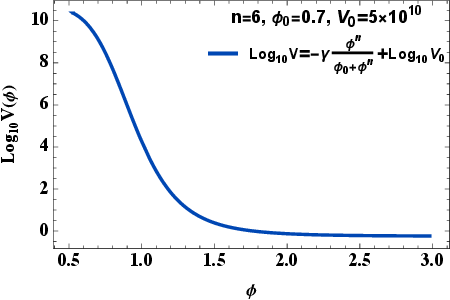}
\caption{Numerically evolved MSE potential~\eqref{eq:potMSE} for  $V_0=5\times 10^{10}\rho_{\rm c 0}$, $n=6$, $\phi_0=0.7$ and $\Om_{\m0}=0.3$. Initial conditions are $\phi_i=\phi_0-0.1 \phi_0$ and $\phi_i'=0.1$.}
\label{fig:MSEpot}
\end{figure}

\begin{figure}[t]
\centering
\includegraphics[scale=.93]{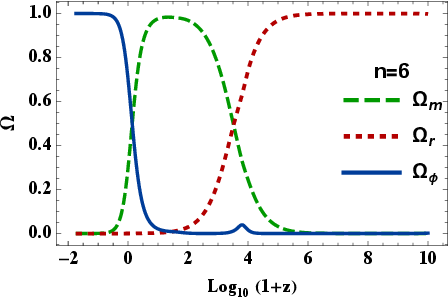} ~
\includegraphics[scale=.93]{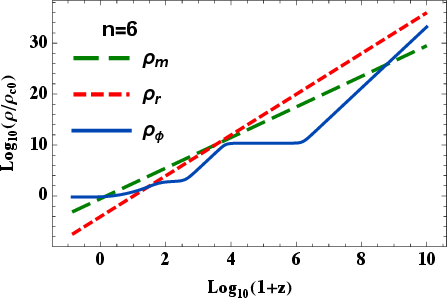}
\includegraphics[scale=.93]{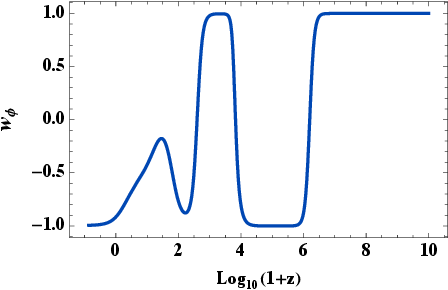}~
\includegraphics[scale=.93]{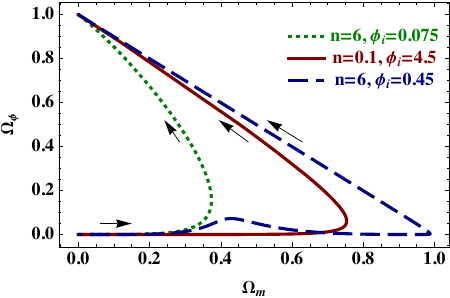}
\caption{Evolution of th background quantities for the MSE potential~\eqref{eq:potMSE} for $V_0=5\times 10^{10}\rho_{\rm c 0}$, $n=6$, $\phi_0=0.7$ and $\Om_{\m0}=0.3$. For the upper and bottom left figures the initial conditions are $\phi_i=\phi_0-0.1 \phi_0$ and $\phi_i'=0.1$. For the bottom right figure the lines are for different initial conditions as mentioned in the figure.}
\label{fig:dyn}
\end{figure}

As we have already discussed that for $\phi/\Mpl\ll \phi_0^{1/n}$, {\it i.e.}, at higher redshifts (early times), the MSE potential~\eqref{eq:potMSE} reduces to the SE potential~\eqref{eq:pot}. So, the cosmological dynamics of the scalar field will be similar to the SE potential at higher redshifts or early times. Also, when $\phi/\Mpl\sim \phi_0^{1/n}$ the MSE potential is still very steep which can be understood from Fig.~\ref{fig:MSEpot} which shows the numerical evolution of the MSE potential for $n=6$ and $\phi_0=0.7$. For the chosen $\phi_0$ we have $\phi_0^{1/n}\approx 0.94$ where the potential is very steep. Note that the potential has been shown in log scale in the Fig.~\ref{fig:MSEpot}. So, around $\phi_0$ we will have the effect of the steep potential which can lead to multiple frozen epochs in the scalar field dynamics as discussed in the previous section (Sec.~\ref{sec:SE}) and has been shown in the upper figure of Fig.~\ref{fig:SE_scaling}. For $\phi/\Mpl>\phi_0^{1/n}$, from Eq.~\eqref{eq:potlargephi}, we see that
\begin{eqnarray}
    V(\phi)\approx V_0\e^{-\gam}\e^{\gam\phi_0\Mpl^n/\phi^n}\, .
    \label{eq:MSE_track}
\end{eqnarray}
The above equation is very crucial to understand the late time dynamics of the scalar field. The MSE potential~\eqref{eq:potMSE} behaves like the potential~\eqref{eq:MSE_track} before becoming a flat potential {\it i.e.}, during the recent past. The potential~\eqref{eq:MSE_track} is a generalization of the potential $V(\phi)\sim \e^{\Mpl/\phi}$ which gives rise to tracker solution \cite{Zlatev:1998tr,Steinhardt:1999nw}. So, the potential~\eqref{eq:MSE_track} would also give rise to a tracker solution. In tracker behaviour the scalar field dynamics does not exactly follow the background energy density like in scaling behaviour but decays a little slower than the background. Because of this slower decay process the scalar field energy density eventually takes over the background unlike the scaling behaviour in which the scalar field energy density follows the background forever. This tracker behaviour in the MSE potential~\eqref{eq:potMSE} brings a slight difference in the scalar field dynamics compared to the SE potential~\eqref{eq:pot}. Now when $\phi/\Mpl\gg \phi_0^{1/n}$ the MSE potential~\eqref{eq:potMSE} behaves like a CC which makes the scalar field to exit from the tracker behaviour and the scalar field EoS $w_\phi$ becomes very close to $-1$. So, the MSE potential not only can give rise to viable cosmology but also give interesting dynamics along with relating two different energy scales and thereby we do not need to invoke any CC in the theory. The steep nature of the potential gives multiple frozen epochs in the evolution of the scalar field energy density which can be very useful to achieve viable EDE solutions which we discuss in the next section (Sec.~\ref{sec:EDE}). In Fig.~\ref{fig:dyn} we have shown the cosmological evolution of different cosmological parameters for the MSE potential~\eqref{eq:potMSE}.In the top right figure of Fig.~\ref{fig:dyn} we see that we have two frozen epochs in the evolution of the scalar field energy density which is followed by a tracker behaviour. The bottom figure of Fig.~\ref{fig:dyn} shows that the scalar field EoS $w_\phi$ is very close to $-1$ at $z=0$. The top left figure of the Fig.~\ref{fig:dyn} shows the evolution of the density parameters. The bottom right figure of Fig.~\ref{fig:dyn} shows the stability of solutions in the $\Om_\m-\Omega_\phi$ plane. The three lines in the figure represent three different initial conditions and all the line moving towards a common future stable de Sitter solution, $\Om_\m=0$ and $\Om_\phi=1$. $\Om_\m=\Om_\phi=0$ represents radiation dominated era while $\Om_\m=1$ and $\Om_\phi=0$ represents matter dominated era. Both are either unstable or saddle points.

\section{Early Dark Energy}
\label{sec:EDE}
When the scalar field is rolling down a steep potential its kinetic energy dominates over potential energy which makes $w_\phi\approx 1$ and $\rho_\phi\sim a^{-6}$. Because of this $\rho_\phi$ becomes subdominant than the background energy density and the scalar field experiences large Hubble damping. This makes the scalar field frozen. During this frozen period $w_\phi=-1$ and $\rho_\phi=$ constant. During the frozen period $\rho_\phi$ eventually becomes comparable to the background energy density and at some redshift starts rolling again as the Hubble damping reduces eventually. During this rolling period if the potential is still steeper than an exponential potential the scalar field energy density can repeat the previous dynamics. Now this kind of dynamics can be very useful to get EDE. If the frozen period ends when $\rho_\phi$ becomes comparable to background energy density but do not take over it then we can have a small but finite energy density parameter corresponding to the scalar field {\it i.e.}, $\Om_\phi$ at the redshift where the scalar field just starts rolling again. We call that redshift as $z_{\rm c}$. For $z> z_{\rm c}$ but close to $z_{\rm c}$ the scalar field is frozen and $w_\phi=-1$. So, the small but finite $\Om_\phi(z=z_{\rm c})$ is the density parameter of a scalar field with $w_\phi=-1$ {\it i.e.}, similar to a CC or dark energy. We call this early dark energy (EDE) as long as $z_{\rm c}$ is large. Also, $\Om_\phi(z=z_{\rm c})$ is the energy density parameter of EDE at $z_{\rm c}$,{\it i.e.}, $\Om_{\rm EDE}(z_{\rm c})$ which is generally denoted by $f_{\rm EDE}$ in the literature of EDE \cite{Poulin:2018cxd,Poulin:2018dzj,Poulin:2023lkg}.

It has been shown that EDE solutions can alleviate $H_0$ tension to some extent \cite{Poulin:2018cxd,Poulin:2018dzj,Poulin:2023lkg,Agrawal:2019lmo,Kodama:2023fjq,Lin:2019qug,Niedermann:2019olb,Niedermann:2020dwg,Berghaus:2019cls,Berghaus:2022cwf,Karwal:2021vpk} if we consider $z_{\rm c}\approx z_{\rm eq}$, where $z_{\rm eq}$ is the redshift of matter-radiation equality. To understand this lets consider the definition of the angle subtended by the comoving sound horizon $r_{\rm s}$ at the decoupling epoch $z_\star$ given by \cite{Kamionkowski:2022pkx} 
\begin{eqnarray}
    \theta_{\rm s}=\frac{r_{\rm s}(z_\star)}{(1+z_\star)D_{\rm A}(z_\star)} \, ,
    \label{eq:thet}
\end{eqnarray}
which is precisely measured by the CMB observations \cite{Planck:2018vyg,Planck:2013pxb}. In the above equation $D_{\rm A}(z)$ is the angular diameter distance and $r_{\rm s}(z)$ and $D_{\rm A}(z_\star)$ are given by
\begin{eqnarray}
    r_{\rm s}(z) &=& \int_{z}^{\infty} \frac{c_s(z)}{H(z)}{\rm d}z = \frac{1}{H_\star}\int_{z}^{\infty} \frac{c_s(z)}{H(z)/H_\star}{\rm d}z \, ,~~~ 
    \label{eq:rs}\\
    D_{\rm A}(z) &=& \frac{c}{1+z}\int_{0}^{z_\star}\frac{dz}{H(z)} =  \frac{c}{H_0(1+z)}\int_{0}^{z}\frac{dz}{E(z)} \, ,~~~~
     \label{eq:DA}
\end{eqnarray}
where the Hubble parameter at the decoupling $z_\star$ is $H_\star$, $E(z)=H(z)/H_0$ with $H_0$ as the present value of the Hubble parameter, $c$ is the velocity of light and $c_{\rm s}(z)$ is the sound speed of the photon-baryon fluid which is given by
\begin{eqnarray}
    c_{\rm s}(z) &=& \frac{c}{\sqrt{3(1+\frac{R_{\rm b}}{1+z})}} \, , \\
    R_{\rm b} &=& \frac{3\omega_{\rm b}}{4\omega_\gamma} \, , 
    \label{eq:Rb}\\
    \omega_\r &=& \(1+\frac{7}{8}N_{\rm eff}\(\frac{4}{11}\)^{4/3}\)\omega_\gamma \, ,
\end{eqnarray}
where $100 {\rm km}\; {\rm s}^{-1}{\rm Mpc}^{-1}h=H_0$, photon density and baryon density at present is $\Om_{\gamma 0}$ and $\Om_{\rm b 0}$ respectively. $\omega_{\gamma} =\Om_{\gamma 0} h^2= 2.47\times10^{-5}$ and $\omega_{\rm b}=\Om_{\rm b 0}h^2=0.0224\pm 0.0001$, $100\theta_{\rm s}=1.0411\pm 0.0003$ and $N_{\rm eff}=3.06$ \cite{Planck:2018vyg,Riess:2021jrx}. In terms of the above parameters the decoupling redshift $z_\star$ can be parametrized as 
\begin{eqnarray}
    z_\star &=& 1048\(1+0.00124\omega_{\rm b}^{-0.738}\)\(1+g_1\om_\m^{g_2}\) \\
    g_1 &=& \frac{0.0783\om_{\rm b}^{-0.238}}{1+39.5\om_{\rm b}^{0.763}}\\
    g_2 &=& \frac{0.560}{1+21.1\om_{\rm b}^{1.81}} \; .
\end{eqnarray}

If we modify the cosmic history only around $z_\star$ or $z_{\rm eq}$ by keeping it same as standard model for $z<z_\star$ then the comoving sound horizon $r_{\rm s}$ will be changed. So, $D_{\rm A}$ remains same as standard model. As $\theta_{\rm s}$ is precisely measured it can not be changed. So, $H_0$ has to be changed to keep $\theta_{\rm s}$ fixed. By injecting EDE around $z_\star$ or $z_{\rm eq}$ we can increase the expansion rate due to which $r_{\rm s}(z_\star)$ decreases and it forces $H_0$ to increase so that $\theta_{\rm s}$ remains fixed. 

\begin{table*}[t]
\begin{center}
\caption{Value of $H_0$ in the unit of ${\rm km}\; {\rm s}^{-1} {\rm Mpc}^{-1}$ for different $n$ by maintaining $\Om_{\m0}\approx 0.31$, $f_{\rm EDE}\approx 0.07$, $z_\star \approx 1091$ and $z_{\rm eq}=3300$ and the redshift of drag epoch $z_{\rm d}=1060$. To calculate $H_0$ we have added a CC with SE~\eqref{eq:pot}, axionlike~\eqref{eq:ax} and power law~\eqref{eq:pl} potentials. Here $\lam$ signifies $\mu$, $\phi_0$ and $f$.}
\label{tab:H0}
\resizebox{\textwidth}{!}{%
\begin{tabular}{c c c c c c c c c c} \hline \hline
 Potential & ~~~~~$n$~~~~~ & ~~~~~$\lambda$~~~~~ & ~~~~~$V_0/\rho_{\rm c0}$~~~~~ & ~~~~~$H_0$~~~~~ & ~~$r_{\rm d}$ (Mpc)~~ & ~~$r_{\rm s}$ (Mpc)~~ & ~~$\phi_i(\Mpl)$~~ & ~~$\phi_i'(\Mpl)$~~ & ~~~~$z_{\rm c}$~~~~ \\ 
\hline\hline
& 4 & 48.5 & $7\times 10^{9}$ & 68.74 & 142.30 & 139.53 & 0.3 & $10^{-15}$ & 3544.69 \\
SE & 6 & 5.78 & $ 10^{10}$ & 68.75 & 142.24 & 139.48 & 0.7 & $10^{-15}$ &  3532.89  \\
  & 10 & $0.492$ & $10^{10}$& 68.72 & 142.42 & 139.66 & 1.01 & $10^{-15}$ & 3774.5\\
\hline
\hline
& 4 & 0.263 & $8.37\times 10^{9}$ & 67.17 & 145.65 & 142.82 & 0.2367 & $10^{-15}$ & 3605 \\
MSE & 6 & 0.498 & $8.37\times 10^9$ & 67.63 & 144.67 & 141.86 & 0.4482 & $10^{-15}$ & 3624.14 \\
  & 10 & 0.6887 & $8.37\times 10^9$ & 68.14 & 143.74 & 140.96 & 0.6198 & $10^{-15}$ & 3962.47 \\
\hline
\hline
 & 2 & 0.0265 & $3.87\times 10^{10}$ & 68.53 & 141.50 & 138.76 & 0.083 & $10^{-15}$ & 10933.2 \\
Axionlike & 3 & 0.033 & $3.87\times 10^{10}$ & 68.60 & 142.62 & 139.92 & 0.1036 & $10^{-15}$ & 13100.6 \\
  & 6 & 0.0488 & $2.48\times 10^{9}$ & 68.49 & 143.45 & 140.74 & 0.1532 & $10^{-15}$ & 10956.1 \\
\hline
\hline
 & 2 & 0.172 & $1.56\times 10^7$ & 68.53 & 142.09 & 139.34 & 0.538 & $10^{-15}$ & 2232.91  \\
Power law & 3 & 0.231 & $6.22\times 10^7$ & 68.76 & 142.40 & 139.64 & 0.723 & $10^{-15}$ & 5993.41 \\
 & 6 & 0.435 & $6.6\times 10^4$ & 68.82 & 141.92 & 139.17 & 1.362 & $10^{-15}$ & 5664.94 \\
\hline
\hline
$\Lambda$CDM & $-$ & $-$ & $-$ & $68.02$ & $143.92$ & $141.16$ & $-$ & $-$ & $-$ \\
\hline\hline
\end{tabular}}
\end{center}
\end{table*}

Using Eqs.~\eqref{eq:thet}, \eqref{eq:rs} and \eqref{eq:DA} we can have \cite{Kamionkowski:2022pkx}
\begin{eqnarray}
H_0 &=& \sqrt{3}H_{\star}\theta_{\rm s}\frac{\int_{0}^{z_{\star}}\frac{{\rm d}z}{E(z)}}{\int_{z_{\star}}^{\infty}\frac{E(z_{\star})}{E(z)}(1 +R)^{-1/2}{\rm d}z }
\label{eq:h0}
\end{eqnarray}
where
\begin{eqnarray}
    H_{\star} &=&  \omega_\r^{1/2} (1+z_{\rm ls})^2 \Big(\frac{\omega_\m}{\omega_\r}\times\frac{1}{1+z_{\star}} + 1\Big)^{1/2} \times 100 {\rm km~sec^{-1} Mpc^{-1}} \, .  
\end{eqnarray}
Eq.~\eqref{eq:h0} gives us an estimate of $H_0$. We have listed values of $H_0$ for different parameter values for the SE~\eqref{eq:pot} and MSE~\eqref{eq:potMSE} potentials in Tab.~\ref{tab:H0} along with the values of $H_0$ for the axion-like and power law potentials given by
\begin{eqnarray}
    V(\phi) &=& V_0\(1-\cos\(\frac{\phi}{f_{\rm pl}}\)\)^n \, , 
    \label{eq:ax}\\
    V(\phi) &=& V_0 \(\frac{\phi}{f_{\rm pl}}\)^{2n} \, ,
    \label{eq:pl}
\end{eqnarray}
respectively by keeping the EDE density parameter at the redshift $z_{\rm c}$, $f_{\rm EDE}$ fixed at $0.07$ as small amount of $f_{\rm EDE}(<0.1)$ is allowed by the observation \cite{Poulin:2018cxd,Poulin:2018dzj,Poulin:2023lkg,Agrawal:2019lmo,Qu:2024lpx}. In the above equations $f_{\rm pl}=f\Mpl$ where $f$ is a constant. One should note that to estimate $H_0$ for using the Eq.~\eqref{eq:h0} we are using the bounds from CMB observations on the parameters $\om_\r$, $\om_\gam$, $\om_\m$, $\om_{\rm b}$ and $\theta_{\rm s}$ as mentioned above. So, this is just an estimation. The proper analysis should be done by doing observational data analysis by considering different cosmological observations. From Tab.~\ref{tab:H0} we see that the MSE potential~\eqref{eq:potMSE} can improve the value of $H_0$ compared to $\Lambda$CDM but SE potential~\eqref{eq:pot} with a CC improves the value even more. This happens because of the presence of tracker behaviour in the MSE potentials during the late time while we have scaling behaviour in the SE potential during the late time. Due to tracker behaviour the present value of $w_\phi$ can be slightly higher than $-1$ which can reduce the effect of EDE. While in SE potential, as we are adding a CC, the present value of $w_\phi$ is $-1$ which can have maximum effect of EDE for $w_\phi\geq 0$. It is also should be noted from Tab.~\ref{tab:H0} that the EDE in SE potential can give us better result in terms of the value of $H_0$ compared to the axion-like potential. To get the proper estimation of $H_0$ we do observational data analysis in the next section.

\section{Growth of structures}
\label{sec:PT}
In this section we analyse the growth of structures in the models under consideration. We consider the
following metric in the Newtonian gauge.

\begin{equation}
\d s^2=-(1+2\Phi)d\tau^2+a(t)^2(1-2\Psi)\d\vec{x}^2 \, ,
\label{eq:pmetric}
\end{equation}
where, $\Phi$ and $\Psi$ are the Bardeen’s potentials \cite{Bardeen:1980kt} in the Newtonian gauge.

In the subhorizon ($k^2\gg a^2H^2$, $k$ is the wavenumber of fluctuation) and quasistatic ($|\ddot\phi|\lesssim H|\dot\phi|\ll k^2|\phi|$)  approximations, we have

\begin{equation}
 \ddot\delta+2H\dot\delta-4\pi G\bar\rho_m\delta=0 \, ,
\label{eq:den_con_evo}
\end{equation}
where, $\bar \rho_\m$ is the unperturbed (background) matter energy density and $\delta=(\rho_\m-\bar\rho_\m)/\bar\rho_\m$ is the density contrast. The solution of Eq.~(\ref{eq:den_con_evo}) can be written as the linear superposition of the growing $D_+$ and decaying $D_-$ modes,
\begin{eqnarray}
 \del(t,\vec k)=c_+ D_+(t)\del(\vec k,0)+c_- D_-(t)\del(\vec k,0) \, ,
 \label{eq:del_sol}
\end{eqnarray}
where $c_+$ and $c_-$ are the constants and $\del(\vec k,0)$ is the primordial density fluctuation. For gorwth of structures we are mainly interested about the growing mode $D_+$ and using that we can define the growth factor $f$ as
\begin{equation}
 f=\frac{\d \ln D_+}{\d \ln a} \, .
 \label{eq:growth}
\end{equation}

Fourier transform of the two point correlation function, {\it i.e.}, the power spectrum $\P(t,k)$ of the scalar perturbation is given by 
\begin{equation}
\Big<\del(t,\vck)\del(t,\vck')\Big>=\del^{(3)}(\vck+\vck')\P(t,k) \, ,
\end{equation}
where $\big<\dots\big>$ represents the ensemble average. $\P(t,k)$ only Depends on $|\vck|$ and not on the vector $\vck$ which is a consequence of the assumption of statistical homogeneity and isotropy of the initial fluctuations. In terms of the growing mode $D_+$The power spectrum can be written as \cite{Eisenstein:1997jh,Duniya:2015nva}
\begin{eqnarray}
 \P(t,k)&\propto& A_{\rm H}^2\Big|D_+(\t a)\Big|^2 T^2(k)\(\frac{k}{H_0}\)^{n_\s} \, ,
 \label{eq:PS}
 \end{eqnarray}
 where, $A_{\rm H}$ is a normalization factor and $n_\s$ is the spectral index of scalar perturbation during inflation. $T(k)$ represents the transfer function \cite{Eisenstein:1997ik} and it relates the primordial curvature perturbation with the comoving matter perturbation. Eisenstein-Hu fitting formula has been used for the transfer function \cite{Eisenstein:1997ik}.

 The root mean square amplitude of mass fluctuations $\sig_{\rm R}$ in sphere of radius $Rh^{-1}$Mpc is given by
\begin{eqnarray}
 \sig_{\rm R}^2=\int_0^\infty \d k \frac{k^2}{2\pi^2}\P(t,k)\big|W_{\rm win}(kR)\big|^2 \, ,
\end{eqnarray}
where, $W_{\rm win}(kR)$ is the window function of size $R$. Using the window function ($W_{\rm win}(kR)$) we define a smoothed density field
\begin{equation}
 \del(\vec{x};R)=\int\del(\vec{x}') W_{\rm win}(\vec{x}-\vec{x}';R) \d^3x' \, .
\end{equation}
The above relation is a convolution. Therefore, the Fourier transform of the smoothed density field is a product of $\del(\vck)$ and $W_{\rm win}(kR)$. For the window function we choose a spherical top-hat function which is given by
\begin{eqnarray}
 W_{\rm win}(kR)=\frac{3}{(kR)^3}\Big(\sin(kR)-kR \cos(kR)\Big) \, ,
\end{eqnarray}
The smoothing scale at which $\sig_{\rm R}\sim 1$ is the scale where the linear perturbation theory breaks down and nonlinear effects become important. In this regard $R=8h^{-1}\rm Mpc$ is a relevant scale. From Planck 2018 results we have $\sig_8=0.811\pm 006$ \cite{Planck:2018vyg} at $z=0$. Using the best fit value of $\sig_8$ from Planck 2018 results we fix the normalization factor $A_{\rm H}$ of the power spectrum ~(\ref{eq:PS}). $\sig_8(z)$ can be represented by the growing mode $D_+(z)$
\begin{equation}
 \sig_8(z)=\sig_8(0)\frac{D_+(z)}{D_+(0)} \, .
\end{equation}
To fix the normalization we fix the initial value of $D_{+}(z)$ at the matter dominated era which is essentially same as in the $\Lam$CDM model.

\begin{figure}[t]
\centering
\includegraphics[scale=.8]{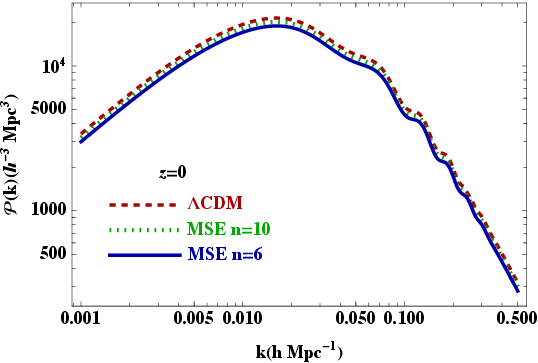} ~~~
\includegraphics[scale=.8]{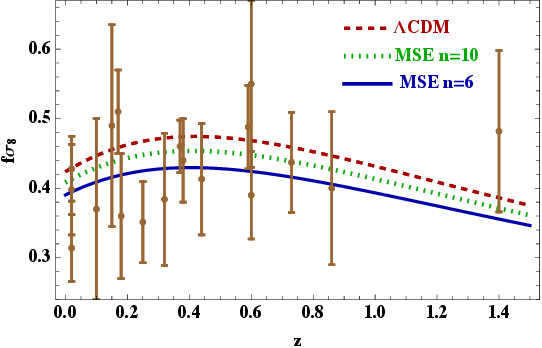}
\caption{Left: Power spectrum is shown for the MSE potential~\eqref{eq:potMSE} with $n=6$ (solid blue) and $n=10$ (dotted green) at $z=0$ along with the power spectrum for the $\Lambda$CDM (dashed red) model. Right: Evolution of $f\sig_8(z)$, for the similar cases as the left figure, has been shown. The brown dots are the observational data of $f\sig_8$ along with their $1\sig$ error bars \cite{Nesseris:2017vor}. For both the figures $\Om_{\m0}=0.31$.}
\label{fig:PS}
\end{figure}

Left figure of Fig.~\ref{fig:PS} shows the matter power spectrum for MSE potential~\eqref{eq:potMSE} for $n=6$ and $10$ along with the $\Lambda$CDM case.We can see that the power spectrum is getting suppressed as we decrease the value of $n$ and the linear perturbation level the MSE potential gives different predictions than the $\Lambda$CDM model. This nature is also represented in the right figure of Fig.~\ref{fig:PS} where the evolution of $f\sig_8(z)$ has been shown. We can  see clear difference in the evolution of $f\sig_8(z)$ for the MSE potential than the standard $\Lambda$CDM model. For the other three potential (Eqs.~\eqref{eq:pot}, \eqref{eq:ax} and \eqref{eq:pl}) we don't see any difference in the behaviour of power spectrum and $f\sig_8(z)$ which clearly tells us that the difference between the MSE potential and $\Lambda$CDM does not come from the EDE solution but rather it comes from the late time behaviour of the scalar field. For SE~\eqref{eq:pot}, axion-like~\eqref{eq:ax} and power law~\eqref{eq:pl} potentials we have considered a CC along with the potential itself which is responsible foe the late time behavior. But, for the MSE potential~\eqref{eq:potMSE} there is no need of additional CC term and viable late time dynamics is governed by the potential itself. Also,  the late time behaviour of the scalar field in the MSE potential has tracker nature. This difference in the late time behaviour in MSE potential and $\Lambda$CDM model is responsible for the difference in the power spectrum and the evolution of $f\sig_8(z)$ in the two cases. On the other hand, for the other three EDE potentials there is no difference in the late time behaviour with the $\Lambda$CDM model and thereby we do not observe any difference in the power spectrum and the evolution of $f\sig_8(z)$.

\begin{figure}[t]
\centering
\includegraphics[scale=.8]{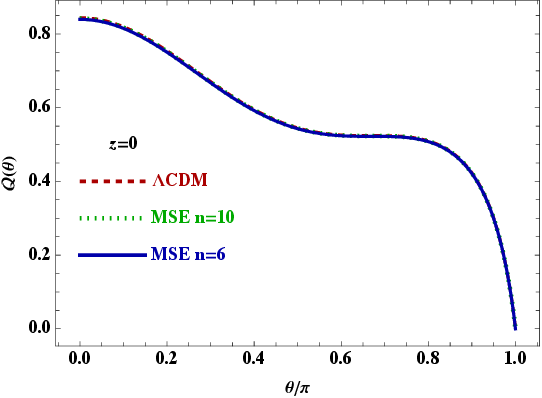} ~~~
\includegraphics[scale=.8]{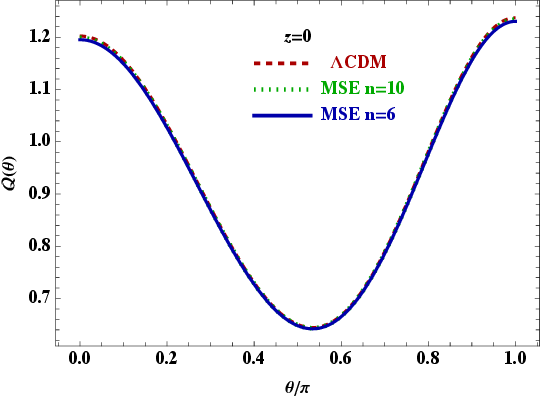}
\caption{Reduced bispectrum, as a function of the angle $\theta$, at $z=0$, is shown for $k=k'=0.01~\rm hMpc^{-1}$ (left figure) and $5k=k'=0.05~\rm hMpc^{-1}$ (right figure). In both the figure the blue (solid), green (dotted) and red (dashed) lines represent the bispectrum for MSE potential with $n=6,\; 10$, and $\Lam$CDM respectively.}
\label{fig:BS}
\end{figure}

Now, to study mildly nonlinear density perturbation we have to study the matter bispectrum which is related to the three point function through a Fourier transform. To study the matter bispectrum for the MSE potential we follow the method discussed in \cite{Hossain:2017ica}. Matter bispectrum is given by the relation
\begin{eqnarray}
 \big<\del(t,\vck)\del(t,\vck')\del(t,\vck'')\big>=\del^{(3)}(\vck+\vck'+\vck'')\B(t,k,k'),~~~~~
\end{eqnarray}
with
\begin{eqnarray}
\del(a,\vck) &=& \del_1(a,\vck)+\frac{1}{2}\del_2(a,\vck) \nn \\* 
  &=& D_+(a) \del_1(\vck)+\int \d^3k_1\d^3k_2 \del^{(3)}(\vck-\vck_1-\vck_2)\nn \\* && \times F_2(a,\vck_1,\vck_2) \del_1(a,\vck_1)\del_1(a,\vck_2) \, ,\\
 \B(t,k,k') &=& 2F_2(\vck,\vck')\P(t,k)\P(t,k')+\rm cyc \, .
\end{eqnarray}
where, $\del_1$ and $\del_2$ are the first and second order density contrasts. $F_2(\vck,\vck')$ denotes the second-order kernel which can be represented in terms of the growing ($D_+(a)$) and decaying ($D_-(a)$) modes as
\begin{eqnarray}
 F_2(a,\vck_1,\vck_2) &=& \int_{a_\m}^{a}\d\t a' ~\frac{\D(a,a')\K(a',\vck_1,\vck_2)}{2a'^4 H^2(a')W_\r(a')} \,,\\
 \D(a,a') &=& \frac{D_+^2(a')}{D_+^2(a)} \Big(D_-(a)D_+(a')  -D_+(a)D_-(a')\Big) \, ,
\end{eqnarray}
where, $a_\m$ is an initial scale factor which can be fixed during the matter dominated era, $\K(a',\vck_1,\vck_2)$ is the symmetrized kernel which is related the Fourier transform of the inhomogeneous part of the evolution equation of the second order density contrast \cite{Hossain:2017ica} and $W_\r(a')$ is the Wronskian given by \cite{Hossain:2017ica}
\begin{eqnarray}
 W_\r (a) &=& D_+(a) \frac{\d D_-(a)}{\d a}- D_-(a) \frac{\d D_+(a)}{\d a}\nn \\ &=& -\frac{5}{2}\frac{H_0}{a^3 H(z)}\, .
\end{eqnarray}

Instead of working with the bispectrum it is more convenient to define the reduced bispectrum $\Q$ as
\begin{eqnarray}
\Q= \frac{\B(t,k,k')}{\P(t,k)\P(t,k')+\P(t,k')\P(t,k'')+\dots} \, ,
\end{eqnarray}
because it is scale and time independent to lowest order in nonlinear perturbation theory. Fig.~\ref{fig:BS} shows the nature of the reduced bispectrum at $z=0$ for the triangle configuration given by $k/k'=1$ with $k=0.01 h{\rm Mpc}^{-1}$ (left figure) and $k/k'=0.2$ with $k=0.01h{\rm Mpc}^{-1}$ (right figure) as a
function of the angle $\theta$ between the wave vectors $\vec k$ and $\vec k'$ given by $\hat k . \hat k'=\cos \theta$. We don't see any difference in the reduced bispectrum for the MSE potential and $\Lambda$CDM model.

\section{Observational Constraints}
\label{sec:obs}

\subsection{Cosmological Data}

We consider the following cosmological data.
\subsubsection{CMB}
The CMB distance prior utilizes the positions of acoustic peak to determine the cosmological distance at the fundamental level.This prior is commonly incorporated using several key parameters: shift parameter(R), acoustic scale ($l_{\rm A}$), the baryon energy density ($\omega_{\rm b}$), spectral index($n_{\rm s}$) \cite{2019JCAP}.
\begin{eqnarray}
R(z_\star) &=& (1 + z_\star)D_A(z_\star) \frac{{\sqrt{\Omega_{\m0} H_0^2}}}{{c}}  \\ &=& \sqrt{\Om_{\m 0}} \int_0^{z_\star} \frac{\d z}{E(z)}\, \\
l_{\rm A} &=& \frac{\pi}{\theta_\star} = \frac{{\pi(1 + z_\star) D_A(z_\star)}}{{r_s(z_\star)}}  \ , 
 \end{eqnarray}
where, $z_\star$ is the decoupling redshift, $r_s$ signifies the sound horizon as defined in Eq.~\eqref{eq:rs} and $D_A$ corresponds to the angular diameter distance defined in Eq.~\eqref{eq:DA}. The baryon-photon ratio $R_{\rm b}$, defined in Eq.~\eqref{eq:Rb}, can be related to CMB temperature $T_{\rm CMB}=2.7255K$ as
\begin{eqnarray}
    R_{\rm b}=31500\omega_{\rm b}\(\frac{T_{\rm CMB}}{2.7K}\)^{-4} \, .
\end{eqnarray}
The present values of radiation and matter energy density parameters, $\Om_{\r0}$ and $\Om_{\m0}$ respectively, are related as
\begin{eqnarray}
    \Om_{\r0}=\frac{\Om_{\m0}}{1+z_{\rm eq}} \, ,
\end{eqnarray}
where  matter-radiation equality redshift $z_{\rm eq}$ is given by
\begin{eqnarray}
    z_{\rm eq}= 25000\omega_\m\(\frac{T_{\rm CMB}}{2.7K}\)^{-4}\, ,
\end{eqnarray}
where $\omega_\m=\Om_{\m0}h^2$.

\subsubsection{BAO}
BAO measurements are generally quoted as the ratios $D_{\rm M}/r_{\rm d}$, $D_{\rm H}/r_{\rm d}$ and $D_{\rm V}/r_{\rm d}$. Here $D_{\rm M}=(1+z)D_{\rm A}$ is the comoving angular diameter distance, $D_{\rm H}=c/H(z)$ and $D_{\rm V}$ is defined as
\begin{eqnarray}
    D_{\rm V}(z)=\(zD_{\rm M}^2(z)D_{\rm H}(z)\)^{1/3} \, .
\end{eqnarray}
$r_{\rm d}$ is the sound horizon at the drag epoch $z_{\rm d}\approx1060$ \cite{Planck:2018vyg} {\it i.e.}, $r_{\rm d}=r_{\rm s}(z_{\rm d})$. In this paper we consider the BAO results from the observations before DESI and from DESI separately. 

{\bf BAO before DESI:} We use BAO data points via the SDSS DR7-MGS survey (z = 0.15)\cite{10.1093/mnras/}, the Isotropic BAO observations by 6dF galaxy survey (z = 0.106)\cite{2011MNRAS} and the SDSS DR14-eBOSS quasar samples (z = 1.52)\cite{eBOSS:2018cab}. BAO measurement with Ly$\alpha$ samples in conjunction with quasar samples from the SDSS DR12(z=2.4)\cite{2017A&A...608A.130D}. We additionally take into account anisotropic BAO data at redshifts 0.38, 0.51, and 0.61 from the BOSS DR12 galaxy collection\cite{2017MNRAS.470.2617A}. We call this data as BAO.

\vskip5pt
\noindent
{\bf BAO from DESI:} DESI DR1 BAO measurements \cite{DESI:2024mwx,DESI:2024uvr,DESI:2024lzq} contains the measurements from
the galaxy, quasar and Lyman-$\al$ forest tracers within the redshift range
$0.1< z< 4.2$. These include the bright galaxy sample (BGS) within the redshift range $0.1<z<0.4$, luminous red galaxy sample
(LRG) in $0.4<z<0.6$ and $0.6<z<0.8$, emission line galaxy sample (ELG) in $1.1<z<1.6$, combined LRG and ELG sample in  $0.8<z<1.1$, the quasar
sample (QSO) in $0.8<z<2.1$ \cite{DESI:2024uvr} and the Lyman-$\al$ Forest
Sample (Ly-$\al$) in $1.77<z<4.16$ \cite{DESI:2024lzq}. We call this data DBAO.

\subsubsection{Type-Ia Supernova}
We consider distance moduli measurements from the Pantheon$+$ sample of Type-Ia supernovae (SNeIa), which consists of 1550 SneIa luminosity distance measurements within  the redshift range $0.001 < z < 2.26$ \cite{Scolnic:2021amr,Brout:2022vxf}. We also use the 22 binned data points of the Union3 \cite{Rubin:2023ovl} compilation of 2087 cosmologically useful SNeIa and the Dark Energy Survey Y5 (D5) analysis \cite{DES:2024tys} of 1635 SNeIa within the redshift range $0.10< z < 1.13$ along with 194 low redshift sample ($0.025<z<0.10$) of SNeIa which sum up to 1829 data points.

\subsubsection{Observational Hubble Data}
We analyze observational data for the Hubble parameter measured at various redshifts
within the range of 0.07 to 1.965. We focus on a collection of 31 $H(z)$ measurements derived using the cosmic chronometric method \cite{2018JCAP...04..051G}.

\subsection{Data combinations}

For a detail analysis and comparison with other scenario we perform the data analysis for the SE~\eqref{eq:pot}, MSE~\eqref{eq:potMSE}, axionlike~\eqref{eq:ax} and power law~\eqref{eq:pl} potentials along with the standard $\Lambda$CDM model. The different data combinations are CMB+BAO$_i$+SNIa$_j$+Hubble, where subscript BAO$_i$ stands for BAO and DBAO while SNIa$_j$ stands for Pantheon$+$, Union3 and D5. So, we have six combinations of different observational data. We call CMB+DBAO+Pantheon$+$\\+Hubble as DESIPanth and similarly CMB+BAO+Pantheon$+$+Hubble as BAOPanth, CMB\\+DBAO+D5+Hubble as DESID5, CMB+BAO+D5+Hubble as BAOD5, CMB+DBAO \\+Union3+Hubble as DESIUnion and CMB+BAO+Union3+Hubble as BAOUnion. We also fix the value of the parameter $n$ for all the four scalar field potentials under consideration and perform the analysis for two values of $n$. We perform Markov Chain Monte Carlo (MCMC) analysis for the above mentioned observational data combinations to constrain the model
parameters. We use
the publicly available code {\tt EMCEE} \cite{Foreman-Mackey:2012any} for this purpose. For analysing the results and plotting the contours of the model parameters we use the publicly available python package {\tt GetDist} \cite{Lewis:2019xzd}.

\subsection{Parameters and priors}

\subsubsection{$\Lambda$CDM}
 For DESID5, BAOD5, DESIUnion and BAOUnion we have four parameters $\{\Om_{\m0}, h,\; \omega_{\rm b}, r_{\rm d}h\}$ while for DESIPanth and BAOPanth we have to add another parameter namely the absolute magnitude $M$. So, for the latter case the parameters are $\{\Om_{\m0}, h,\; \omega_{\rm b}, r_{\rm d}h,M\}$.
 Priors of the parameters are given in the Tab.~\ref{tab:para_lcdm}.
\begin{table}[ht]
\begin{center}
\caption{Prior for the parameters.}
\label{tab:para_lcdm}
\begin{tabular}{c c}\hline \hline
  Parameter ~~~~~~~~~~~~~~~~~~ & ~~~~~~~~~~~~~~~~~~ Prior   \\ 
\hline \hline
 $\Om_{\m 0}$ ~~~~~~~~~~~~~~~~~~& ~~~~~~~~~~~~~~~~~~ $[0.2,0.5]$  \\
  $h$ ~~~~~~~~~~~~~~~~~~&~~~~~~~~~~~~~~~~~~ $[0.5,0.8]$  \\
 $\omega_{\rm b}$ ~~~~~~~~~~~~~~~~~~&~~~~~~~~~~~~~~~~~~ $[0.005,0.05]$   \\
 $r_{\rm d}h/$Mpc ~~~~~~~~~~~~~~~~~~&~~~~~~~~~~~~~~~~~~ $[60, 140]$ \\
 $M$ ~~~~~~~~~~~~~~~~~~&~~~~~~~~~~~~~~~~~~ $[-22, -15]$ \\
\hline\hline
\end{tabular}
\end{center}
\end{table}

\subsubsection{SE potential}

Apart from the parameters mentioned in Tab.~\ref{tab:para_lcdm} for $\Lambda$CDM the model parameters for the SE potential~\eqref{eq:pot} are $\{\mu,\;V_0\}$. We take $\log_{10}V_0$ as the model parameter instead of $V_0$. $V_0$ sets the scale at which EDE starts decaying after reaching its maximum value. In the literature this scale is generally represented as the redshift $z_{\rm c}$. $\mu$ determines the peak value of EDE at $z_{\rm c}$. $\log_{10}V_0$ also has some effect on this. The fractional energy density at the peak value of EDE at $z_{\rm c}$ is denoted by $f_{\rm EDE}$ in the literature. In our analysis instead of using the standard $f_{\rm EDE}$ and $z_{\rm c}$ parameters we use the parameters $\mu$ and $\log_{10}V_0$. Also, we do not use any parametrisation of the scalar field energy density as suggested in Ref.~\cite{Poulin:2018dzj} rather we completely rely on solving the system numerically. The priors for the common parameters with the $\Lambda$CDM model are same as Tab.~\ref{tab:para_lcdm}. The priors for the parameters $\Big\{\mu,\; \log_{10}V_0\Big\}$ are $\Big\{[0.1,5.0],\; [-2.0,18.0]\Big\}$. For $\mu=0$ and $V_0\sim \rho_{\rm DE0}$ the SE potential behaves as a CC. Practically, very small values of $\mu$ makes the potential similar to the CC. So, smaller value of $\mu$ and $V_0$ can mimic  a CC. This has been checked numerically.

\subsubsection{MSE potential}
For the MSE potential~\eqref{eq:potMSE} the extra parameters, other than the ones we have for $\Lambda$CDM model, would be $\phi_0$ and $V_0$. Here also we take $\log_{10}V_0$ as the model parameter instead of $V_0$. The prior on the parameters $\Big\{\phi_0,\; \log_{10}V_0\Big\}$ are $\Big\{[0.25,5],\; [-2,18]\Big\}$ for $n=6$ and $\Big\{[0.5,5],\; [-2,18]\Big\}$ for $n=10$. $\phi_0$ changes the amount of EDE. We have chosen the prior of $\phi_0$ such a way so that $f_{\rm EDE}$ can have maximum value of $\sim 0.5$. For larger values of $n$, {\it e.g.}, $n=6~{\rm or}~10$ the MSE potential behaves like a CC during the present time. Also, for very small values of $V_0$ the dynamics can be very similar to $\Lambda$CDM.

\subsubsection{Axionlike potential}

Similar to the SE and MSE potential for the axionlike potential~\eqref{eq:ax} also we have two extra parameter compared to the $\Lambda$CDM model. The extra parameters are $\Big\{f,\; \log_{10}V_0\Big\}$ and their priors are $\Big\{[0.01,0.3],\; [-2,18]\Big\}$. 
The value of $f_{\rm EDE}$ depends on $f$. Similar to the MSE potential for the axionlike potential also we have chosen the prior such a so that $f_{\rm EDE}$ can have maximum value of $\sim 0.5$. Also, similar to the MSE potential axionlike potential also gives a dynamics similar to the $\Lambda$CDM for smaller values of $V_0$ and smaller value of $f$.

\subsubsection{Power law potential}

For the power potential~\eqref{eq:pl} also we have two more parameters than the $\Lambda$CDM model and they are $\Big\{f,\; \log_{10}V_0\Big\}$. The priors of these two parameters are $\Big\{[0.2,3]; [-5,18]\Big\}$. In this case also $f$ determines the value of $f_{\rm EDE}$ and within this prior range the maximum value of $f_{\rm EDE}$ can be $0.5$. This prior range can give a $\Lambda$CDM like behaviour for smaller value of $V_0$ and smaller value of $f$.



{\footnotesize    
\begin{longtable}{l c c c c c c c}

\caption{Constraints on the parameters for SE~\eqref{eq:pot}, MSE~\eqref{eq:potMSE}, axionlike~\eqref{eq:ax} and power law~\eqref{eq:pl} potentials. The initial conditions to solve the Eq.~\eqref{eq:eom_phi} are $\{\phi_i,\phi'_i\}\equiv \{1.13,0.15\},\{0.9 \phi_0,10^{-10}\},\{\pi f-0.01 f,10^{-15}\}\; {\rm and}\; \{f,10^{-15}\}$ for the SE, MSE, axionlike and power law potential respectively.}
\label{tab:constraint}\\
 \hline\hline
 Parameter &DESIPanth & BAOPanth & DESID5 & BAOD5 & DESIUnion & BAOUnion\\ 
\hline\hline
&&&$\Lambda$CDM&&&\\
\hline
$\Omega_{\m0}$ & $0.3136\pm 0.0068$ & $0.3155\pm 0.0070$ & $0.2780\pm 0.0022$ & $0.2779\pm 0.0022$ & $0.3119\pm 0.0068$ & $0.3138\pm 0.0072$ \\ 

 $h$ & $0.6752\pm 0.0051$ & $0.6738\pm 0.0053$ & $0.7048 \pm 0.0017$ & $0.7049\pm 0.0017$ & $0.6765\pm 0.0051$ &$0.6751\pm 0.0054$ \\ 

 $\omega_{b}$ & $0.02240\pm 0.00014$ & $0.02237\pm 0.00014$ & $0.02296\pm 0.00010$ & $0.02296\pm 0.00010$ & $0.02242\pm 0.00014$ & $0.02240\pm 0.00014$ \\ 

 $r_{\rm d}h$ & $100.41\pm 0.70$ & $100.08\pm 0.80$ & $103.23\pm 0.52$ & $102.40\pm 0.72$ & $100.53\pm 0.70$ & $100.18\pm 0.81$ \\  

 ${M}$ & $-19.435\pm 0.015$ & $-19.439\pm 0.015$ & --- & --- & --- & --- \\ 
 \hline\hline
&&&  SE~\eqref{eq:pot} for $n=6$ &&&\\
\hline
$\Om_{\m 0}$ & $0.3147\pm{0.0070}$ & $0.3165\pm{0.0073}$ & $0.3150\pm{0.0076}$ & $0.3148^{+0.0092}_{-0.0059}$ & $0.3136^{+0.0067}_{-0.0076}$ & $0.3155\pm{0.0074}$ \\ 

 $h$ & $0.6757\pm{0.0063}$ & $0.6739\pm{0.0062} $ & $0.6997\pm{0.0020}$ & $0.6992^{+0.0023}_{-0.0019}$ & $0.6796^{+0.0059}_{-0.0098}$ & $0.6760^{+0.0063}_{-0.0073}$ \\ 

 $\mu$ & $>2.25$ & $>2.33$ & $0.76^{+0.29}_{-0.23}$ & $0.85^{+0.20}_{-0.34}$ & $2.5\pm 1.4$ & $>2.04$ \\ 

 $\log_{10} V_0$ & unconstrained & unconstrained & $11.0^{+1.0}_{-2.6}$ & $10.8^{+1.4}_{-2.3}$ & unconstrained & unconstrained \\

 $\omega_{\rm b}$ & $0.02240 \pm 0.00014$ & $0.02238\pm 0.00014$ & $0.02238\pm{0.0.00015}$ & $0.022445^{+0.000055}_{-0.00025}$ & $0.02241\pm{0.0.00014}$ & $0.02239\pm 0.00014$ \\ 

 $r_{\rm d}h$ & $100.28\pm{0.75}$ & $99.98\pm{0.84}$ & $100.30\pm{0.75}$ & $100.1\pm{1.4}$ & $100.34\pm{0.81}$ & $100.02\pm{0.87}$ \\  

 $M$ & $-19.432^{+0.016}_{-0.018}$ & $-19.438\pm{0.018}$ & $-$ & $-$ & $-$ & $-$ \\ 
 \hline
&&& SE~\eqref{eq:pot} for $n=10$ &&& \\
\hline 
$\Om_{\m0}$ & $0.3140\pm 0.0069$ & $0.3165\pm 0.0073$ & $0.3133^{+0.0088}_{-0.0057}$ & $0.3151^{+0.0099}_{-0.0057}$ & $0.3128\pm 0.0070$ & $0.3146 \pm 0.0075$   \\
$h$  & $0.6756\pm 0.0057$ & $0.6739\pm 0.0059$ & $0.7000^{+0.0019}_{-0.0022}$ & $0.6996^{+0.0018}_{-0.0024}$ & $0.6782^{+0.0051}_{-0.0077}$ & $0.6757^{+0.0058}_{-0.0066}$  \\
$\mu$  & $>1.97$ & $>1.94$ & $0.495^{+0.024}_{-0.26}$ & $0.516^{+0.014}_{-0.31}$ & unconstrained & $>1.79$ \\
$\log_{10}V_0$  & unconstrained & unconstrained & $10.46^{+0.90}_{-2.3}$ & $10.6^{+1.2}_{-2.4}$ & unconstrained & unconstrained \\
$\omega_{\rm b}$  & $0.02240\pm 0.00014$ & $0.02237\pm 0.00014$ & $0.02240^{+0.00012}_{-0.00018}$ & $0.02238^{+0.00012}_{-0.00019}$ & $0.02242\pm 0.00014$ & $0.02240\pm 0.00014$ \\
$r_{\rm d}h$  & $100.36\pm 0.69$ & $99.98\pm 0.81$ & $100.43^{+0.61}_{-0.89}$ & $100.12^{+0.77}_{-1.0}$ & $100.44\pm 0.71$ & $100.09\pm 0.83$ \\
$M$  & $-19.433\pm 0.017$ & $-19.438\pm 0.017$ & $-$ & $-$ & $-$ & $-$ \\

\hline \hline

&&& MSE~\eqref{eq:potMSE} for $n=6$ &&& \\
\hline
$\Om_{\m 0}$ & $0.3158^{+0.0067}_{-0.0087}$ & $0.3194^{+0.0055}_{-0.011}$ & $0.2783\pm{0.0023}$ & $0.27970^{+0.00074}_{-0.0039}$ & $0.3139^{+0.0068}_{-0.0080}$ & $0.3164^{+0.0074}_{-0.0086}$ \\ 

 $h$ & $0.6727^{+0.0076}_{-0.0049}$ & $0.6707^{+0.0083}_{-0.0047} $ & $0.7043^{+0.0020}_{-0.0017}$ & $0.7031^{+0.0033}_{-0.00068}$ & $0.6741^{+0.0070}_{-0.0052}$ &$0.6721^{+0.0079}_{-0.0055}$ \\
 
 $\phi_0$ & $3.44^{+1.1}_{-0.90}$ & $3.46^{+1.2}_{-0.80}$ & $3.65^{+1.2}_{-0.48}$ & $3.57^{+1.3}_{-0.65}$ & $3.37^{+1.4}_{-0.79}$ & $3.34^{+1.1}_{-0.89}$ \\ 

 $\log_{10} V_0$ & $7.2^{+3.8}_{-8.6}$ & $7.0^{+3.5}_{-8.6}$ & $6.8^{+3.7}_{-8.5}$ & $7.1^{+3.6}_{-8.7}$ & $6.9^{+6.2}_{-8.6}$ & $7.3^{+9.9}_{-9.0}$ \\

 $\omega_{\rm b}$ & $0.02238 \pm{0.00017}$ & $0.022444^{+0.000082}_{-0.00022}$ & $0.02298\pm{0.00010}$ & $0.023005^{+0.000078}_{-0.00014}$ & $0.02243\pm 0.00014$ & $0.02241\pm 0.00014$ \\ 

 $r_{\rm d}h$ & $100.1^{+1.0}_{-0.66}$ & $99.83^{+1.0}_{-0.89}$ & $103.14\pm{0.56}$ & $102.3\pm{1.2}$ & $100.2^{+1.0}_{-0.71}$ & $99.8^{+1.1}_{-0.88}$ \\  

 $M$ & $-19.442^{+0.022}_{-0.014}$ & $-19.444\pm{0.059}$ & $-$ & $-$ & $-$ & $-$ \\ 
  \hline
 &&& MSE~\eqref{eq:potMSE} for $n=10$ &&& \\ 
\hline
$\Om_{\m 0}$ & $0.3140\pm{0.0069}$ & $0.3159\pm{0.0087}$ & $0.2781\pm{0.0036}$ & $0.2790^{+0.0012}_{-0.0033}$ & $0.3123\pm{0.0071}$ & $0.3150^{+0.0071}_{-0.0080}$ \\ 

 $h$ & $0.6748\pm{0.0053}$ & $0.6732\pm{0.0056} $ & $0.7039^{+0.0026}_{-0.00091}$ & $0.7045^{+0.0021}_{-0.0015}$ & $0.6759\pm{0.0057}$ &$0.6737^{+0.0067}_{-0.0053}$ \\ 

$\phi_0$ & $3.35\pm{0.98}$ & $3.32\pm{0.99}$ & $3.41\pm{0.97}$ & $3.3\pm{1.0}$ & $3.3^{+1.5}_{-1.3}$ & $3.2^{+1.6}_{-1.3}$ \\ 

 $\log_{10} V_0$ & $7.6\pm{5.9}$ & $7.6\pm{5.8}$ & $7.7\pm{5.8}$ & $7.7^{+8.2}_{-9.2}$ & $7.8\pm{5.9}$ & $7.8\pm{5.8}$ \\ 

 $\omega_{\rm b}$ & $0.02240\pm{0.00014}$ & $0.02238\pm{0.00017}$ & $0.023051^{+0.000027}_{-0.00019}$ & $0.02296\pm 0.00022$ & $0.02243\pm 0.00014$ & $0.02241\pm 0.00031$ \\ 

 $r_{\rm d}h$ & $100.34\pm{0.74}$ & $99.94^{+0.90}_{-0.75}$ & $103.36^{+0.41}_{-0.66}$ & $102.54^{+0.61}_{-0.87}$ & $100.43^{+0.79}_{-0.70}$ & $100.02\pm{0.94}$ \\ 

 $M$ & $-19.436\pm 0.015$ & $-19.441\pm{0.017}$ & $-$ & $-$ & $-$ & $-$ \\ 
 \hline\hline

&&& Axionlike~\eqref{eq:ax} for $n=3$ &&& \\
 \hline

$\Om_{\m 0}$ & $0.3188^{+0.0072}_{-0.011}$ & $0.3246^{+0.0059}_{-0.016}$ & $0.3189^{+0.0076}_{-0.0089}$ & $0.3233^{+0.0081}_{-0.010}$ & $0.3167^{+0.0065}_{-0.011}$ & $0.3179^{+0.0071}_{-0.0097}$ \\ 

 $h$ & $0.6800^{+0.0097}_{-0.0015}$ & $0.6760^{+0.0018}_{-0.0015} $ & $0.6987^{+0.0026}_{-0.0018}$ & $0.6981^{+0.0026}_{-0.0021}$ & $0.6838^{+0.0098}_{-0.00115}$ &$0.683^{+0.0011}_{-0.0016}$ \\ 
 
 $f$ & $0.062^{+0.018}_{-0.046}$ & $0.073^{+0.035}_{-0.053}$ & $0.090^{+0.014}_{-0.030}$ & $0.098^{+0.016}_{-0.038}$ & $0.056^{+0.017}_{-0.040}$ & $0.060^{+0.018}_{-0.044}$ \\ 

 $\log_{10} V_0$ & $7.2^{+3.8}_{-2.9}$ & $8.4^{+8.8}_{-5.2}$ & $10.1^{+7.6}_{-3.1}$ & $8.9^{+1.2}_{-1.8}$ & $8.6^{+5.5}_{-4.3}$ & $9.4^{+8.3}_{-4.9}$ \\ 

 $\omega_{\rm b}$ & $0.02240 \pm 0.00016$ & $0.022561^{+0.000016}_{-0.00037}$ & $0.02240^{+0.00013}_{-0.00017}$ & $0.02237\pm{0.00019}$ & $0.02240^{+0.000016}_{-0.00034}$ & $0.02239\pm 0.00029$ \\ 

 $r_{\rm d}h$ & $100.07\pm{0.87}$ & $99.5^{+1.3}_{-0.83}$ & $99.99\pm{0.89}$ & $99.61\pm{0.90}$ & $100.22^{+0.87}_{-0.76}$ & $99.99\pm{1.1}$ \\ 

 $M$ & $-19.418^{+0.028}_{-0.049}$ & $-19.410^{+0.032}_{-0.071}$ & $-$ & $-$ & $-$ & $-$ \\ 
  \hline
 &&& Axionlike~\eqref{eq:ax} for $n=6$ &&& \\ 
\hline
$\Om_{\m 0}$ & $0.3159^{+0.0068}_{-0.0087}$ & $0.3199^{+0.0070}_{-0.011}$ & $0.3181^{+0.0074}_{-0.0087}$ & $0.3195^{+0.0076}_{-0.0087}$ & $0.3163^{+0.0069}_{-0.010}$ & $0.3164\pm{0.0080}$ \\ 

 $h$ & $0.6798^{+0.0083}_{-0.013}$ & $0.682^{+0.012}_{-0.016} $ & $0.6991^{+0.0023}_{-0.0019}$ & $0.6988\pm{0.0022}$ & $0.6831^{+0.0080}_{-0.016}$ &$0.6837^{+0.0094}_{-0.014}$ \\ 
 
 $f$ & $0.097^{+0.035}_{-0.077}$ & $0.129^{+0.061}_{-0.089}$ & $0.157^{+0.028}_{-0.059}$ & $0.157^{+0.027}_{-0.064}$ & $0.070^{+0.018}_{-0.052}$ & $0.111^{+0.039}_{-0.088}$ \\ 

 $\log_{10} V_0$ & $8.5^{+6.5}_{-5.5}$ & $9.4^{+6.2}_{-5.1}$ & $9.0^{+3.5}_{-2.9}$ & $9.6^{+3.9}_{-3.5}$ & $8.3^{+5.3}_{-4.2}$ & $9.6^{+7.8}_{-4.7}$ \\ 

 $\omega_{\rm b}$ & $0.02240\pm{0.00014}$ & $0.02237\pm{0.00016}$ & $0.02239^{+0.00014}_{-0.00016}$ & $0.02235\pm 0.00015$ & $0.02240\pm 0.00015$ & $0.02238\pm 0.00014$ \\ 

 $r_{\rm d}h$ & $100.23\pm{0.76}$ & $99.83\pm{0.93}$ & $100.07\pm{0.79}$ & $99.91\pm{0.85}$ & $100.26\pm{0.84}$ & $100.04\pm{0.83}$ \\ 

 $M$ & $-19.420^{+0.025}_{-0.039}$ & $-19.412^{+0.036}_{-0.049}$ & $-$ & $-$ & $-$ & $-$ \\ 
 \hline\hline

&&& Power law~\eqref{eq:pl} for $n=3$ &&& \\ 
\hline
$\Omega_{\m 0}$ & $0.3155\pm 0.0075$ & $0.3205^{+0.0068}_{-0.011}$ &$0.3151^{+0.0072}_{-0.0064}$ & $0.3172\pm 0.0078$ & $0.3144^{+0.0069}_{-0.0084}$ & $0.3165^{+0.0067}_{-0.0095}$ \\ 

 $h$ & $0.6773^{+0.0064}_{-0.001}$ & $0.6713^{+0.0092}_{-0.008}$ & $0.6997\pm 0.0019$ & $0.6992\pm 0.0021$ & $0.684^{+0.0011}_{-0.0014}$ &$0.6782^{+0.0095}_{-0.0012}$ \\
 
 $f$ & $0.69^{+0.14}_{-0.48}$ & $0.66^{+0.098}_{-0.44}$ & $1.29^{+0.36}_{-0.49}$ & $1.30^{+0.26}_{-0.52}$ & $0.83^{+0.23}_{-0.58}$ & $0.88^{+0.20}_{-0.66}$ \\ 

 $\log_{10} V_0$ & $7.4^{+9.5}_{-4.4}$ & $6.0\pm 5.6$ & $13\pm 3.6$ & $12.9\pm 3.4$ & $10^{+7.5}_{-2.9}$ & $7.4^{+8}_{-9.1}$ \\ 

 $\omega_{\rm b}$ & $0.02240\pm 0.00014$ & $0.02243^{+0.00011}_{-0.00019}$ & $0.02237^{+0.00013}_{-0.00015}$ & $0.02235\pm 0.00017$ & $0.02240\pm 0.00015$ & $0.02239\pm 0.00014$ \\ 

 $r_{\rm d}h$ & $100.23\pm 0.76$ & $99.8\pm1.7$ & $100.28^{+0.66}_{-0.77}$ & $99.95\pm 0.81$ & $100.26^{+0.86}_{-0.67}$ & $99.9^{+1.0}_{-0.72}$ \\ 

 ${M}$ & $-19.428^{0.018}_{-0.030}$ & $-19.439^{+0.019}_{-0.029}$ & $-$ & $-$ & $-$ & $-$ \\ 
 \hline
 &&& Power law~\eqref{eq:pl} for $n=6$ &&& \\ 
\hline
$\Om_{\m 0}$ & $0.3172^{+0.0059}_{-0.0096}$ & $0.3191^{+0.0064}_{-0.0095}$ & $0.313^{+0.011}_{-0.0065}$ & $0.316^{+0.010}_{-0.0059}$ & $0.3138^{+0.0066}_{-0.0081}$ & $0.3161^{+0.0067}_{-0.0084}$ \\ 

 $h$ & $0.6735^{+0.0078}_{-0.0062}$ & $0.6731\pm{0.0096} $ & $0.6993^{+0.0027}_{-0.0018}$ & $0.6994^{+0.0019}_{-0.0024}$ & $0.6820^{+0.0081}_{-0.0011}$ &$0.6783^{+0.0081}_{-0.001}$ \\ 

 $f$ & $0.92^{+0.17}_{-0.71}$ & $1.04^{+0.23}_{-0.81}$ & $1.67^{+0.45}_{-0.38}$ & $1.82^{+0.34}_{-0.44}$ & $1.40^{+0.49}_{-1.1}$ & $1.25^{+0.38}_{-0.99}$ \\ 

 $\log_{10} V_0$ & $6^{+11}_{-10}$ & $7.1\pm 6.5$ & $10.7\pm 2.3$ & $10.5^{+2.4}_{-1.6}$ & $9.0^{+8.2}_{-12}$ & $8.2^{+9}_{-11}$ \\ 

 $\omega_{\rm b}$ & $0.022457^{+0.000093}_{-0.0002}$ & $0.02240^{+0.00012}_{-0.00017}$ & $0.022486^{+0.000071}_{-0.00027}$ & $0.02238^{+0.00012}_{-0.00019}$ & $0.02241\pm 0.00014$ & $0.02240\pm 0.00014$ \\ 

 $r_{\rm d}h$ & $100.07^{+0.92}_{-0.67}$ & $99.9\pm{1.6}$ & $100.51^{+0.64}_{-1.1}$ & $100.02^{+0.74}_{-1.0}$ & $100.31^{+0.88}_{-0.65}$ & $99.93^{+1.0}_{-0.77}$ \\ 

 $M$ & $-19.433^{+0.016}_{-0.023}$ & $-19.436^{+0.02}_{-0.025}$ & $-$ & $-$ & $-$ & $-$ \\ 
 \hline\hline

\end{longtable}}


{\footnotesize    
\begin{longtable}{l c c c c c c c}
\caption{Model comparison for $\Lambda$CDM, SE~\eqref{eq:pot}, MSE~\eqref{eq:potMSE}, axionlike~\eqref{eq:ax} and power law~\eqref{eq:pl} potentials. Here $\Lambda$CDM with the minimum AIC and BIC is considered as the reference model. $\log_{10}V_0=7,\; 7,\; 11\; {\rm and}\; 13$ has been considered for the SE potential for DESIPanth, BAOPanth, DESIUnion and BAOUnion respetively as this parameter is unconstrained for these data sets.}
\label{tab:AIC}\\
 \hline\hline
 Parameter &DESIPanth & BAOPanth & DESID5 & BAOD5 & DESIUnion & BAOUnion\\ 
\hline\hline
&&&$\Lambda$CDM&&&\\
\hline
$\chi^2_{\rm min}$ & $1434.03$ & $1435.08$ & $1719.33$ & $1721.39$ & $59.00$ & $60.13$ \\ 

 $\chi^2_{\rm red}$ & $0.88$ & $0.88$ & $0.92$ & $0.92$ & $0.95$ & $0.94$ \\ 
 
 AIC & $1446.03$ & $1447.07$ & $1729.33$ & $1731.39$ & $69.00$ & $70.13$ \\  
 
 $\Delta$AIC & $0$ & $0$ & $0$ & $0$ & $0$ & $0$ \\ 

 BIC & $1478.42$ & $1479.48$ & $1757.01$ & $1759.04$ & $80.02$ & $81.30$ \\ 

 $\Delta$BIC & $0$ & $0$ & $0$ & $0$ & $0$ & $0$ \\  

 \hline\hline
&&& SE~\eqref{eq:pot} for $n=6$ &&& \\
\hline
$\chi^2_{\rm min}$ & $1437.82$ & $1441.73$ & $1713.78$ & $1677.56$ & $58.30$ & $61.88$ \\ 

 $\chi^2_{\rm red}$ & $0.88$ & $0.88$ & $0.92$ & $0.90$ & $0.96$ & $0.98$ \\ 
 
 AIC & $1451.82$ & $1455.73$ & $1725.78$ & $1689.56$ & $70.30$ & $73.88$ \\  

 $\Delta$AIC & $5.80$ & $8.66$ & $-3.54$ & $-41.83$ & $1.30$ & $3.75$ \\

 BIC & $1489.61$ & $1493.54$ & $1758.99$ & $1722.78$ & $83.53$ & $87.29$ \\ 

 $\Delta$BIC & $11.20$ & $14.06$ & $1.98$ & $-36.26$ & $3.51$ & $5.99$ \\

  \hline
 &&& SE~\eqref{eq:pot} for $n=10$ &&& \\ 
\hline
$\chi^2_{\rm min}$ & $1435.89$ & $1438.21$ & $1676.02$ & $1679.73$ & $58.58$ & $60.14$ \\ 

 $\chi^2_{\rm red}$ & $0.88$ & $0.88$ & $0.90$ & $0.90$ & $0.96$ & $0.95$ \\ 
 
 AIC & $1449.89$ & $1452.21$ & $1688.02$ & $1691.73$ & $70.58$ & $72.14$ \\  
 
 $\Delta$AIC & $3.86$ & $5.14$ & $-41.30$ & $-39.65$ & $1.58$ & $2.01$ \\ 

 BIC & $1487.69$ & $1490.02$ & $1721.23$ & $1724.95$ & $83.81$ & $85.54$ \\ 

 $\Delta$BIC & $9.27$ & $10.53$ & $-35.77$ & $-34.08$ & $3.79$ & $4.24$ \\  
 
 \hline\hline

&&&  MSE~\eqref{eq:potMSE} for $n=6$ &&&\\
\hline
$\chi^2_{\rm min}$ & $1435.34$ & $1441.59$ & $1720.04$ & $1724.48$ & $65.25$ & $68.87$ \\ 

 $\chi^2_{\rm red}$ & $0.88$ & $0.88$ & $0.92$ & $0.92$ & $1.07$ & $1.09$ \\ 
 
 AIC & $1449.34$ & $1455.59$ & $1732.03$ & $1736.47$ & $77.24$ & $80.87$ \\  
 
 $\Delta$AIC & $3.31$ & $8.51$ & $2.70$ & $5.08$ & $8.24$ & $10.74$ \\ 

 BIC & $1487.14$ & $1493.39$ & $1765.25$ & $1769.70$ & $90.47$ & $94.27$ \\ 

 $\Delta$BIC & $8.72$ & $13.91$ & $8.24$ & $10.66$ & $10.45$ & $12.97$ \\  

 \hline
&&& MSE~\eqref{eq:potMSE} for $n=10$ &&& \\
\hline 
$\chi^2_{\rm min}$ & $1434.13$ & $1435.82$ & $1734.22$ & $1723.45$ & $59.96$ & $63.06$ \\ 

 $\chi^2_{\rm red}$ & $0.88$ & $0.88$ & $0.93$ & $0.92$ & $0.98$ & $1.00$ \\ 
 
 AIC & $1448.13$ & $1449.82$ & $1746.22$ & $1735.45$ & $71.96$ & $75.06$ \\  
 
 $\Delta$AIC & $2.10$ & $2.75$ & $16.89$ & $4.06$ & $2.96$ & $4.93$ \\ 

 BIC & $1485.93$ & $1487.63$ & $1779.43$ & $1768.67$ & $85.19$ & $88.47$ \\ 

 $\Delta$BIC & $7.51$ & $8.15$ & $22.42$ & $9.63$ & $5.17$ & $7.17$ \\  

\hline \hline

 &&& Axionlike~\eqref{eq:ax} for $n=3$ &&& \\ 
\hline
$\chi^2_{\rm min}$ & $1448.79$ & $1469.22$ & $1684.22$ & $1677.70$ & $58.01$ & $59.14$ \\ 

 $\chi^2_{\rm red}$ & $0.89$ & $0.90$ & $0.90$ & $0.90$ & $0.95$ & $0.94$ \\ 

 AIC & $1463.00$ & $1483.22$ & $1696.22$ & $1689.70$ & $70.01$ & $71.14$ \\  
 
 $\Delta$AIC & $16.97$ & $36.15$ & $-33.10$ & $-41.68$ & $1.01$ & $1.01$ \\
 
 BIC & $1500.79$ & $1521.03$ & $1729.44$ & $1722.92$ & $83.24$ & $84.54$ \\ 

 $\Delta$BIC & $22.37$ & $41.55$ & $-27.57$ & $-36.11$ & $3.22$ & $3.24$ \\  
 
 \hline\hline

 &&& Axionlike~\eqref{eq:ax} for $n=6$ &&& \\ 
\hline
$\chi^2_{\rm min}$ & $1434.45$ & $1436.72$ & $1713.87$ & $1693.91$ & $59.85$ & $61.30$ \\ 

 $\chi^2_{\rm red}$ & $0.88$ & $0.88$ & $0.92$ & $0.91$ & $0.98$ & $0.97$ \\ 
 
 AIC & $1448.45$ & $1450.72$ & $1725.87$ & $1705.91$ & $71.85$ & $73.30$ \\  
 
 $\Delta$AIC & $2.42$ & $3.64$ & $-3.46$ & $-25.47$ & $2.85$ & $3.17$ \\

 BIC & $1486.25$ & $1488.52$ & $1759.08$ & $1739.13$ & $85.08$ & $86.71$ \\ 

 $\Delta$BIC & $7.83$ & $9.04$ & $2.07$ & $-19.90$ & $5.06$ & $5.41$ \\  
 
 \hline\hline

 &&& Power law~\eqref{eq:pl} for $n=3$ &&& \\ 
\hline
$\chi^2_{\rm min}$ & $1434.37$ & $1441.03$ & $1695.48$ & $1693.24$ & $58.29$ & $74.70$ \\ 

 $\chi^2_{\rm red}$ & $0.88$ & $0.88$ & $0.90$ & $0.90$ & $0.95$ & $1.18$ \\ 
 
 AIC & $1448.37$ & $1455.03$ & $1707.48$ & $1705.24$ & $70.29$ & $86.70$ \\  
 
 $\Delta$AIC & $2.34$ & $7.96$ & $-21.84$ & $-26.15$ & $1.29$ & $16.57$ \\ 

 BIC & $1486.16$ & $1492.84$ & $1740.70$ & $1738.46$ & $83.52$ & $100.11$ \\ 

 $\Delta$BIC & $7.74$ & $13.35$ & $-16.31$ & $-20.58$ & $3.50$ & $18.81$ \\  

 \hline\hline

 &&& Power law~\eqref{eq:pl} for $n=6$ &&& \\ 
\hline
$\chi^2_{\rm min}$ & $1438.94$ & $1437.13$ & $1716.83$ & $1700.29$ & $57.99$ & $59.48$ \\ 

 $\chi^2_{\rm red}$ & $0.88$ & $0.88$ & $0.92$ & $0.91$ & $0.95$ & $0.94$ \\ 
 
 AIC & $1452.94$ & $1451.13$ & $1728.83$ & $1712.29$ & $69.99$ & $71.48$ \\  
 
 $\Delta$AIC & $6.91$ & $4.06$ & $-0.50$ & $-19.10$ & $0.99$ & $1.35$ \\ 
 
 BIC & $1490.74$ & $1488.94$ & $1762.04$ & $1745.51$ & $83.21$ & $84.88$ \\ 

 $\Delta$BIC & $12.32$ & $9.46$ & $5.03$ & $-13.53$ & $3.20$ & $3.58$ \\  
 \hline\hline
\end{longtable}}

{\footnotesize
\begin{longtable}{c c c c c c c c}
\caption{Theoretical values of $f_{\rm EDE}$, $z_{\rm c}$ and $H_0$ using the constraint values of the parameters of the SE~\eqref{eq:pot}, MSE~\eqref{eq:potMSE}, axionlike~\eqref{eq:ax} and power law~\eqref{eq:pl} potentials. $H_0=100h$ is calulated from Eq.~\eqref{eq:h0}. Maximum value of $f_{\rm EDE}$ is estimated and with this the value of $h$ is estimated using the 1$\sig$ bound of $\Om_{\m0}$. Note that the bound given in this table for $h$ are not the 1$\sig$ confidence level rather these are just the estimation of $h$ for the largest value of $f_{\rm EDE}$ with 1$\sig$ bound of $\Om_{\m0}$ from Eq.~\eqref{eq:h0}. The initial conditions are $\{\phi_i,\phi'_i\}\equiv \{1.13,0.15\},\{0.9 \phi_0,10^{-10}\},\{\pi f-0.01 f,10^{-15}\}\; {\rm and}\; \{f,10^{-15}\}$ for the SE, MSE, axionlike and power law potential respectively.}
\label{tab:fEDE}\\
 \hline\hline
 Parameter & DESIPanth & BAOPanth & DESID5 & BAOD5 & DESIUnion & BAOUnion\\ 
\hline\hline
&&&$\Lambda$CDM&&&\\
\hline
$f_{\rm EDE}$ & $-$ & $-$ & $-$ & $-$ & $-$ & $-$ \\
$z_{\rm c}$ & $-$ & $-$ & $-$ & $-$ & $-$ & $-$
\\ 
$h$ & $0.6768_{-0.0062}^{+0.0064}$ & $0.6750_{-0.0063}^{+0.0066}$ & $0.7129\pm 0.0024$ & $0.7130\pm 0.0024$ & $0.6784_{-0.0063}^{+0.0065}$ & $0.6766_{-0.0066}^{+0.0068}$ \\
 \hline\hline

&&&  SE~\eqref{eq:pot} for $n=6$ &&&\\
\hline
$f_{\rm EDE}$ & $<0.0252$ & $<0.024$ & $< 0.132$ & $< 0.117$ & $<0.078$ & $<0.03$ \\
$z_{\rm c}$ & $\sim 3690$ & $\sim 3666$ & $\sim 5500$ & $\sim 4958$ & $\sim3630$ & $\sim 3741$ \\
$h$ & $0.6777_{-0.0022}^{+0.0023}$ & $0.6760_{-0.0023}^{+0.0024}$& $0.6996_{-0.0024}^{+0.0025}$ & $0.6981_{-0.0029}^{+0.0019}$ & $0.6797_{-0.0021}^{+0.0024}$ & $0.6780\pm 0.0024$ \\ \hline

&&&  SE~\eqref{eq:pot} for $n=10$ &&&\\
\hline
$f_{\rm EDE}$ & $<0.0134$ & $<0.013$ & $<0.25$ & $<0.288$ & $-$ & $<0.0154$ \\
$z_{\rm c}$ & $\sim 5095$  & $\sim 5093$ & $\sim 3988$ & $\sim 4173$ & $-$ & $\sim 5121$ \\
$h$ & $0.6765_{-0.0022}^{+0.0023}$ & $0.6748_{-0.0022}^{+0.0024}$ & $71.2_{-0.0030}^{+0.0019}$ & $0.7156_{-0.0034}^{+0.0018}$ & $-$ & $0.6766_{-0.0023}^{+0.0026}$\\

\hline \hline

&&& MSE~\eqref{eq:potMSE} for $n=6$ &&& \\
\hline
$f_{\rm EDE}$ & $\sim 0$ & $\sim 0$ & $\sim 0$ & $\sim 0$ & $\sim 0$ & $\sim 0$ \\
$z_{\rm c}$ & $-$ & $-$ & $-$ & $-$ & $-$ & $-$
 \\ 
 $h$ & $0.6732_{-0.0021}^{+0.0029}$ & $0.6712_{-0.0017}^{+0.0036}$ & $0.7040\pm 0.001$& $0.7008_{-0.0003}^{+0.0015}$ & $0.6746_{-0.0022}^{+0.0026}$ & $0.6728_{-0.0023}^{+0.0028}$ \\
  \hline
  
 &&& MSE~\eqref{eq:potMSE} for $n=10$ &&& \\ 
\hline
$f_{\rm EDE}$ & $\sim 0$ & $\sim 0$ & $\sim 0$ & $\sim 0$ & $\sim 0$ & $\sim 0$  \\
$z_{\rm c}$ & $-$ & $-$ & $-$ & $-$ & $-$ & $-$
 \\ 
$h$ & $0.6750_{-0.0022}^{+0.0023}$ & $0.6735_{-0.0027}^{+0.0028}$ & $0.7042\pm -0.0014$ & $0.7074_{-0.0005}^{+0.0013}$& $0.6761_{-0.0023}^{+0.0024}$ & $0.6741_{-0.0022}^{+0.0026}$ \\

 \hline\hline

&&& Axionlike~\eqref{eq:ax} for $n=3$ &&& \\
 \hline
$f_{\rm EDE}$ & $<0.195$ & $<0.32$ & $<0.3$ & $<0.35$ & $<0.17$ & $<0.19$ \\
$z_{\rm c}$ & $\sim 12192$ & $\sim 10174$ & $\sim 10571$ & $\sim 9977$ & $\sim 12876$ & $\sim 12394$
 \\ 
 $h$ & $0.6910_{-0.0024}^{+0.0004}$ & $0.6990_{-0.002}^{+0.0005}$ & $0.7112_{-0.0026}^{+0.0032}$  & $0.7157_{-0.0022}^{+0.0027}$ & $0.6916_{-0.0020}^{+0.0034}$ & $0.6928_{-0.0024}^{+0.0033}$ \\
  \hline
  
 &&& Axionlike~\eqref{eq:ax} for $n=6$ &&& \\ 
\hline
$f_{\rm EDE}$ & $<0.23$ & $<0.34$ & $<38$ & $<0.38$ & $<0.11$ & $<0.28$ \\
$z_{\rm c}$ & $\sim 10814$ & $\sim 9300$ & $\sim 8893$ & $\sim 8921$ & $\sim 7458$ & $\sim 9152$
 \\ 
 $h$ & $0.6915_{-0.0022}^{+0.030}$ & $0.7038_{-0.002}^{+0.0033}$ & $0.7204_{-0.0024}^{+0.0029}$ & $0.7193_{-0.0024}^{+0.0028}$ & $0.6861_{-0.0022}^{+0.0033}$ & $0.7002_{-0.0024}^{+0.0030}$ \\
 \hline\hline

&&& Power law~\eqref{eq:pl} for $n=3$ &&& \\ 
\hline
$f_{\rm EDE}$ & $<0.086$ & $<0.072$ & $<0.33$ & $<0.3$ & $<0.134$ & $<0.14$ \\
$z_{\rm c}$ & $\sim 3468$ & $\sim 3642$ & $\sim 13840$ & $\sim 14407$ & $\sim 4043$ & $\sim 3981$
 \\ 
 $h$ & $0.6852_{-0.0024}^{+0.0025}$ & $0.6791_{-0.0021}^{+0.0036}$ & $0.7117_{-0.0024}^{+0.0022}$ & $0.7076_{-0.0025}^{+0.0026}$ & $0.6955_{-0.0022}^{+0.0028}$ & $0.6925_
 {-0.0022}^{+0.0032}$ \\
 \hline
 &&& Power law~\eqref{eq:pl} for $n=6$ &&& \\ 
\hline
$f_{\rm EDE}$ & $<0.04$ & $<0.058$ & $<0.16$ & $<0.165$ & $<0.13$ & $<0.097$ \\
$z_{\rm c}$ & $\sim 4075$ & $\sim 3692$ & $\sim 5100$ & $\sim 5037$ & $\sim 4515$ & $\sim 4926$
 \\ 
 $h$ & $0.6778_{-0.0019}^{+0.0032}$ & $0.6782_{-0.0020}^{+0.0031}$ & $0.7065_{-0.0035}^{+0.0022}$ & $0.7060_{-0.0032}^{+0.0020}$ & $0.6930_{-0.0022}^{+0.0028}$ & $0.6867_{-0.0022}^{+0.0028}$ \\
 \hline\hline
 
\end{longtable}}

\subsection{Results}

\begin{figure}[ht]
\centering
\includegraphics[scale=.28]{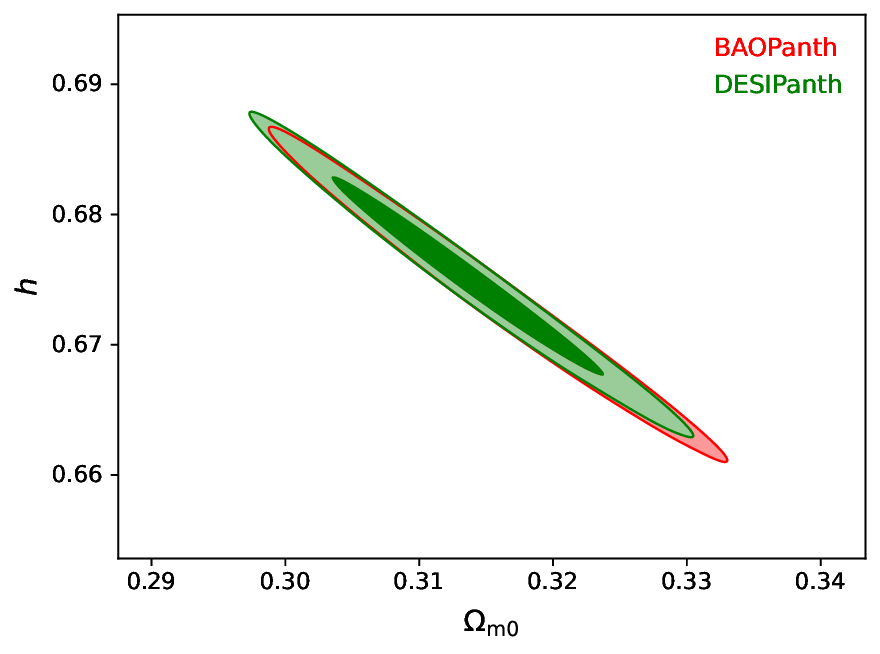} 
\includegraphics[scale=.28]{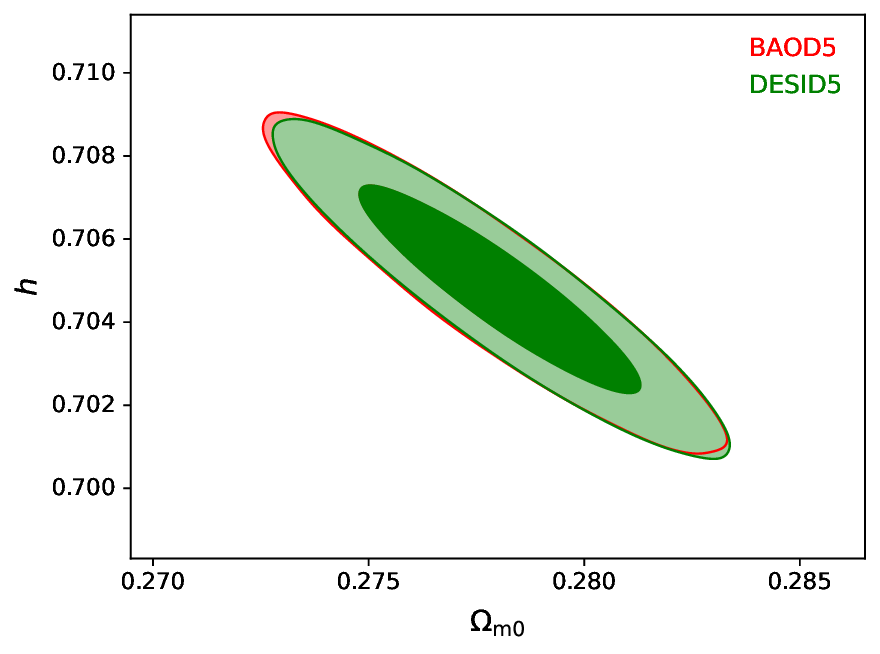}
\includegraphics[scale=.28]{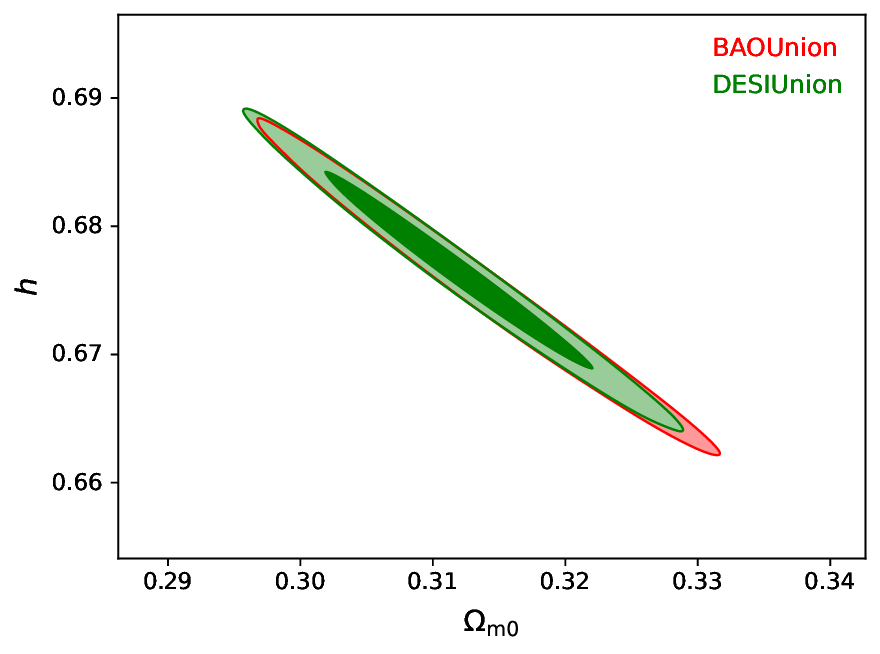}
\caption{1$\sig$ and 2$\sig$ confidence levels on the $\Om_{\m 0}-h$ plane for the considered data combinations for $\Lambda$CDM model.}
\label{fig:cont_lcdm}
\end{figure}

The observational constraints of the model parameters of all four potentials and the $\Lambda$CDM model are shown in Tab.~\ref{tab:constraint}. Corresponding values of Akaike information criterion (AIC) and Bayesian Information Criterion (BIC) \cite{Akaike74,Liddle:2006tc,Trotta:2017wnx,Shi:2012ma} are listed in Tab.~\ref{tab:AIC}  along with the minimum chi-squared ($\chi_{\rm min}^2$) and reduced chi-squared ($\chi_{\rm red}^2=\chi_{\rm min}^2/\nu$), where $\nu=k-N$ is the degree of freedom while $k$ and $N$ are the total number of data points and the total number of model parameters respectively. AIC and BIC are defined as 
\begin{eqnarray}
   \rm AIC &=& 2N - 2\ln\mathcal{L}_{\rm max} 
 = 2N + \chi^2_{\rm min} \, , \\
   \rm BIC &=& N \ln K - 2\ln\mathcal{L}_{\rm max} = N \ln K + \chi_{\rm min}^2 \, ,
\end{eqnarray}
where, $\mathcal{L}_{\rm max}$ is the maximum likelihood. We also compute the difference in AIC ($\Delta$AIC) and BIC ($\Delta$BIC) between the EDE and $\Lambda$CDM models such that
\begin{eqnarray}
    \Delta \rm AIC &=& \rm AIC_{\rm EDE} - AIC_{\rm \Lambda CDM}  \\ 
    \Delta \rm BIC &=& \rm BIC_{\rm EDE} - BIC_{\rm \Lambda CDM} \, .
\end{eqnarray}
$\Delta$AIC and $\Delta$BIC tell us about the preference of the model by the observational data in comparison to a reference model. In our case, we have considered $\Lambda$CDM model as the reference model which also has less number of free parameters than the EDE models. From Tab.~\ref{tab:AIC} we can infer from the values of $\chi^2_{\rm min}$ and $\chi^2_{\rm red}$ that the data prefers the $\Lambda$CDM model more than the EDE models as the values of $\chi^2_{\rm min}$ and $\chi^2_{\rm red}$ are minimum for $\Lambda$CDM except for the data combinations with DESY5. In this regard, it should be noted that even though we have some constraint on the parameters $\{\mu,\;  \phi_0,\; f\}$ the constraints are not very stringent and one has to take care of it while calculating AIC and BIC. In Tab.~\ref{tab:AIC} the values of AIC and BIC are listed. EDE models also have similar values of $\chi^2_{\rm min}$ and $\chi^2_{\rm red}$, except for the data combinations with DESY5, but they have more free parameters. So, for our case it is safe to choose the standard $\Lambda$CDM model as the reference model. Now, larger the values of $\Delta$AIC and $\Delta$BIC lesser is the probability that the data will prefer our model over the standard $\Lambda$CDM model. More precisely, if $0<\Delta\rm AIC,\Delta \rm BIC<2$ then the observational data can not distinguish between the models. If $\Delta\rm AIC,\Delta \rm BIC>10$ then the deviation of the considered model from the reference model is decisive. In between these limits the deviation is significant \cite{Shi:2012ma}. For the data combinations with Pantheon$+$ the data prefer MSE potential \eqref{eq:potMSE} with $n=10$ compared to the other EDE potentials. But, for the data combinations with Union3 the data prefer power law potential~\eqref{eq:pl} with $n=6$ over the other EDE potentials. Even though there are slight preferences on one EDE potential over other by some data combinations the preference is not very significant. For the data combinations with DESY5 the data show preference towards EDE potentials specially the SE potential~\eqref{eq:pl} with $n=10$ and power law potential~\eqref{eq:pl} with $n=6$. For DESY5 only MSE potential behaves closer to the $\Lambda$CDM model.

\begin{figure*}[ht]
\centering
\includegraphics[scale=.5]{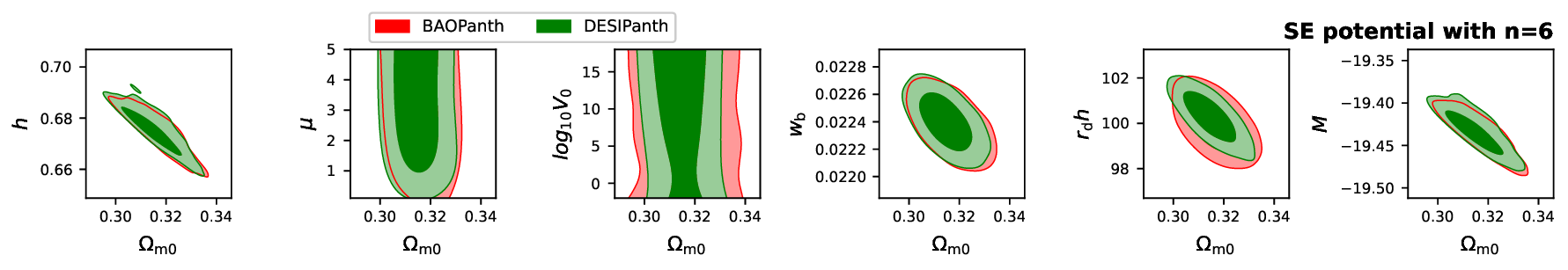} 
\includegraphics[scale=.5]{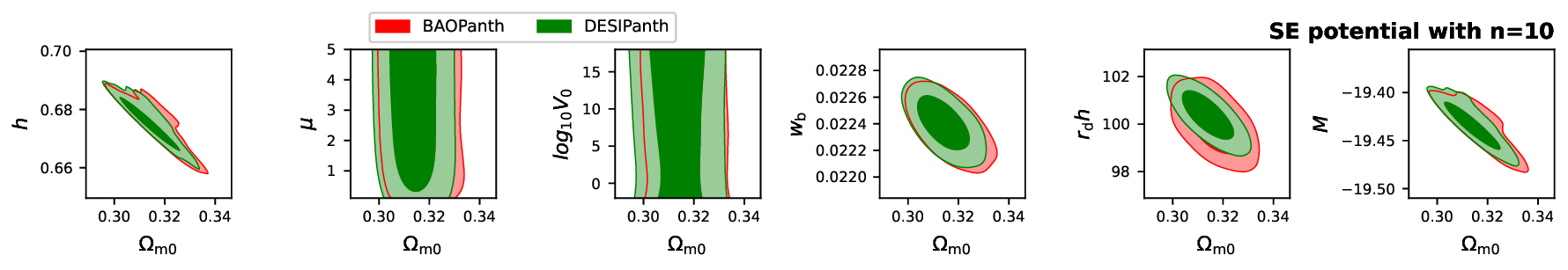}
\includegraphics[scale=.5]{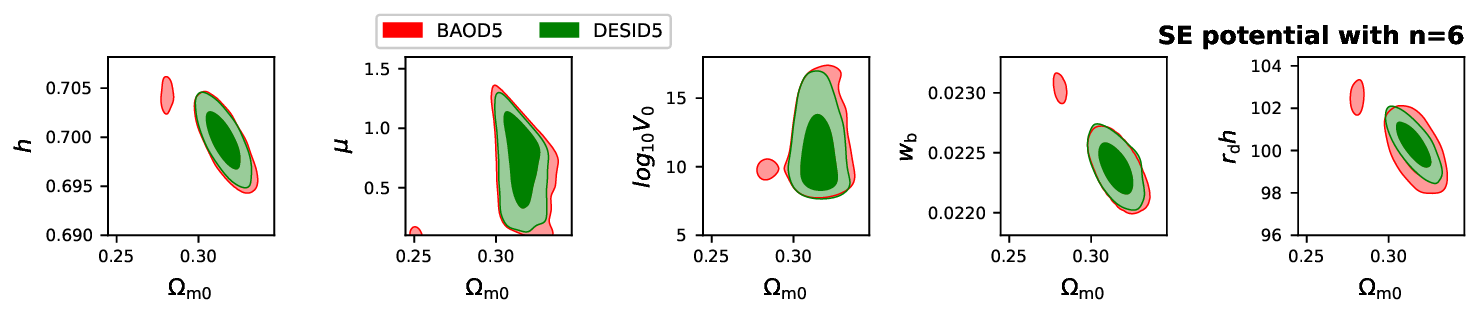}
\includegraphics[scale=.5]{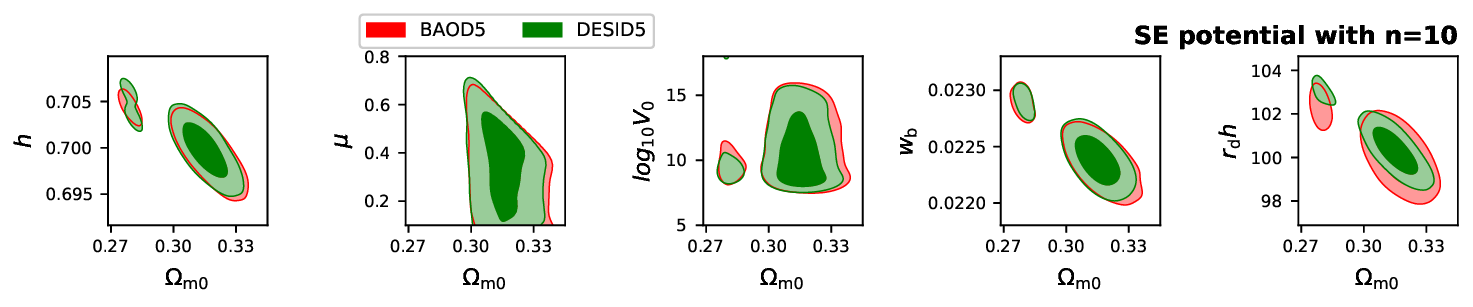}
\includegraphics[scale=.5]{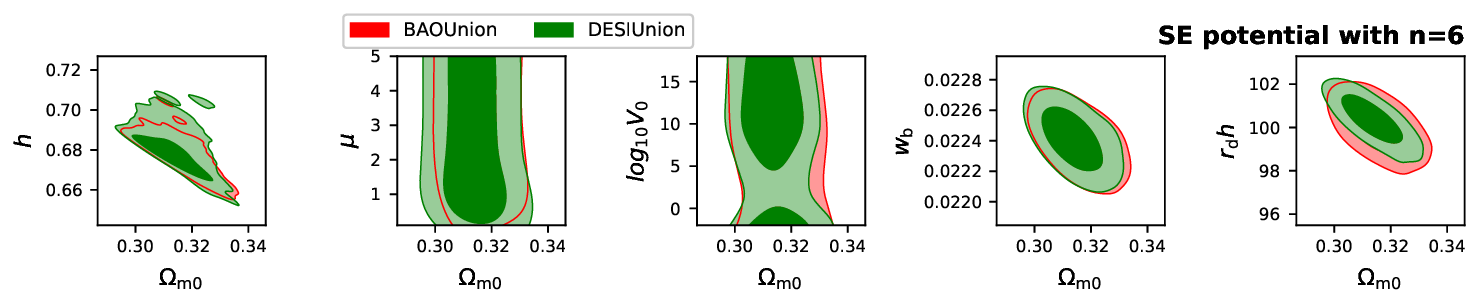}
\includegraphics[scale=.5]{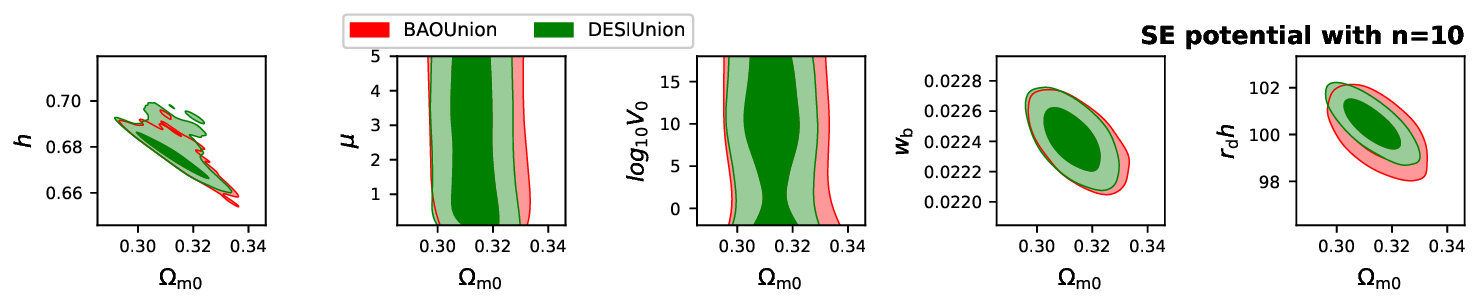}
\caption{1$\sig$ and 2$\sig$ confidence levels of the parameters of the SE potential~\eqref{eq:pot} for the considered data combinations.}
\label{fig:cont_se}
\end{figure*}

\begin{figure*}[ht]
\centering
\includegraphics[scale=.5]{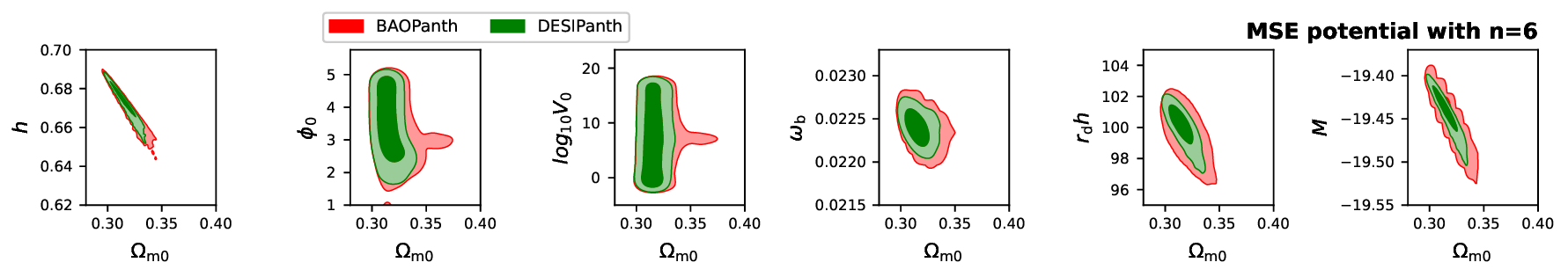} 
\includegraphics[scale=.5]{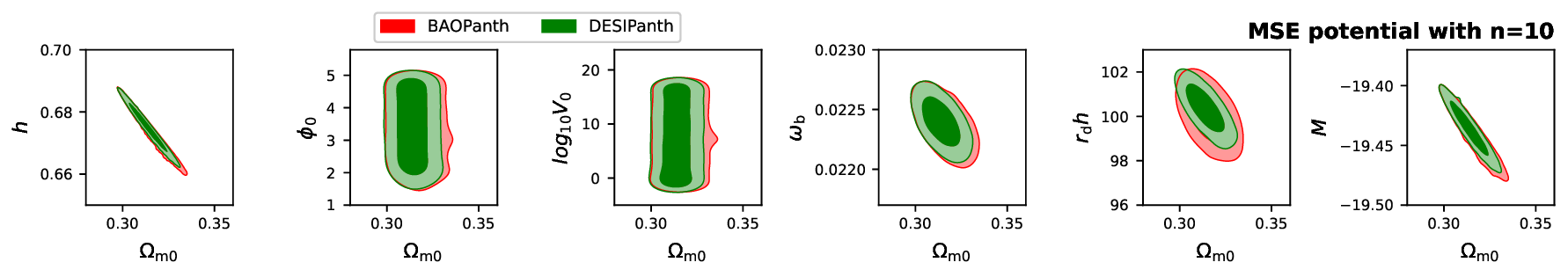}
\includegraphics[scale=.5]{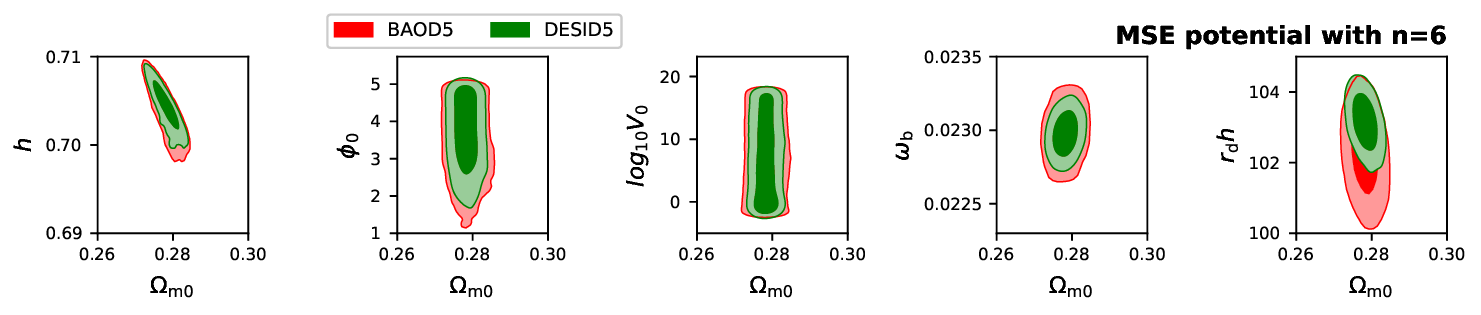}\\
\includegraphics[scale=.5]{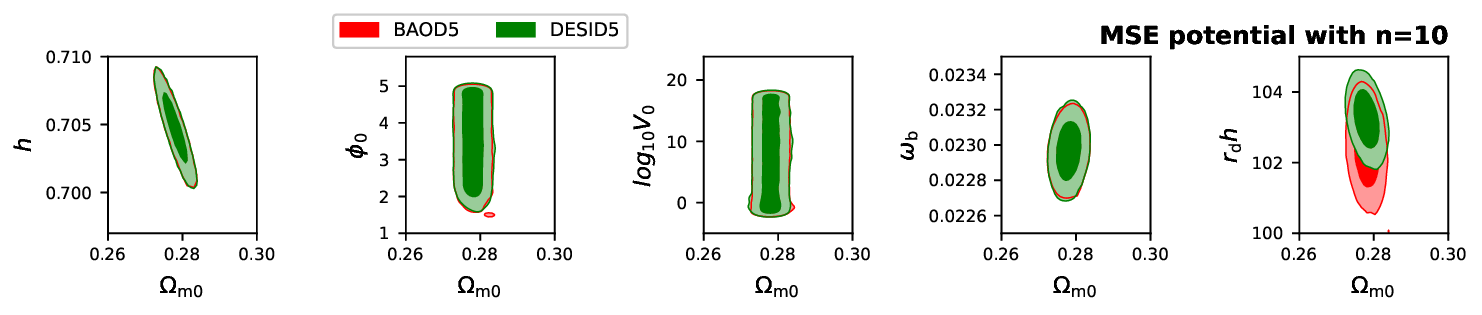}\\
\includegraphics[scale=.5]{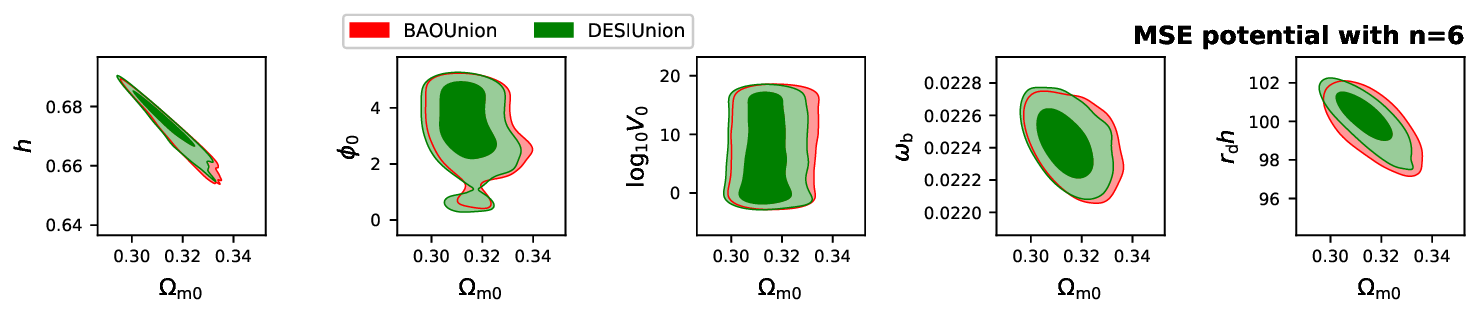}\\
\includegraphics[scale=.5]{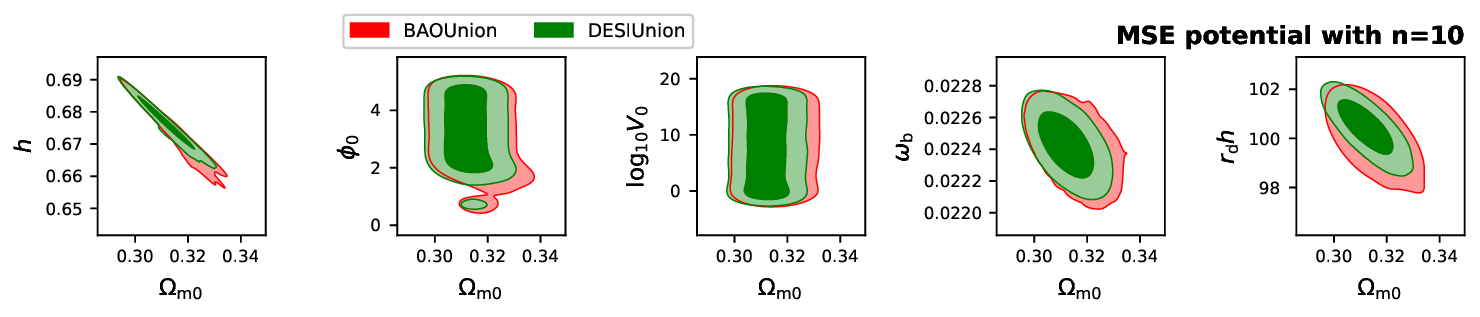}
\caption{1$\sig$ and 2$\sig$ confidence levels of the parameters of the MSE potential~\eqref{eq:potMSE} for the considered data combinations.}
\label{fig:cont_mse}
\end{figure*}

\begin{figure*}[ht]
\centering
\includegraphics[scale=.5]{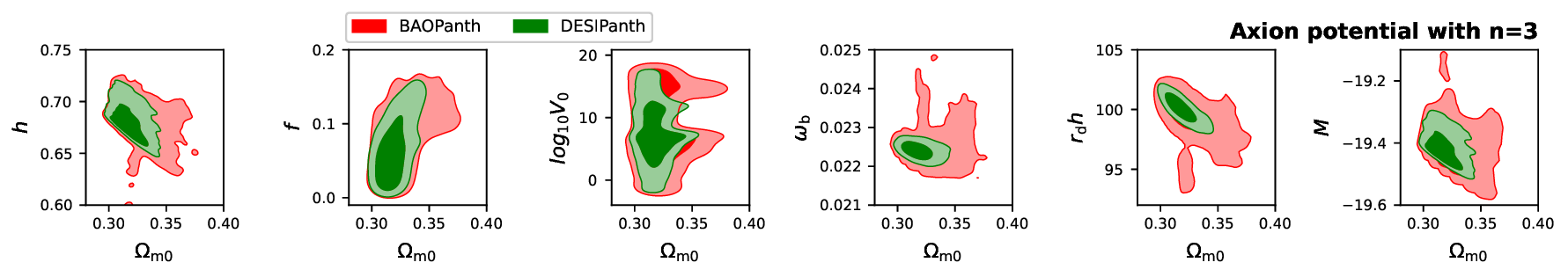} 
\includegraphics[scale=.5]{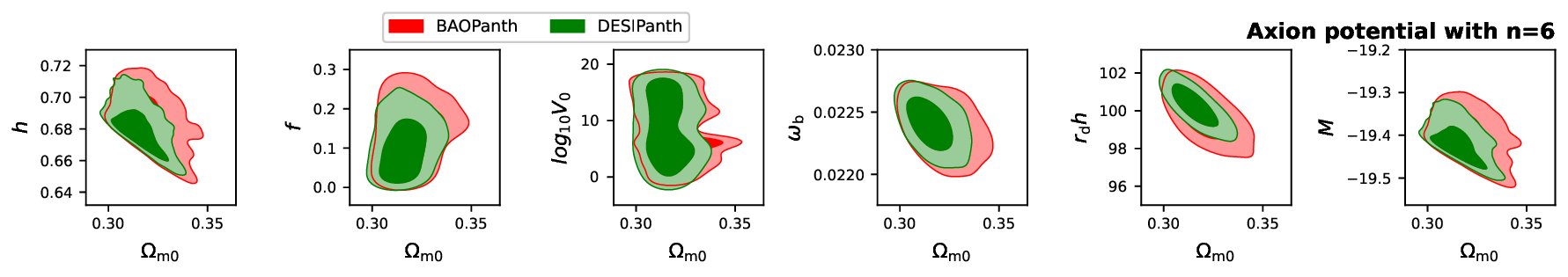}
\includegraphics[scale=.5]{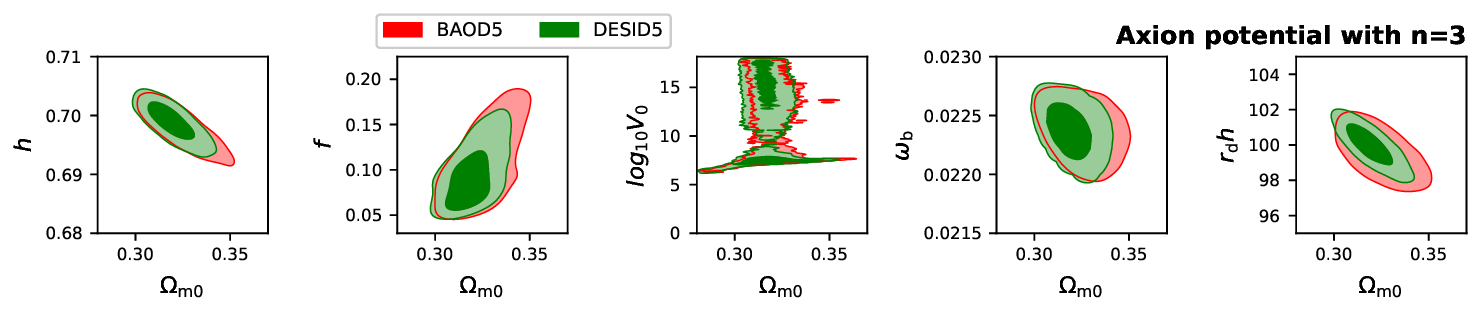}
\includegraphics[scale=.5]{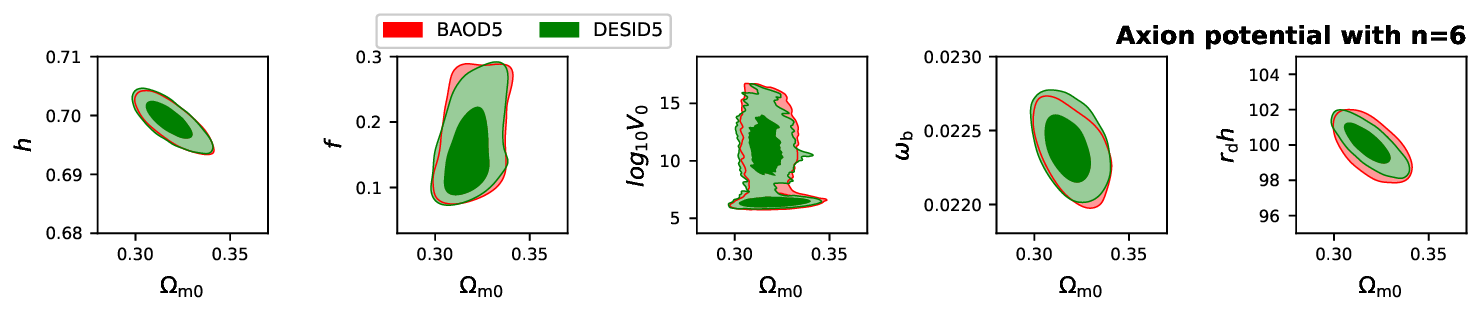}
\includegraphics[scale=.5]{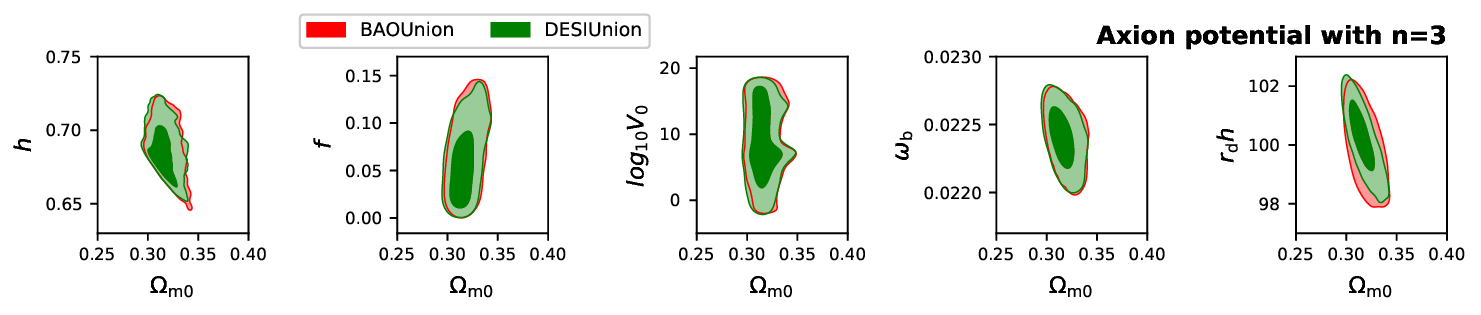}
\includegraphics[scale=.5]{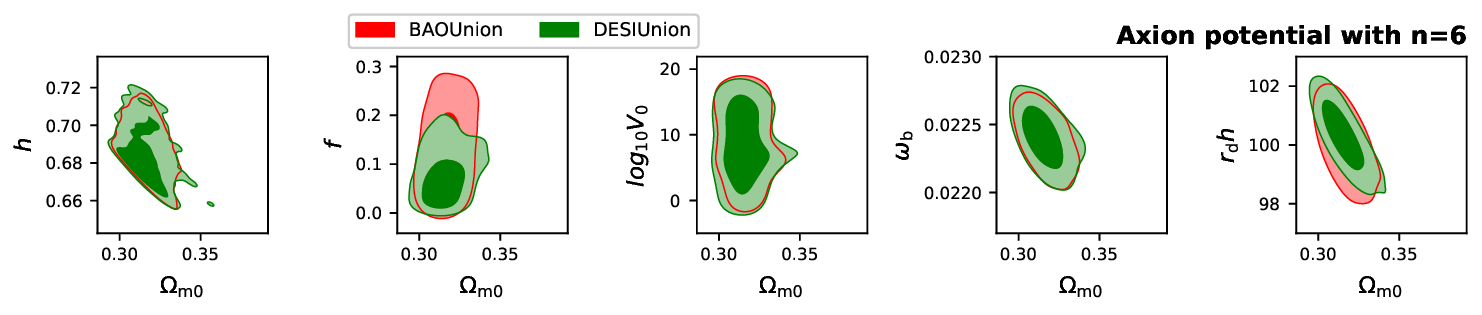}
\caption{1$\sig$ and 2$\sig$ confidence levels of the parameters of the axionlike potential~\eqref{eq:ax} for the considered data combinations.}
\label{fig:cont_ax}
\end{figure*}

\begin{figure*}[ht]
\centering
\includegraphics[scale=.5]{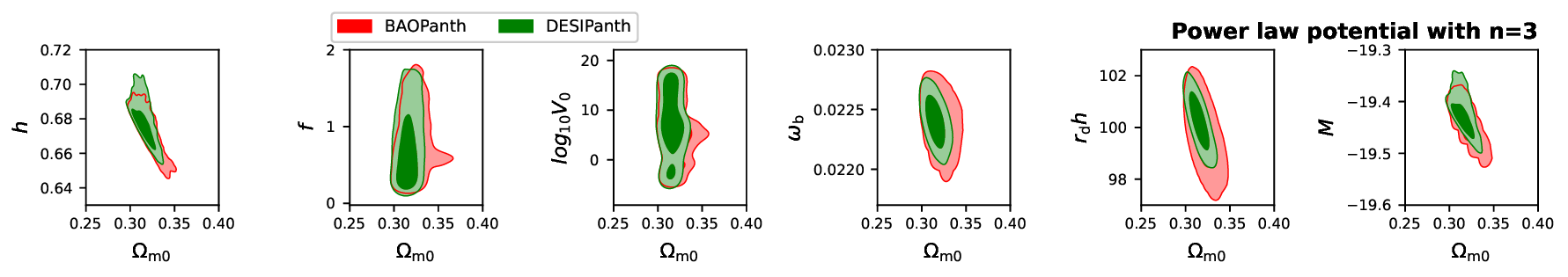} 
\includegraphics[scale=.5]{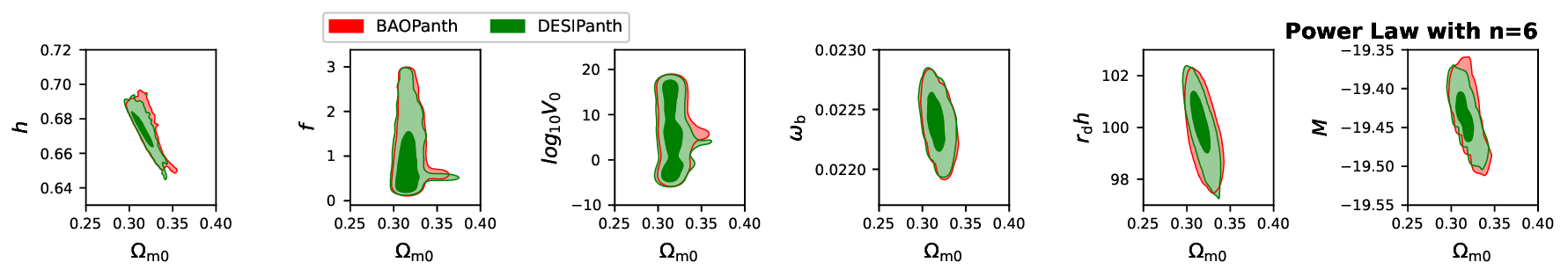}
\includegraphics[scale=.5]{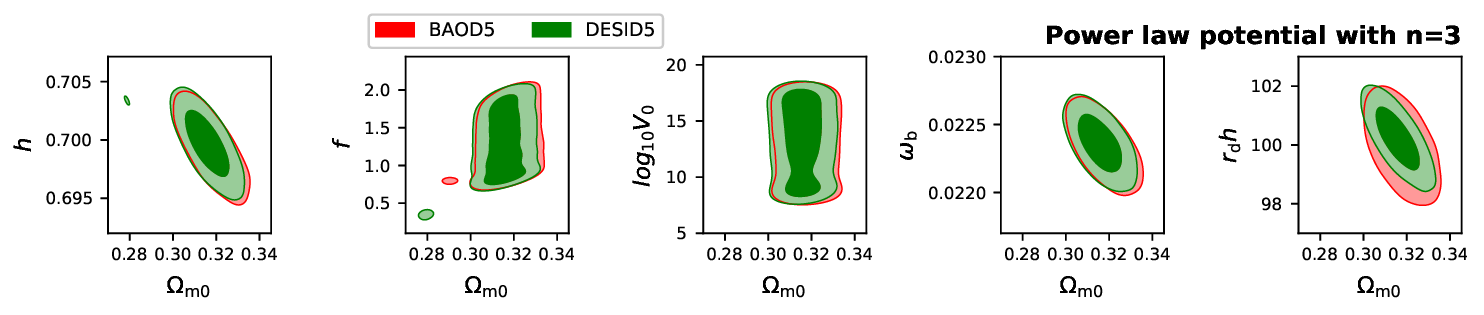}
\includegraphics[scale=.5]{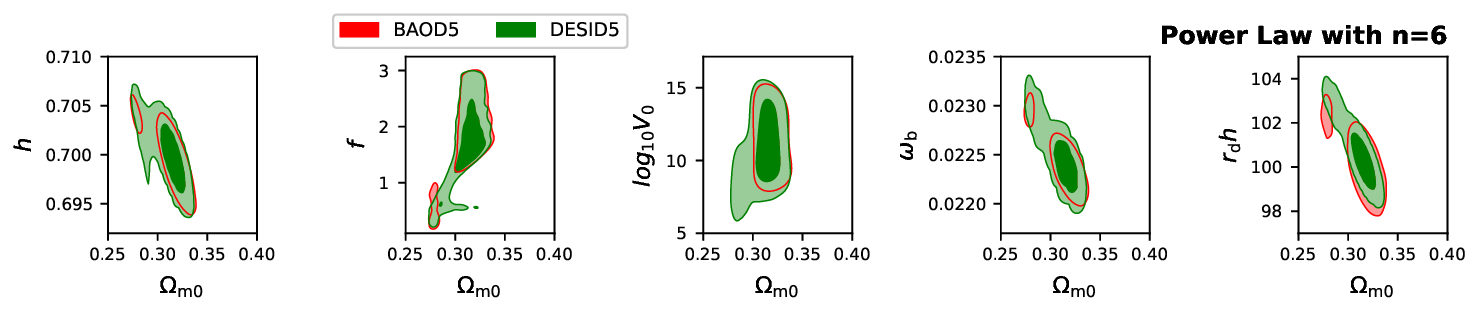}
\includegraphics[scale=.5]{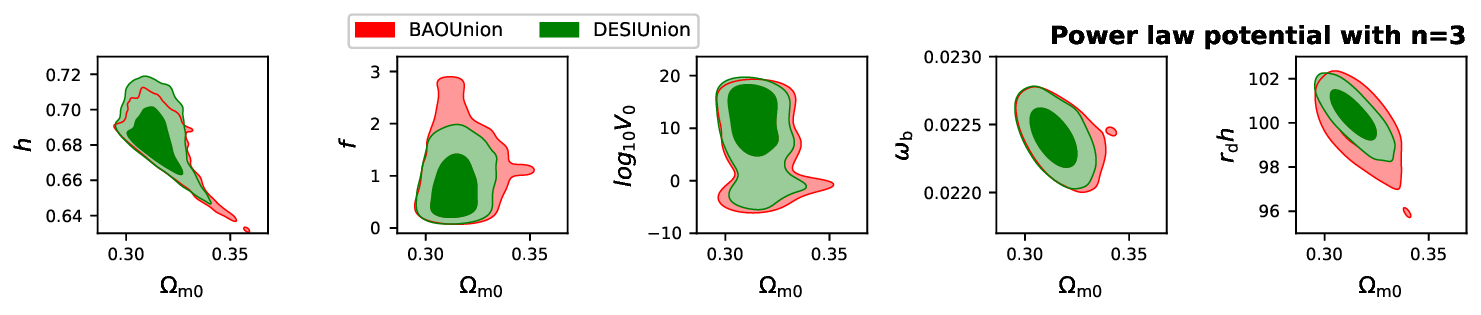}
\includegraphics[scale=.5]{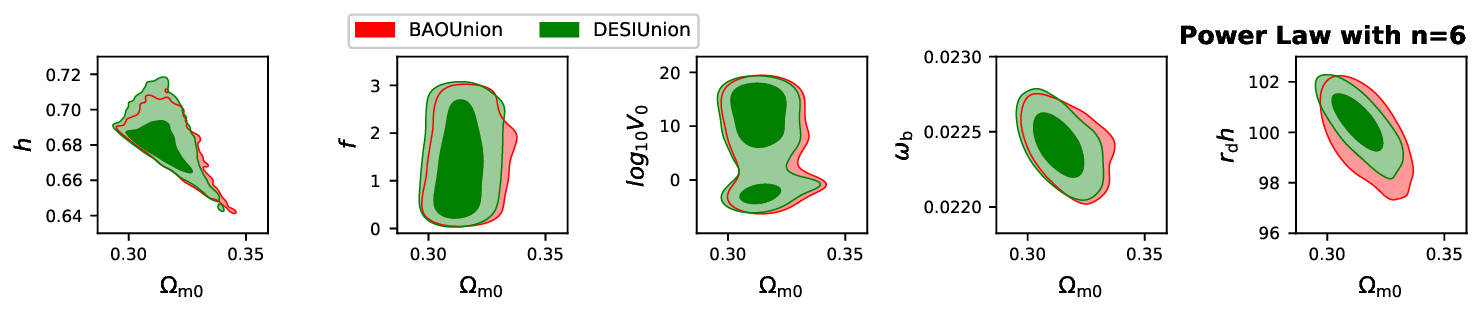}
\caption{1$\sig$ and 2$\sig$ confidence levels of the parameters of the power law potential~\eqref{eq:pl} for the considered data combinations.}
\label{fig:cont_pl}
\end{figure*}

In Tab.~\ref{tab:fEDE} we have listed the theoretically calculated values of $f_{\rm EDE}$ and $z_{\rm c}$ by solving the Eq.~\eqref{eq:eom_phi} and $h$ from the Eq.~\eqref{eq:h0}. For Tab.~\ref{tab:fEDE} we have used the constraints values of the parameter from Tab.~\ref{tab:constraint}. We have estimated the largest possible value of $f_{\rm EDE}$ and the corresponding $z_{\rm c}$ around the redshift $10^3$ to $10^4$, {\it i.e.}, around the matter-radiation equality.The values of $h$ in Tab.~\ref{tab:fEDE} are calculated from Eq.~\eqref{eq:h0} for the largest estimated values of $f_{\rm EDE}$ and the observational constraint of $\Om_{\m 0}$. So, the bounds on $h$ given in Tab.~\ref{tab:fEDE} are not the 1$\sig$ observational bounds rather they just correspond to the bounds of $\Om_{\m0}$ for the largest values of $f_{\rm EDE}$. Since the values of $h$ are calculated from the Eq.~\eqref{eq:h0} the Tab.~\ref{tab:fEDE} also shows validity of Eq.~\eqref{eq:h0} and we see that, except for some values for axionlike potential, the Eq.~\eqref{eq:h0} gives almost same values of $h$ as the constraint values in Tab.~\ref{tab:constraint} as long as we use the constraint values of the parameter from Tab.~\ref{tab:constraint}. From Tab.~\ref{tab:constraint} we see that there is no significant improvement in  the value of $h$ for the SE and MSE potentials compared to the axionlike and power law potentials. The $1\sig$ and $2\sig$ contours for the data combinations considered are shown in the Fig.~\ref{fig:cont_lcdm} for $\Lambda$CDM, Fig.~\ref{fig:cont_se} for SE potential~\eqref{eq:pot}, Fig.~\ref{fig:cont_mse} for MSE potential~\eqref{eq:potMSE}, Fig.~\ref{fig:cont_ax} for axionlike~\eqref{eq:ax} and Fig.~\ref{fig:cont_pl} for the power law~\eqref{eq:pl} potential.  

The constraints on the parameters for the SE potential~\eqref{eq:pot}, shown in Tab.~\ref{tab:constraint}, are for $n=6$ and 10. It has been checked that the larger values of $n$ can give rise to larger values in the parameter $h$ (see Tab.~\ref{tab:H0}) and that is why we have chosen larger values of $n$. Smaller values of $n$ can lead to scaling behaviour in the scalar field dynamics just after the redshift $z_{\rm c}$ where $\rho_\phi$ scales the background energy density \cite{Geng:2015fla}. This behaviour forces the maximum value of EDE density at $z_{\rm c}$ to be very small so that $\rho_\phi$ remains small when it scales the background. But we need the EDE energy density to decay very fast just after $z_{\rm c}$ so that it does not affect the late time dynamics of the universe. This can easily be achieved for larger values of $n$ as depicted in the upper figure of Fig.~\ref{fig:SE_scaling}. Similar to the SE potential we have chosen larger values $n$ for MSE potential also. One should note that the MSE potential leads to tracker behaviour in the scalar field dynamics during the late time. This tracker behaviour can lead to  even less amount of EDE at the redshift $z_{\rm c}$ which may lead to lesser values in $h$ compared to SE potential. Tab.~\ref{tab:constraint} and \ref{tab:fEDE} show these behaviour in the values of $f_{\rm EDE}$ and $h$. In fact, from Tab.~\ref{tab:fEDE}, we see that, for MSE potential, the observationally allowed value of $f_{\rm EDE}$ is almost zero even though theoretically non-zero EDE is possible in MSE potential (see Fig.~\ref{fig:dyn} and Tab.~\ref{tab:H0}). For SE potential the allowed value of $f_{\rm EDE}$ are non zero but small which gives very slight improvement in $h$ compared to MSE potential. But the improvement is not much and the constrained values are similar to the $\Lambda$CDM model. Similar to the SE and MSE potential we also don't see any improvement in the values of $h$ for the power law potential. On the other hand the axionlike gives rise to higher values in $h$. Here, one should note that, for the oscillatory potentials (axionlike and power law), for the considered data combinations, even though the values of $h$ are larger than that of $\Lambda$CDM model for different data combinations the values are significantly lower than the values obtained in Ref.~\cite{Poulin:2018cxd,Poulin:2018dzj,Smith:2019ihp,Agrawal:2019lmo} but very similar to the constraint obtained in \cite{Efstathiou:2023fbn} for the axionlike potential without considering the SH0ES prior \cite{Riess:2021jrx}. In our analysis we have not considered the SH0ES prior. Also, we have considered a wide range of the prior for the parameter $\log_{10}V_0$ so that we can accommodate the $\Lambda$CDM model within the models with SE, MSE, axionlike and power law potential. We see that the considered observational data has a tendency to go towards $\Lambda$CDM model. The maximum deviation we get for the data combination of DESIUnion. In general the data combination with DESI BAO data prefers larger value of $h$ compared to the other data combinations.


\section{Summary and Conclusion}
\label{sec:conclusion}

We examine a quintessential EDE model where both EDE and LDE arise from a single scalar field with the MSE potential~\eqref{eq:potMSE}. The MSE potential refines the SE potential~\eqref{eq:pot}, initially proposed in \cite{Geng:2015fla} in the context of quintessential inflation~\cite{WaliHossain:2014usl} scenario. While the SE potential follows a scaling solution during the late time (Fig.~\ref{fig:SE_scaling}), the MSE potential exhibits tracker dynamics (Fig.~\ref{fig:dyn}). Both allows a transient EDE phase but unlike SE potential MSE potential ensures late time acceleration without additional interactions.

We compare MSE with SE~\eqref{eq:pot}, axionlike~\eqref{eq:ax}, and power law~\eqref{eq:pl} potentials, constraining parameters using cosmological data. Although, theoretically, all four potentials predict a larger $H_0$ than $\Lambda$CDM (Tab.~\ref{tab:H0}), the MSE potential, with its tracker behaviour, yields a more conservative increase, and for lower $n$, even predicts $H_0$ values below $\Lambda$CDM. This occurs for lower values of $n$ as the scalar field exhibits tracker dynamics even during the recent past making the scalar field EoS larger than $-1$. The MSE potential also affects perturbations, suppressing the linear power spectrum (Fig.~\ref{fig:PS}, left) and lowering $\sigma_8$. This suppression is visible in the $f\sigma_8(z)$ evolution (Fig.~\ref{fig:PS}, right), particularly for small $n$, though the reduced bispectrum remains close to $\Lambda$CDM (Fig.~\ref{fig:BS}). We see this suppression only for the MSE potential {\it i.e.}, the scalar field with tracker dynamics. 

Observational constraints (Tab.~\ref{tab:constraint}) reveal that among the tested EDE potentials, the axionlike model shows the largest deviation in $h$, whereas SE, MSE, and power law potentials remain closer to $\Lambda$CDM. Notably, the DESIUnion data combination exhibits the strongest deviation from $\Lambda$CDM. However, our analysis finds no significant statistical preference for any EDE model over $\Lambda$CDM. While some data combinations favour MSE over other EDE potentials (Tab.~\ref{tab:AIC}), the improvement is marginal. Crucially, despite theoretical motivations for EDE models to reconcile the Hubble tension, observational data provide no compelling evidence supporting their necessity. Our constraints on $\log_{10}V_0$, $\mu$, $\phi_0$, and $f$ (Tab.~\ref{tab:fEDE}) confirm that only a small EDE fraction is viable, except for the axionlike case. Moreover, the $H_0$ values computed using Eq.~\eqref{eq:h0} align closely with direct constraints (Tab.~\ref{tab:constraint}), validating Eq.~\eqref{eq:h0} as a theoretical estimator. These results underscore the fundamental challenge of surpassing $\Lambda$CDM within the EDE framework. While EDE models allow for higher $H_0$ theoretically, the data do not support such an enhancement. This result is crucial for theoretical model building, as it suggests that conventional EDE scenarios, including the MSE potential, fail to resolve the Hubble tension.

In summary, we have introduced a unified model of EDE and LDE by modifying the SE~\eqref{eq:pot} potential into the MSE potential~\eqref{eq:potMSE}, which allows for a CC like behaviour during recent times while exhibiting tracker dynamics during intermediate stages. We have compared this model with observational data alongside the SE, axionlike, and power law potentials. While all four potentials can theoretically produce an EDE solution that permits larger values of $H_0$, our analysis shows that, observationally, none of these models significantly improve the inferred value of $H_0$ compared to $\Lambda$CDM. Even in the case of the axionlike potential, where there is a slight increase in $H_0$, the improvement is not statistically significant. As a result, we find that EDE models, including the MSE potential, do not provide a compelling resolution to the Hubble tension, with $\Lambda$CDM remaining the most favoured model.

\vskip20pt
\noindent
{\bf Note Added:} While working on the observational data analysis part of this paper we came across the recent paper \cite{Mishra:2024qhc} where the authors also have considered the MSE potential~\eqref{eq:potMSE} but in the context of inflation.

\section*{Acknowledgments}
MWH thanks Anjan A. Sen for insightful discussions. The authors thank Shahnawaz A. Adil for discussions regarding observational data analysis. Sonej Alam acknowledges the CSIR fellowship provided by Govt. of India under the CSIR-JRF scheme (file no.:09/0466(12904)/2021). Sonej Alam, SS and MWH also acknowledge the High Performance Computing facility Pegasus at IUCAA, Pune, India. MWH and Shiriny Akthar acknowledge the financial support from ANRF, SERB, Govt of India under the Start-up Research Grant (SRG), file no: SRG/2022/002234. 

\bibliographystyle{elsarticle-num}
\bibliography{references}

\begin{thebibliography}{10}
\expandafter\ifx\csname url\endcsname\relax
  \def\url#1{\texttt{#1}}\fi
\expandafter\ifx\csname urlprefix\endcsname\relax\def\urlprefix{URL }\fi
\expandafter\ifx\csname href\endcsname\relax
  \def\href#1#2{#2} \def\path#1{#1}\fi

\bibitem{SupernovaSearchTeam:1998fmf}
A.~G. Riess, et~al., {Observational evidence from supernovae for an accelerating universe and a cosmological constant}, Astron. J. 116 (1998) 1009--1038.
\newblock \href {http://arxiv.org/abs/astro-ph/9805201} {\path{arXiv:astro-ph/9805201}}, \href {https://doi.org/10.1086/300499} {\path{doi:10.1086/300499}}.

\bibitem{SupernovaCosmologyProject:1998vns}
S.~Perlmutter, et~al., {Measurements of $\Omega$ and $\Lambda$ from 42 high redshift supernovae}, Astrophys. J. 517 (1999) 565--586.
\newblock \href {http://arxiv.org/abs/astro-ph/9812133} {\path{arXiv:astro-ph/9812133}}, \href {https://doi.org/10.1086/307221} {\path{doi:10.1086/307221}}.

\bibitem{Planck:2018vyg}
N.~Aghanim, et~al., {Planck 2018 results. VI. Cosmological parameters}, Astron. Astrophys. 641 (2020) A6, [Erratum: Astron.Astrophys. 652, C4 (2021)].
\newblock \href {http://arxiv.org/abs/1807.06209} {\path{arXiv:1807.06209}}, \href {https://doi.org/10.1051/0004-6361/201833910} {\path{doi:10.1051/0004-6361/201833910}}.

\bibitem{Brout:2022vxf}
D.~Brout, et~al., {The Pantheon+ Analysis: Cosmological Constraints}, Astrophys. J. 938~(2) (2022) 110.
\newblock \href {http://arxiv.org/abs/2202.04077} {\path{arXiv:2202.04077}}, \href {https://doi.org/10.3847/1538-4357/ac8e04} {\path{doi:10.3847/1538-4357/ac8e04}}.

\bibitem{Scolnic:2021amr}
D.~Scolnic, et~al., {The Pantheon+ Analysis: The Full Data Set and Light-curve Release}, Astrophys. J. 938~(2) (2022) 113.
\newblock \href {http://arxiv.org/abs/2112.03863} {\path{arXiv:2112.03863}}, \href {https://doi.org/10.3847/1538-4357/ac8b7a} {\path{doi:10.3847/1538-4357/ac8b7a}}.

\bibitem{Planck:2013pxb}
P.~A.~R. Ade, et~al., {Planck 2013 results. XVI. Cosmological parameters}, Astron. Astrophys. 571 (2014) A16.
\newblock \href {http://arxiv.org/abs/1303.5076} {\path{arXiv:1303.5076}}, \href {https://doi.org/10.1051/0004-6361/201321591} {\path{doi:10.1051/0004-6361/201321591}}.

\bibitem{Martin:2012bt}
J.~Martin, {Everything You Always Wanted To Know About The Cosmological Constant Problem (But Were Afraid To Ask)}, Comptes Rendus Physique 13 (2012) 566--665.
\newblock \href {http://arxiv.org/abs/1205.3365} {\path{arXiv:1205.3365}}, \href {https://doi.org/10.1016/j.crhy.2012.04.008} {\path{doi:10.1016/j.crhy.2012.04.008}}.

\bibitem{Zlatev:1998tr}
I.~Zlatev, L.-M. Wang, P.~J. Steinhardt, {Quintessence, cosmic coincidence, and the cosmological constant}, Phys. Rev. Lett. 82 (1999) 896--899.
\newblock \href {http://arxiv.org/abs/astro-ph/9807002} {\path{arXiv:astro-ph/9807002}}, \href {https://doi.org/10.1103/PhysRevLett.82.896} {\path{doi:10.1103/PhysRevLett.82.896}}.

\bibitem{Steinhardt:1999nw}
P.~J. Steinhardt, L.-M. Wang, I.~Zlatev, {Cosmological tracking solutions}, Phys. Rev. D 59 (1999) 123504.
\newblock \href {http://arxiv.org/abs/astro-ph/9812313} {\path{arXiv:astro-ph/9812313}}, \href {https://doi.org/10.1103/PhysRevD.59.123504} {\path{doi:10.1103/PhysRevD.59.123504}}.

\bibitem{Riess:2021jrx}
A.~G. Riess, et~al., {A Comprehensive Measurement of the Local Value of the Hubble Constant with 1 km s$^{-1}$ Mpc$^{-1}$ Uncertainty from the Hubble Space Telescope and the SH0ES Team}, Astrophys. J. Lett. 934~(1) (2022) L7.
\newblock \href {http://arxiv.org/abs/2112.04510} {\path{arXiv:2112.04510}}, \href {https://doi.org/10.3847/2041-$8213/ac5c5b} {\path{doi:10.3847/2041-$8213/ac5c5b}}.

\bibitem{Kamionkowski:2022pkx}
M.~Kamionkowski, A.~G. Riess, {The Hubble Tension and Early Dark Energy}, Ann. Rev. Nucl. Part. Sci. 73 (2023) 153--180.
\newblock \href {http://arxiv.org/abs/2211.04492} {\path{arXiv:2211.04492}}, \href {https://doi.org/10.1146/annurev-nucl-111422-024107} {\path{doi:10.1146/annurev-nucl-111422-024107}}.

\bibitem{Freedman:2023jcz}
W.~L. Freedman, B.~F. Madore, {Progress in Direct Measurements of the Hubble Constant} (9 2023).
\newblock \href {http://arxiv.org/abs/2309.05618} {\path{arXiv:2309.05618}}.

\bibitem{Bernal:2016gxb}
J.~L. Bernal, L.~Verde, A.~G. Riess, {The trouble with $H_0$}, JCAP 10 (2016) 019.
\newblock \href {http://arxiv.org/abs/1607.05617} {\path{arXiv:1607.05617}}, \href {https://doi.org/10.1088/1475-7516/2016/10/019} {\path{doi:10.1088/1475-7516/2016/10/019}}.

\bibitem{Knox:2019rjx}
L.~Knox, M.~Millea, {Hubble constant hunter\textquoteright{}s guide}, Phys. Rev. D 101~(4) (2020) 043533.
\newblock \href {http://arxiv.org/abs/1908.03663} {\path{arXiv:1908.03663}}, \href {https://doi.org/10.1103/PhysRevD.101.043533} {\path{doi:10.1103/PhysRevD.101.043533}}.

\bibitem{Copeland:2006wr}
E.~J. Copeland, M.~Sami, S.~Tsujikawa, {Dynamics of dark energy}, Int. J. Mod. Phys. D 15 (2006) 1753--1936.
\newblock \href {http://arxiv.org/abs/hep-th/0603057} {\path{arXiv:hep-th/0603057}}, \href {https://doi.org/10.1142/S021827180600942X} {\path{doi:10.1142/S021827180600942X}}.

\bibitem{Sahni:1999gb}
V.~Sahni, A.~A. Starobinsky, {The Case for a positive cosmological Lambda term}, Int. J. Mod. Phys. D 9 (2000) 373--444.
\newblock \href {http://arxiv.org/abs/astro-ph/9904398} {\path{arXiv:astro-ph/9904398}}, \href {https://doi.org/10.1142/S0218271800000542} {\path{doi:10.1142/S0218271800000542}}.

\bibitem{Clifton:2011jh}
T.~Clifton, P.~G. Ferreira, A.~Padilla, C.~Skordis, {Modified Gravity and Cosmology}, Phys. Rept. 513 (2012) 1--189.
\newblock \href {http://arxiv.org/abs/1106.2476} {\path{arXiv:1106.2476}}, \href {https://doi.org/10.1016/j.physrep.2012.01.001} {\path{doi:10.1016/j.physrep.2012.01.001}}.

\bibitem{DESI:2024mwx}
A.~G. Adame, et~al., {DESI 2024 VI: Cosmological Constraints from the Measurements of Baryon Acoustic Oscillations} (4 2024).
\newblock \href {http://arxiv.org/abs/2404.03002} {\path{arXiv:2404.03002}}.

\bibitem{DESI:2024uvr}
A.~G. Adame, et~al., {DESI 2024 III: Baryon Acoustic Oscillations from Galaxies and Quasars} (4 2024).
\newblock \href {http://arxiv.org/abs/2404.03000} {\path{arXiv:2404.03000}}.

\bibitem{DESI:2024lzq}
A.~G. Adame, et~al., {DESI 2024 IV: Baryon Acoustic Oscillations from the Lyman Alpha Forest} (4 2024).
\newblock \href {http://arxiv.org/abs/2404.03001} {\path{arXiv:2404.03001}}.

\bibitem{DESI:2024aqx}
R.~Calderon, et~al., {DESI 2024: Reconstructing Dark Energy using Crossing Statistics with DESI DR1 BAO data} (5 2024).
\newblock \href {http://arxiv.org/abs/2405.04216} {\path{arXiv:2405.04216}}.

\bibitem{Giare:2024smz}
W.~Giar\`e, M.~A. Sabogal, R.~C. Nunes, E.~Di~Valentino, {Interacting Dark Energy after DESI Baryon Acoustic Oscillation measurements} (4 2024).
\newblock \href {http://arxiv.org/abs/2404.15232} {\path{arXiv:2404.15232}}.

\bibitem{Berghaus:2024kra}
K.~V. Berghaus, J.~A. Kable, V.~Miranda, {Quantifying Scalar Field Dynamics with DESI 2024 Y1 BAO measurements} (4 2024).
\newblock \href {http://arxiv.org/abs/2404.14341} {\path{arXiv:2404.14341}}.

\bibitem{Qu:2024lpx}
F.~J. Qu, K.~M. Surrao, B.~Bolliet, J.~C. Hill, B.~D. Sherwin, H.~T. Jense, {Accelerated inference on accelerated cosmic expansion: New constraints on axion-like early dark energy with DESI BAO and ACT DR6 CMB lensing} (4 2024).
\newblock \href {http://arxiv.org/abs/2404.16805} {\path{arXiv:2404.16805}}.

\bibitem{Wang:2024dka}
H.~Wang, Y.-S. Piao, {Dark energy in light of recent DESI BAO and Hubble tension} (4 2024).
\newblock \href {http://arxiv.org/abs/2404.18579} {\path{arXiv:2404.18579}}.

\bibitem{Giare:2024gpk}
W.~Giar\`e, M.~Najafi, S.~Pan, E.~Di~Valentino, J.~T. Firouzjaee, {Robust Preference for Dynamical Dark Energy in DESI BAO and SN Measurements} (7 2024).
\newblock \href {http://arxiv.org/abs/2407.16689} {\path{arXiv:2407.16689}}.

\bibitem{Shlivko:2024llw}
D.~Shlivko, P.~J. Steinhardt, {Assessing observational constraints on dark energy}, Phys. Lett. B 855 (2024) 138826.
\newblock \href {http://arxiv.org/abs/2405.03933} {\path{arXiv:2405.03933}}, \href {https://doi.org/10.1016/j.physletb.2024.138826} {\path{doi:10.1016/j.physletb.2024.138826}}.

\bibitem{Ye:2024ywg}
G.~Ye, M.~Martinelli, B.~Hu, A.~Silvestri, {Non-minimally coupled gravity as a physically viable fit to DESI 2024 BAO} (7 2024).
\newblock \href {http://arxiv.org/abs/2407.15832} {\path{arXiv:2407.15832}}.

\bibitem{Ramadan:2024kmn}
O.~F. Ramadan, J.~Sakstein, D.~Rubin, {DESI Constraints on Exponential Quintessence} (5 2024).
\newblock \href {http://arxiv.org/abs/2405.18747} {\path{arXiv:2405.18747}}.

\bibitem{Jiang:2024xnu}
J.-Q. Jiang, D.~Pedrotti, S.~S. da~Costa, S.~Vagnozzi, {Non-parametric late-time expansion history reconstruction and implications for the Hubble tension in light of DESI} (8 2024).
\newblock \href {http://arxiv.org/abs/2408.02365} {\path{arXiv:2408.02365}}.

\bibitem{Rubin:2023ovl}
D.~Rubin, et~al., {Union Through UNITY: Cosmology with 2,000 SNe Using a Unified Bayesian Framework} (11 2023).
\newblock \href {http://arxiv.org/abs/2311.12098} {\path{arXiv:2311.12098}}.

\bibitem{DES:2024tys}
T.~M.~C. Abbott, et~al., {The Dark Energy Survey: Cosmology Results With \textasciitilde{}1500 New High-redshift Type Ia Supernovae Using The Full 5-year Dataset} (1 2024).
\newblock \href {http://arxiv.org/abs/2401.02929} {\path{arXiv:2401.02929}}.

\bibitem{Copeland:1997et}
E.~J. Copeland, A.~R. Liddle, D.~Wands, {Exponential potentials and cosmological scaling solutions}, Phys. Rev. D 57 (1998) 4686--4690.
\newblock \href {http://arxiv.org/abs/gr-qc/9711068} {\path{arXiv:gr-qc/9711068}}, \href {https://doi.org/10.1103/PhysRevD.57.4686} {\path{doi:10.1103/PhysRevD.57.4686}}.

\bibitem{Barreiro:1999zs}
T.~Barreiro, E.~J. Copeland, N.~J. Nunes, {Quintessence arising from exponential potentials}, Phys. Rev. D 61 (2000) 127301.
\newblock \href {http://arxiv.org/abs/astro-ph/9910214} {\path{arXiv:astro-ph/9910214}}, \href {https://doi.org/10.1103/PhysRevD.61.127301} {\path{doi:10.1103/PhysRevD.61.127301}}.

\bibitem{Sahni:1999qe}
V.~Sahni, L.-M. Wang, {A New cosmological model of quintessence and dark matter}, Phys. Rev. D 62 (2000) 103517.
\newblock \href {http://arxiv.org/abs/astro-ph/9910097} {\path{arXiv:astro-ph/9910097}}, \href {https://doi.org/10.1103/PhysRevD.62.103517} {\path{doi:10.1103/PhysRevD.62.103517}}.

\bibitem{Scherrer:2007pu}
R.~J. Scherrer, A.~A. Sen, {Thawing quintessence with a nearly flat potential}, Phys. Rev. D 77 (2008) 083515.
\newblock \href {http://arxiv.org/abs/0712.3450} {\path{arXiv:0712.3450}}, \href {https://doi.org/10.1103/PhysRevD.77.083515} {\path{doi:10.1103/PhysRevD.77.083515}}.

\bibitem{Hossain:2023lxs}
M.~W. Hossain, A.~Maqsood, {A comparison between axion-like and power law potentials in cosmological background} (11 2023).
\newblock \href {http://arxiv.org/abs/2311.17825} {\path{arXiv:2311.17825}}.

\bibitem{Poulin:2023lkg}
V.~Poulin, T.~L. Smith, T.~Karwal, {The Ups and Downs of Early Dark Energy solutions to the Hubble tension: A review of models, hints and constraints circa 2023}, Phys. Dark Univ. 42 (2023) 101348.
\newblock \href {http://arxiv.org/abs/2302.09032} {\path{arXiv:2302.09032}}, \href {https://doi.org/10.1016/j.dark.2023.101348} {\path{doi:10.1016/j.dark.2023.101348}}.

\bibitem{Perivolaropoulos:2021jda}
L.~Perivolaropoulos, F.~Skara, {Challenges for \ensuremath{\Lambda}CDM: An update}, New Astron. Rev. 95 (2022) 101659.
\newblock \href {http://arxiv.org/abs/2105.05208} {\path{arXiv:2105.05208}}, \href {https://doi.org/10.1016/j.newar.2022.101659} {\path{doi:10.1016/j.newar.2022.101659}}.

\bibitem{Poulin:2018cxd}
V.~Poulin, T.~L. Smith, T.~Karwal, M.~Kamionkowski, {Early Dark Energy Can Resolve The Hubble Tension}, Phys. Rev. Lett. 122~(22) (2019) 221301.
\newblock \href {http://arxiv.org/abs/1811.04083} {\path{arXiv:1811.04083}}, \href {https://doi.org/10.1103/PhysRevLett.122.221301} {\path{doi:10.1103/PhysRevLett.122.221301}}.

\bibitem{Poulin:2018dzj}
V.~Poulin, T.~L. Smith, D.~Grin, T.~Karwal, M.~Kamionkowski, {Cosmological implications of ultralight axionlike fields}, Phys. Rev. D 98~(8) (2018) 083525.
\newblock \href {http://arxiv.org/abs/1806.10608} {\path{arXiv:1806.10608}}, \href {https://doi.org/10.1103/PhysRevD.98.083525} {\path{doi:10.1103/PhysRevD.98.083525}}.

\bibitem{Smith:2019ihp}
T.~L. Smith, V.~Poulin, M.~A. Amin, {Oscillating scalar fields and the Hubble tension: a resolution with novel signatures}, Phys. Rev. D 101~(6) (2020) 063523.
\newblock \href {http://arxiv.org/abs/1908.06995} {\path{arXiv:1908.06995}}, \href {https://doi.org/10.1103/PhysRevD.101.063523} {\path{doi:10.1103/PhysRevD.101.063523}}.

\bibitem{Agrawal:2019lmo}
P.~Agrawal, F.-Y. Cyr-Racine, D.~Pinner, L.~Randall, {Rock \textquoteleft{}n\textquoteright{} roll solutions to the Hubble tension}, Phys. Dark Univ. 42 (2023) 101347.
\newblock \href {http://arxiv.org/abs/1904.01016} {\path{arXiv:1904.01016}}, \href {https://doi.org/10.1016/j.dark.2023.101347} {\path{doi:10.1016/j.dark.2023.101347}}.

\bibitem{Kodama:2023fjq}
T.~Kodama, T.~Shinohara, T.~Takahashi, {Generalized early dark energy and its cosmological consequences}, Phys. Rev. D 109~(6) (2024) 063518.
\newblock \href {http://arxiv.org/abs/2309.11272} {\path{arXiv:2309.11272}}, \href {https://doi.org/10.1103/PhysRevD.109.063518} {\path{doi:10.1103/PhysRevD.109.063518}}.

\bibitem{Lin:2019qug}
M.-X. Lin, G.~Benevento, W.~Hu, M.~Raveri, {Acoustic Dark Energy: Potential Conversion of the Hubble Tension}, Phys. Rev. D 100~(6) (2019) 063542.
\newblock \href {http://arxiv.org/abs/1905.12618} {\path{arXiv:1905.12618}}, \href {https://doi.org/10.1103/PhysRevD.100.063542} {\path{doi:10.1103/PhysRevD.100.063542}}.

\bibitem{Niedermann:2019olb}
F.~Niedermann, M.~S. Sloth, {New early dark energy}, Phys. Rev. D 103~(4) (2021) L041303.
\newblock \href {http://arxiv.org/abs/1910.10739} {\path{arXiv:1910.10739}}, \href {https://doi.org/10.1103/PhysRevD.103.L041303} {\path{doi:10.1103/PhysRevD.103.L041303}}.

\bibitem{Niedermann:2020dwg}
F.~Niedermann, M.~S. Sloth, {Resolving the Hubble tension with new early dark energy}, Phys. Rev. D 102~(6) (2020) 063527.
\newblock \href {http://arxiv.org/abs/2006.06686} {\path{arXiv:2006.06686}}, \href {https://doi.org/10.1103/PhysRevD.102.063527} {\path{doi:10.1103/PhysRevD.102.063527}}.

\bibitem{Berghaus:2019cls}
K.~V. Berghaus, T.~Karwal, {Thermal Friction as a Solution to the Hubble Tension}, Phys. Rev. D 101~(8) (2020) 083537.
\newblock \href {http://arxiv.org/abs/1911.06281} {\path{arXiv:1911.06281}}, \href {https://doi.org/10.1103/PhysRevD.101.083537} {\path{doi:10.1103/PhysRevD.101.083537}}.

\bibitem{Berghaus:2022cwf}
K.~V. Berghaus, T.~Karwal, {Thermal friction as a solution to the Hubble and large-scale structure tensions}, Phys. Rev. D 107~(10) (2023) 103515.
\newblock \href {http://arxiv.org/abs/2204.09133} {\path{arXiv:2204.09133}}, \href {https://doi.org/10.1103/PhysRevD.107.103515} {\path{doi:10.1103/PhysRevD.107.103515}}.

\bibitem{Karwal:2021vpk}
T.~Karwal, M.~Raveri, B.~Jain, J.~Khoury, M.~Trodden, {Chameleon early dark energy and the Hubble tension}, Phys. Rev. D 105~(6) (2022) 063535.
\newblock \href {http://arxiv.org/abs/2106.13290} {\path{arXiv:2106.13290}}, \href {https://doi.org/10.1103/PhysRevD.105.063535} {\path{doi:10.1103/PhysRevD.105.063535}}.

\bibitem{Vagnozzi:2023nrq}
S.~Vagnozzi, {Seven Hints That Early-Time New Physics Alone Is Not Sufficient to Solve the Hubble Tension}, Universe 9~(9) (2023) 393.
\newblock \href {http://arxiv.org/abs/2308.16628} {\path{arXiv:2308.16628}}, \href {https://doi.org/10.3390/universe9090393} {\path{doi:10.3390/universe9090393}}.

\bibitem{Vagnozzi:2021gjh}
S.~Vagnozzi, {Consistency tests of \ensuremath{\Lambda}CDM from the early integrated Sachs-Wolfe effect: Implications for early-time new physics and the Hubble tension}, Phys. Rev. D 104~(6) (2021) 063524.
\newblock \href {http://arxiv.org/abs/2105.10425} {\path{arXiv:2105.10425}}, \href {https://doi.org/10.1103/PhysRevD.104.063524} {\path{doi:10.1103/PhysRevD.104.063524}}.

\bibitem{Vagnozzi:2019ezj}
S.~Vagnozzi, {New physics in light of the $H_0$ tension: An alternative view}, Phys. Rev. D 102~(2) (2020) 023518.
\newblock \href {http://arxiv.org/abs/1907.07569} {\path{arXiv:1907.07569}}, \href {https://doi.org/10.1103/PhysRevD.102.023518} {\path{doi:10.1103/PhysRevD.102.023518}}.

\bibitem{Marsh:2015xka}
D.~J.~E. Marsh, {Axion Cosmology}, Phys. Rept. 643 (2016) 1--79.
\newblock \href {http://arxiv.org/abs/1510.07633} {\path{arXiv:1510.07633}}, \href {https://doi.org/10.1016/j.physrep.2016.06.005} {\path{doi:10.1016/j.physrep.2016.06.005}}.

\bibitem{Hlozek:2014lca}
R.~Hlozek, D.~Grin, D.~J.~E. Marsh, P.~G. Ferreira, {A search for ultralight axions using precision cosmological data}, Phys. Rev. D 91~(10) (2015) 103512.
\newblock \href {http://arxiv.org/abs/1410.2896} {\path{arXiv:1410.2896}}, \href {https://doi.org/10.1103/PhysRevD.91.103512} {\path{doi:10.1103/PhysRevD.91.103512}}.

\bibitem{Efstathiou:2023fbn}
G.~Efstathiou, E.~Rosenberg, V.~Poulin, {Improved Planck Constraints on Axionlike Early Dark Energy as a Resolution of the Hubble Tension}, Phys. Rev. Lett. 132~(22) (2024) 221002.
\newblock \href {http://arxiv.org/abs/2311.00524} {\path{arXiv:2311.00524}}, \href {https://doi.org/10.1103/PhysRevLett.132.221002} {\path{doi:10.1103/PhysRevLett.132.221002}}.

\bibitem{Ratra:1987rm}
B.~Ratra, P.~J.~E. Peebles, {Cosmological Consequences of a Rolling Homogeneous Scalar Field}, Phys. Rev. D 37 (1988) 3406.
\newblock \href {https://doi.org/10.1103/PhysRevD.37.3406} {\path{doi:10.1103/PhysRevD.37.3406}}.

\bibitem{Geng:2015fla}
C.-Q. Geng, M.~W. Hossain, R.~Myrzakulov, M.~Sami, E.~N. Saridakis, {Quintessential inflation with canonical and noncanonical scalar fields and Planck 2015 results}, Phys. Rev. D 92~(2) (2015) 023522.
\newblock \href {http://arxiv.org/abs/1502.03597} {\path{arXiv:1502.03597}}, \href {https://doi.org/10.1103/PhysRevD.92.023522} {\path{doi:10.1103/PhysRevD.92.023522}}.

\bibitem{Wetterich:1987fk}
C.~Wetterich, {Cosmologies With Variable Newton's 'Constant'}, Nucl. Phys. B 302 (1988) 645--667.
\newblock \href {https://doi.org/10.1016/0550-3213(88)90192-7} {\path{doi:10.1016/0550-3213(88)90192-7}}.

\bibitem{Wetterich:1987fm}
C.~Wetterich, {Cosmology and the Fate of Dilatation Symmetry}, Nucl. Phys. B 302 (1988) 668--696.
\newblock \href {http://arxiv.org/abs/1711.03844} {\path{arXiv:1711.03844}}, \href {https://doi.org/10.1016/0550-3213(88)90193-9} {\path{doi:10.1016/0550-3213(88)90193-9}}.

\bibitem{Peebles:1998qn}
P.~J.~E. Peebles, A.~Vilenkin, {Quintessential inflation}, Phys. Rev. D 59 (1999) 063505.
\newblock \href {http://arxiv.org/abs/astro-ph/9810509} {\path{arXiv:astro-ph/9810509}}, \href {https://doi.org/10.1103/PhysRevD.59.063505} {\path{doi:10.1103/PhysRevD.59.063505}}.

\bibitem{WaliHossain:2014usl}
M.~Wali~Hossain, R.~Myrzakulov, M.~Sami, E.~N. Saridakis, {Unification of inflation and dark energy \`a la quintessential inflation}, Int. J. Mod. Phys. D 24~(05) (2015) 1530014.
\newblock \href {http://arxiv.org/abs/1410.6100} {\path{arXiv:1410.6100}}, \href {https://doi.org/10.1142/S0218271815300141} {\path{doi:10.1142/S0218271815300141}}.

\bibitem{deHaro:2021swo}
J.~de~Haro, L.~A. Sal\'o, {A Review of Quintessential Inflation}, Galaxies 9~(4) (2021) 73.
\newblock \href {http://arxiv.org/abs/2108.11144} {\path{arXiv:2108.11144}}, \href {https://doi.org/10.3390/galaxies9040073} {\path{doi:10.3390/galaxies9040073}}.

\bibitem{Bettoni:2021qfs}
D.~Bettoni, J.~Rubio, {Quintessential Inflation: A Tale of Emergent and Broken Symmetries}, Galaxies 10~(1) (2022) 22.
\newblock \href {http://arxiv.org/abs/2112.11948} {\path{arXiv:2112.11948}}, \href {https://doi.org/10.3390/galaxies10010022} {\path{doi:10.3390/galaxies10010022}}.

\bibitem{Ramadan:2023ivw}
O.~F. Ramadan, T.~Karwal, J.~Sakstein, {Attractive proposal for resolving the Hubble tension: Dynamical attractors that unify early and late dark energy}, Phys. Rev. D 109~(6) (2024) 063525.
\newblock \href {http://arxiv.org/abs/2309.08082} {\path{arXiv:2309.08082}}, \href {https://doi.org/10.1103/PhysRevD.109.063525} {\path{doi:10.1103/PhysRevD.109.063525}}.

\bibitem{Chowdhury:2023opo}
D.~Chowdhury, G.~Tasinato, I.~Zavala, {Dark energy, D-branes and pulsar timing arrays}, JCAP 11 (2023) 090.
\newblock \href {http://arxiv.org/abs/2307.01188} {\path{arXiv:2307.01188}}, \href {https://doi.org/10.1088/1475-7516/2023/11/090} {\path{doi:10.1088/1475-7516/2023/11/090}}.

\bibitem{MohseniSadjadi:2024ejb}
H.~Mohseni~Sadjadi, {Early dark energy and scalarization in a scalar-tensor model} (4 2024).
\newblock \href {http://arxiv.org/abs/2404.19695} {\path{arXiv:2404.19695}}.

\bibitem{Copeland:2023zqz}
E.~J. Copeland, A.~Moss, S.~Sevillano Mu\~noz, J.~M.~M. White, {Scaling solutions as Early Dark Energy resolutions to the Hubble tension}, JCAP 05 (2024) 078.
\newblock \href {http://arxiv.org/abs/2309.15295} {\path{arXiv:2309.15295}}, \href {https://doi.org/10.1088/1475-7516/2024/05/078} {\path{doi:10.1088/1475-7516/2024/05/078}}.

\bibitem{Bahamonde:2017ize}
S.~Bahamonde, C.~G. B\"ohmer, S.~Carloni, E.~J. Copeland, W.~Fang, N.~Tamanini, {Dynamical systems applied to cosmology: dark energy and modified gravity}, Phys. Rept. 775-777 (2018) 1--122.
\newblock \href {http://arxiv.org/abs/1712.03107} {\path{arXiv:1712.03107}}, \href {https://doi.org/10.1016/j.physrep.2018.09.001} {\path{doi:10.1016/j.physrep.2018.09.001}}.

\bibitem{Geng:2017mic}
C.-Q. Geng, C.-C. Lee, M.~Sami, E.~N. Saridakis, A.~A. Starobinsky, {Observational constraints on successful model of quintessential Inflation}, JCAP 06 (2017) 011.
\newblock \href {http://arxiv.org/abs/1705.01329} {\path{arXiv:1705.01329}}, \href {https://doi.org/10.1088/1475-7516/2017/06/011} {\path{doi:10.1088/1475-7516/2017/06/011}}.

\bibitem{Ahmad:2017itq}
S.~Ahmad, R.~Myrzakulov, M.~Sami, {Relic gravitational waves from Quintessential Inflation}, Phys. Rev. D 96~(6) (2017) 063515.
\newblock \href {http://arxiv.org/abs/1705.02133} {\path{arXiv:1705.02133}}, \href {https://doi.org/10.1103/PhysRevD.96.063515} {\path{doi:10.1103/PhysRevD.96.063515}}.

\bibitem{Agarwal:2017wxo}
A.~Agarwal, R.~Myrzakulov, M.~Sami, N.~K. Singh, {Quintessential inflation in a thawing realization}, Phys. Lett. B 770 (2017) 200--208.
\newblock \href {http://arxiv.org/abs/1708.00156} {\path{arXiv:1708.00156}}, \href {https://doi.org/10.1016/j.physletb.2017.04.066} {\path{doi:10.1016/j.physletb.2017.04.066}}.

\bibitem{Das:2019ixt}
S.~Das, M.~Banerjee, N.~Roy, {Dynamical System Analysis for Steep Potentials}, JCAP 08 (2019) 024.
\newblock \href {http://arxiv.org/abs/1903.02288} {\path{arXiv:1903.02288}}, \href {https://doi.org/10.1088/1475-7516/2019/08/024} {\path{doi:10.1088/1475-7516/2019/08/024}}.

\bibitem{Lima:2019yyv}
G.~B.~F. Lima, R.~O. Ramos, {Unified early and late Universe cosmology through dissipative effects in steep quintessential inflation potential models}, Phys. Rev. D 100~(12) (2019) 123529.
\newblock \href {http://arxiv.org/abs/1910.05185} {\path{arXiv:1910.05185}}, \href {https://doi.org/10.1103/PhysRevD.100.123529} {\path{doi:10.1103/PhysRevD.100.123529}}.

\bibitem{Rezazadeh:2015dia}
K.~Rezazadeh, K.~Karami, S.~Hashemi, {Tachyon inflation with steep potentials}, Phys. Rev. D 95~(10) (2017) 103506.
\newblock \href {http://arxiv.org/abs/1508.04760} {\path{arXiv:1508.04760}}, \href {https://doi.org/10.1103/PhysRevD.95.103506} {\path{doi:10.1103/PhysRevD.95.103506}}.

\bibitem{Das:2020xmh}
S.~Das, R.~O. Ramos, {Runaway potentials in warm inflation satisfying the swampland conjectures}, Phys. Rev. D 102~(10) (2020) 103522.
\newblock \href {http://arxiv.org/abs/2007.15268} {\path{arXiv:2007.15268}}, \href {https://doi.org/10.1103/PhysRevD.102.103522} {\path{doi:10.1103/PhysRevD.102.103522}}.

\bibitem{Das:2023rat}
S.~Das, S.~Hussain, D.~Nandi, R.~O.~Ramos, R.~Silva, {Stability analysis of warm quintessential dark energy model}, Phys. Rev. D 108~(8) (2023) 083517.
\newblock \href {http://arxiv.org/abs/2306.09369} {\path{arXiv:2306.09369}}, \href {https://doi.org/10.1103/PhysRevD.108.083517} {\path{doi:10.1103/PhysRevD.108.083517}}.

\bibitem{Das:2023nmm}
B.~Das, N.~Jaman, M.~Sami, {Gravitational wave background from quintessential inflation and NANOGrav data}, Phys. Rev. D 108~(10) (2023) 103510.
\newblock \href {http://arxiv.org/abs/2307.12913} {\path{arXiv:2307.12913}}, \href {https://doi.org/10.1103/PhysRevD.108.103510} {\path{doi:10.1103/PhysRevD.108.103510}}.

\bibitem{Bardeen:1980kt}
J.~M. Bardeen, {Gauge Invariant Cosmological Perturbations}, Phys. Rev. D 22 (1980) 1882--1905.
\newblock \href {https://doi.org/10.1103/PhysRevD.22.1882} {\path{doi:10.1103/PhysRevD.22.1882}}.

\bibitem{Eisenstein:1997jh}
D.~J. Eisenstein, W.~Hu, {Power spectra for cold dark matter and its variants}, Astrophys. J. 511 (1997) 5.
\newblock \href {http://arxiv.org/abs/astro-ph/9710252} {\path{arXiv:astro-ph/9710252}}, \href {https://doi.org/10.1086/306640} {\path{doi:10.1086/306640}}.

\bibitem{Duniya:2015nva}
D.~G.~A. Duniya, D.~Bertacca, R.~Maartens, {Probing the imprint of interacting dark energy on very large scales}, Phys. Rev. D 91 (2015) 063530.
\newblock \href {http://arxiv.org/abs/1502.06424} {\path{arXiv:1502.06424}}, \href {https://doi.org/10.1103/PhysRevD.91.063530} {\path{doi:10.1103/PhysRevD.91.063530}}.

\bibitem{Eisenstein:1997ik}
D.~J. Eisenstein, W.~Hu, {Baryonic features in the matter transfer function}, Astrophys. J. 496 (1998) 605.
\newblock \href {http://arxiv.org/abs/astro-ph/9709112} {\path{arXiv:astro-ph/9709112}}, \href {https://doi.org/10.1086/305424} {\path{doi:10.1086/305424}}.

\bibitem{Nesseris:2017vor}
S.~Nesseris, G.~Pantazis, L.~Perivolaropoulos, {Tension and constraints on modified gravity parametrizations of $G_{\textrm{eff}}(z)$ from growth rate and Planck data}, Phys. Rev. D 96~(2) (2017) 023542.
\newblock \href {http://arxiv.org/abs/1703.10538} {\path{arXiv:1703.10538}}, \href {https://doi.org/10.1103/PhysRevD.96.023542} {\path{doi:10.1103/PhysRevD.96.023542}}.

\bibitem{Hossain:2017ica}
M.~W. Hossain, {First and second order cosmological perturbations in light mass Galileon models}, Phys. Rev. D 96~(2) (2017) 023506.
\newblock \href {http://arxiv.org/abs/1704.07956} {\path{arXiv:1704.07956}}, \href {https://doi.org/10.1103/PhysRevD.96.023506} {\path{doi:10.1103/PhysRevD.96.023506}}.

\bibitem{2019JCAP}
L.~{Chen}, Q.-G. {Huang}, K.~{Wang}, {Distance priors from Planck final release}, jcap 2019~(2) (2019) 028.
\newblock \href {http://arxiv.org/abs/1808.05724} {\path{arXiv:1808.05724}}, \href {https://doi.org/10.1088/1475-7516/2019/02/028} {\path{doi:10.1088/1475-7516/2019/02/028}}.

\bibitem{10.1093/mnras/}
A.~J. Ross, L.~Samushia, C.~Howlett, W.~J. Percival, A.~Burden, M.~Manera, \href{https://doi.org/10.1093/mnras/stv154}{{The clustering of the SDSS DR7 main Galaxy sample – I. A 4 per cent distance measure at z = 0.15}}, Monthly Notices of the Royal Astronomical Society 449~(1) (2015) 835--847.
\newblock \href {http://arxiv.org/abs/https://academic.oup.com/mnras/article-pdf/449/1/835/13767551/stv154.pdf} {\path{arXiv:https://academic.oup.com/mnras/article-pdf/449/1/835/13767551/stv154.pdf}}, \href {https://doi.org/10.1093/mnras/stv154} {\path{doi:10.1093/mnras/stv154}}.
\newline\urlprefix\url{https://doi.org/10.1093/mnras/stv154}

\bibitem{2011MNRAS}
F.~{Beutler}, C.~{Blake}, M.~{Colless}, D.~H. {Jones}, L.~{Staveley-Smith}, L.~{Campbell}, Q.~{Parker}, W.~{Saunders}, F.~{Watson}, {The 6dF Galaxy Survey: baryon acoustic oscillations and the local Hubble constant}, mnras 416~(4) (2011) 3017--3032.
\newblock \href {http://arxiv.org/abs/1106.3366} {\path{arXiv:1106.3366}}, \href {https://doi.org/10.1111/j.1365-2966.2011.19250.x} {\path{doi:10.1111/j.1365-2966.2011.19250.x}}.

\bibitem{eBOSS:2018cab}
P.~Zarrouk, et~al., {The clustering of the SDSS-IV extended Baryon Oscillation Spectroscopic Survey DR14 quasar sample: measurement of the growth rate of structure from the anisotropic correlation function between redshift 0.8 and 2.2}, Mon. Not. Roy. Astron. Soc. 477~(2) (2018) 1639--1663.
\newblock \href {http://arxiv.org/abs/1801.03062} {\path{arXiv:1801.03062}}, \href {https://doi.org/10.1093/mnras/sty506} {\path{doi:10.1093/mnras/sty506}}.

\bibitem{2017A&A...608A.130D}
d.~et~al., {Baryon acoustic oscillations from the complete SDSS-III Ly{\ensuremath{\alpha}}-quasar cross-correlation function at z = 2.4}, aap 608 (2017) A130.
\newblock \href {http://arxiv.org/abs/1708.02225} {\path{arXiv:1708.02225}}, \href {https://doi.org/10.1051/0004-6361/201731731} {\path{doi:10.1051/0004-6361/201731731}}.

\bibitem{2017MNRAS.470.2617A}
S.~A. et~al., {The clustering of galaxies in the completed SDSS-III Baryon Oscillation Spectroscopic Survey: cosmological analysis of the DR12 galaxy sample}, mnras 470~(3) (2017) 2617--2652.
\newblock \href {http://arxiv.org/abs/1607.03155} {\path{arXiv:1607.03155}}, \href {https://doi.org/10.1093/mnras/stx721} {\path{doi:10.1093/mnras/stx721}}.

\bibitem{2018JCAP...04..051G}
A.~{G{\'o}mez-Valent}, L.~{Amendola}, {H$_{0}$ from cosmic chronometers and Type Ia supernovae, with Gaussian Processes and the novel Weighted Polynomial Regression method}, jcap 2018~(4) (2018) 051.
\newblock \href {http://arxiv.org/abs/1802.01505} {\path{arXiv:1802.01505}}, \href {https://doi.org/10.1088/1475-7516/2018/04/051} {\path{doi:10.1088/1475-7516/2018/04/051}}.

\bibitem{Foreman-Mackey:2012any}
D.~Foreman-Mackey, D.~W. Hogg, D.~Lang, J.~Goodman, {emcee: The MCMC Hammer}, Publ. Astron. Soc. Pac. 125 (2013) 306--312.
\newblock \href {http://arxiv.org/abs/1202.3665} {\path{arXiv:1202.3665}}, \href {https://doi.org/10.1086/670067} {\path{doi:10.1086/670067}}.

\bibitem{Lewis:2019xzd}
A.~Lewis, {GetDist: a Python package for analysing Monte Carlo samples} (10 2019).
\newblock \href {http://arxiv.org/abs/1910.13970} {\path{arXiv:1910.13970}}.

\bibitem{Akaike74}
H.~{Akaike}, {A New Look at the Statistical Model Identification}, IEEE Transactions on Automatic Control 19 (1974) 716--723.

\bibitem{Liddle:2006tc}
A.~R. Liddle, P.~Mukherjee, D.~Parkinson, {Cosmological model selection} (8 2006).
\newblock \href {http://arxiv.org/abs/astro-ph/0608184} {\path{arXiv:astro-ph/0608184}}.

\bibitem{Trotta:2017wnx}
R.~Trotta, {Bayesian Methods in Cosmology} (1 2017).
\newblock \href {http://arxiv.org/abs/1701.01467} {\path{arXiv:1701.01467}}.

\bibitem{Shi:2012ma}
K.~Shi, Y.~Huang, T.~Lu, {A comprehensive comparison of cosmological models from latest observational data}, Mon. Not. Roy. Astron. Soc. 426 (2012) 2452--2462.
\newblock \href {http://arxiv.org/abs/1207.5875} {\path{arXiv:1207.5875}}, \href {https://doi.org/10.1111/j.1365-2966.2012.21784.x} {\path{doi:10.1111/j.1365-2966.2012.21784.x}}.

\bibitem{Mishra:2024qhc}
S.~S. Mishra, V.~Sahni, {New models of Quintessential Inflation featuring plateau and hilltop potentials} (2 2024).
\newblock \href {http://arxiv.org/abs/2402.04316} {\path{arXiv:2402.04316}}.

\end{thebibliography}

\end{document}